\documentclass[preprint]{aastex}
\usepackage{emulateapj5,apjfonts}
\journalid{}{}
\articleid{}{}
\slugcomment{}

\begin{document}

\def\plottwo#1#2{\centering \leavevmode
\epsfxsize=0.33\columnwidth \epsfbox{#1} \hfil
\epsfxsize=0.33\columnwidth \epsfbox{#2}}

\def\plot4#1#2#3#4{\centering \leavevmode
\epsfxsize=0.49\columnwidth \epsfbox{#1} \hfil
\epsfxsize=0.49\columnwidth \epsfbox{#2}\\
\epsfxsize=0.49\columnwidth \epsfbox{#3} \hfil
\epsfxsize=0.49\columnwidth \epsfbox{#4}}

\def\plotthree#1#2#3{\centering \leavevmode
\epsfxsize=0.30\columnwidth \epsfbox{#1} \hfil
\epsfxsize=0.30\columnwidth \epsfbox{#2} \hfil
\epsfxsize=0.30\columnwidth \epsfbox{#3}}

\title{Optical and Near-Infrared Imaging of the IRAS 1-Jy Sample of 
Ultraluminous Infrared Galaxies. II. The Analysis}

\author{S. Veilleux\altaffilmark{1,2}, D.-C. Kim\altaffilmark{3,4}, 
and D. B. Sanders\altaffilmark{2,5}}

\altaffiltext{1}{Department of Astronomy, University of Maryland,
College Park, MD 20742; E-mail: veilleux@astro.umd.edu}

\altaffiltext{2}{Visiting Astronomer, W. M. Keck Observatory, jointly
operated by the California Institute of Technology and the University
of California}

\altaffiltext{3}{Institute of Astronomy and Astrophysics, Academia
Sinica, P.O. Box 1-87, Nankang, Taipei, 115 Taiwan}

\altaffiltext{4}{Current address: School of Earth and Environmental
Sciences (BK21), Seoul National University, Seoul, Korea; E-mail:
dckim@astro.snu.ac.kr}

\altaffiltext{5}{Institute for Astronomy, University of Hawaii, 2680
Woodlawn Drive, Honolulu, HI 96822, and Max-Planck Institut for
Extraterrestrische Physik, D-85740, Garching, Germany; E-mail:
sanders@ifa.hawaii.edu}

\begin{abstract}
An $R$ \& $K^\prime$ atlas of the {\em IRAS} 1-Jy sample of 118
ultraluminous infrared galaxies (ULIGs) was presented in a companion
paper (Kim, Veilleux, \& Sanders 2002; Paper I). The present paper
discusses the results from the analysis of these images supplemented
with new spectroscopic data obtained at Keck.  All but one object in
the 1-Jy sample show signs of a strong tidal
interaction/merger. Multiple mergers involving more than two galaxies
are seen in no more than 5 of the 118 ($<$ 5\%) systems.  None of the
1-Jy sources is in the first-approach stage of the interaction, and
most (56\%) of them harbor a single disturbed nucleus and are
therefore in the later stages of a merger.  Seyfert galaxies
(especially those of type 1), warm ULIGs ($f_{25}/f_{60} > 0.2$) and
the more luminous systems ($>$ 10$^{12.5}$ $L_\odot$) all show a
strong tendency to be advanced mergers with a single nucleus.

The individual galaxies in the binary systems of the 1-Jy sample show
a broad distribution in host magnitudes (luminosities) with a mean of
--21.02 $\pm$ 0.76 mag. (0.85 $\pm$ $^{0.86}_{0.43}$ $L^\ast$) at $R$
and --23.98 $\pm$ 1.25 mag. (0.90 $\pm$ $^{1.94}_{0.61}$ $L^\ast$) at
$K^\prime$, and a $R$- or $K^\prime$-band luminosity ratio generally
less than $\sim$ 4.  Single-nucleus ULIGs also show a broad
distribution in host magnitudes (luminosities) with an average of
--21.77 $\pm$ 0.92 mag. (1.69 $\pm$ $^{2.25}_{0.97}$ $L^\ast$) at $R$
and --25.03 $\pm$ 0.94 mag. (2.36 $\pm$ $^{3.24}_{1.38}$) at
$K^\prime$.  These distributions overlap considerably with those of
quasars. The same statement applies to $R - K^\prime$ colors in ULIG
and quasar hosts.

An analysis of the surface brightness profiles of the host galaxies in
single-nucleus sources reveals that about 73\% of the $R$ and
$K^\prime$ surface brightness profiles are fit adequately by an
elliptical-like $R^{1/4}$-law.  These elliptical-like 1-Jy systems
have luminosity and $R$-band axial ratio distributions that are
similar to those of normal (inactive) intermediate-luminosity
ellipticals and follow with some scatter the same $\mu_e - r_e$
relation, giving credence to the idea that some of these objects may
eventually become intermediate-luminosity elliptical galaxies if they
get rid of their excess gas or transform this gas into stars.  These
elliptical-like hosts are most common among merger remnants with
Seyfert 1 nuclei (83\%), Seyfert 2 optical characteristics (69\%) or
mid-infrared ($ISO$) AGN signatures (80\%).  The mean half-light
radius of these ULIGs is $4.80 \pm 1.37$ kpc at $R$ and 3.48 $\pm$
1.39 kpc at $K^\prime$, typical of intermediate-luminosity
ellipticals. These values are in excellent agreement with recent
quasar measurements obtained at $H$ with $HST$, but are systematically
lower than other $HST$ measurements derived at $R$. The reason for
this discrepancy between the two quasar datasets is not known.

In general, the results from the present study are consistent with the
merger-driven evolutionary sequence ``cool ULIGs $\rightarrow$ warm
ULIGs $\rightarrow$ quasars.'' However, many exceptions appear to
exist to this simple picture (e.g., 46\% of the 41 advanced mergers
show no obvious signs of Seyfert activity).  This underlines the
importance of using a large homogeneous sample like the 1-Jy sample to
draw statistically meaningful conclusions; the problems of small
sample size and/or inhomogeneous selection criteria have plagued many
studies of luminous infrared galaxies in the past.
\end{abstract}


\section{Introduction}

Local ultraluminous infrared galaxies (ULIGs; log [L$_{\rm
IR}$/$L_\odot$] $\ge$ 12; $H_0$ = 75 km s$^{-1}$ Mpc$^{-1}$ and $q_0$
= 0) represent some of the best laboratories to study in detail the
violent aftermaths of galaxy collisions and their possible connection
with quasars and normal (inactive) elliptical galaxies. Recent deep
surveys with the Infrared Space Observatory ($ISO$) and sub-mm
ground-based facilities have revealed several distant ($z \approx$ 0.5
-- 4.0) infrared-luminous galaxies which appear to share many of the
properties of local ULIGs ($ISO:$ Kawara et al. 1998; Puget et
al. 1999; Matsuhara et al. 2000; Efstathiou et al. 2000; Serjeant et
al. 2001; Sanders et al. 2002, in prep.; $SCUBA:$ Smail, Ivison, \&
Blain 1997; Hughes et al. 1998; Blain et al. 1999; Eales et al. 1999;
Barger, Cowie, \& Sanders 1999). This extragalactic population of
high-$z$ infrared bright galaxies appears to dominate the far-infrared
extragalactic background and is probably a major contributor to the
overall star formation and metal enrichment history of the universe
(e.g., Smail et al. 1997; Hughes et al. 1998; Barger et al. 1998;
Genzel \& Cesarsky 2000; Rupke, Veilleux, \& Sanders 2002; Blain et
al. 2002).

Two crucial questions need to be answered to properly address the
issue of the origin and evolution of ULIGs: (1) What is the dominant
energy source in ULIGs: Starbursts or active galactic nuclei (AGNs)?
(2) Is the dominant energy source a function of the interaction/merger
phase in these systems?  Considerable progress has been made in recent
years in answering the first of these questions. Ground-based optical
and near-infrared spectroscopic survey of the 1-Jy sample has shown
that at least 25 -- 30\% of ULIGs show genuine signs of AGN activity
(Kim, Veilleux, \& Sanders 1998; Veilleux, Kim, \& Sanders 1999a;
Veilleux, Sanders, \& Kim 1997, 1999b; see also Goldader et al. 1995;
Murphy et al. 2001b). This fraction increases to 35 -- 50\% among the
objects with log [$L_{\rm IR}$/$L_\odot$] $\ge$ 12.3. Comparisons of
the dereddened emission-line luminosities of the BLRs detected at
optical or near-infrared wavelengths in the ULIGs of the 1-Jy sample
with those of optical quasars indicate that the AGN/quasar in ULIGs is
the main source of energy in at least 15 -- 25\% of all ULIGs in the
1-Jy sample. This fraction is closer to 30 -- 50\% among ULIGs with
$L_{\rm IR} > 10^{12.3}\ L_\odot$. These results are compatible with
those from recent mid-infrared spectroscopic surveys carried out with
{\it ISO} (e.g., Genzel et al. 1998; Lutz et al. 1998; Rigopoulou et
al. 1999; Tran et al. 2001). Indeed, a detailed object-by-object
comparison of the optical and mid-infrared classifications shows an
excellent agreement between the two classification schemes (Lutz,
Veilleux, \& Genzel 1999).  These results suggest that strong nuclear
activity, once triggered, quickly breaks the obscuring screen at least
in certain directions, thus becoming detectable over a wide wavelength
range.

Much of the research effort now focusses on answering the second, more
difficult question of a possible dependence of the energy source on
the interaction/merger phase. A large dataset already exists in the
literature on the morphology of luminous and ultraluminous infrared
galaxies. Optical studies have shown that the fraction of strongly
interacting/merger systems increases with increasing infrared
luminosities, reaching $>$ 95\% among the ultraluminous systems (e.g.,
Sanders et al. 1988a; Melnick \& Mirabel 1990; Murphy et al. 1996;
Clements et al. 1996; although see Lawrence et al. 1989; Zou et
al. 1991; Leech et al. 1994). The improved angular resolution and
lower optical depth in the near-infrared as compared to the optical
provides a cleaner view of the nuclear stellar distribution of
ULIGs. The infrared morphologies of ULIGs often look significantly
different from the optical images (e.g., Carico et al. 1990; Graham et
al. 1990; Eales et al. 1990; Majewski et al. 1993).  Over the past few
years, high-resolution imaging with {\em HST} has contributed
significantly to our knowledge of ULIGs (e.g., Surace et al. 1998;
Zheng et al. 1999; Scoville et al. 2000; Borne et al. 2000; Cui et
al. 2001; Farrah et al. 2001; Colina et al. 2001; Bushouse et
al. 2002). These studies have had great success characterizing the
central cores and massive stellar clusters in these objects, but the
images often do not reach faint enough flux limits to fully
characterize the host galaxies and associated tidal features. Adaptive
optics imaging with ground-based telescopes offers a promising new way
to acquire deep high-resolution of ULIGs, but this technique has so
far been used for only a limited number of objects (e.g., Surace \&
Sanders 1999, 2000; Surace, Sanders, \& Evans 2000, 2001).

In Kim, Veilleux, \& Sanders (2002; Paper I), we presented an $R$ \&
$K^\prime$ atlas of the {\em IRAS} 1-Jy sample of 118 ULIGs.  This
large and homogeneous dataset on the nearest (median redshift = 0.145)
and brightest such objects in the universe is particularly well suited
to address the issue of the origin and evolution of ULIGs (see Kim
1995 and Kim \& Sanders 1998 for a detailed description of the 1-Jy
sample; note that Galactic extinction is negligible for the 1-Jy
sample since $\vert b \vert$ $>$ 30$^\circ$ for all these
sources). Results derived from the 1-Jy sample should serve as a good
local baseline for studies on distant luminous infrared galaxies
planned in the years to come with $SIRTF$, $SOFIA$, and other
ground-based infrared and submm facilities.  In the present paper, we
analyze these images and combine the results of this analysis with
those derived from our optical and near-infrared spectroscopic data to
look for possible trends with optical and near-infrared morphological
and spectrophotometric parameters.  When possible, the results from
our analysis of the ground-based data on the 1-Jy sample are combined
with published results obtained at other wavelengths and compared with
the quickly growing set of high-quality data on optical and
infrared-bright QSOs (e.g., McLure et al. 1999; M\'arquez et al. 2000;
McLeod \& McLeod 2001; Canalizo \& Stockton 2000a, 2000b, 2001; Surace
et al. 2001; Dunlop et al. 2002 and references therein; see Stockton
1999 for a review of the data before 1999).  Given the high frequency
of mergers among ULIGs, the data on the 1-Jy sample also allow us to
examine the disk-disk merger scenario for elliptical galaxy formation
(e.g., Toomre 1977; Schweizer 1982; Wright et al. 1990; Scoville et
al. 1990; Stanford \& Bushouse 1991; Kormendy \& Sanders 1992; Doyon
et al. 1994; Genzel et al. 2001). The results of our analysis are
described in \S\S 2 -- 4. These results are then used in \S 5 to
discuss the origin and evolution of ULIGs and their possible
evolutionary link with quasars and elliptical galaxies. A summary is
presented in \S 6. Each object in the 1-Jy sample is described in
detail in an Appendix. Preliminary results of this study were
presented in Veilleux (2001).

We adopt $H_0$ = 75 km s$^{-1}$ Mpc$^{-1}$ and $q_0$ = 0.0 throughout this paper
and have converted the results from published papers to this cosmology
to facilitate comparisons. We also compare our absolute magnitudes
with the magnitudes corresponding to a $L^\ast$ galaxy in a Schechter
function description of the local field galaxy luminosity function.
The reader should be cautious when comparing luminosities from various
papers as they may be based on different definitions of $L^\ast$. The
values of the absolute magnitudes of a $L^\ast$ galaxy in various
bands, $M^\ast$, are now better constrained than in the past thanks to
large-scale surveys such as the 2dF Galaxy Redshift Survey
(2dFGRS). Here we adopt $M^\ast_R = -21.2$ mag. and $M^\ast_{K^\prime}
= -24.1$ mag.  The value of $M^\ast_R$ was derived using a $L^\ast$
magnitude of $M_r = -20.9$ mag. based on the results from the Las
Campanas Redshift Survey (Lin et al. 1996) and assuming a typical $r -
R$ = 0.3 mag.  for a moderately old stellar population at $z \approx
0.1$ (Surace 2002, private communication). A similar value for
$M^\ast_R$ is derived using $M^\ast_B = -20.1$ mag. (e.g, Mobasher et
al. 1993; Zucca et al. 1997; Ratcliffe et al. 1998; Folkes et
al. 1999) and adopting $B - R \approx 1.0$ mag. for typical field
galaxies at $z \approx 0.1$ (e.g. Fukugita et al. 1995), but keeping
in mind that this color term is strongly dependent on morphological
type and more specifically on the level of star formation activity in
the galaxy. The $K^\prime$ magnitude is based directly on the results
of Cole et al. (2001) from an infrared-selected subsample of the
2dFGRS. These results are also consistent to within $\pm$ $\sim$ 0.2
mag. with the numbers quoted from optically selected samples (e.g.,
Mobasher, Sharples, \& Ellis 1993;
Loveday 2000) and from the K-band surveys by Glazebrook et al. (1995)
and Gardner et al. (1997), after taking into account the various $k-$
and photometric corrections mentioned in Table 3 of Cole et
al. (2001). 

\section{Global and Nuclear Properties}

\subsection{Absolute Magnitudes}

The integrated (nucleus + host) $R$ and $K^\prime$ absolute magnitudes
and $R - K^\prime$ colors of the 1-Jy ULIGs were discussed briefly in
Paper I (\S 3.3). Figure 1 presents the global and nuclear (4-kpc
diameter) color -- magnitude diagrams of the 1-Jy sample.  As
mentioned in Paper I, the galaxies in our sample show a broad range of
integrated $R$ and $K^\prime$ luminosities $\sim$ 0.4 -- 120
$L^\ast$. There is no significant trend between global or nuclear $R$
magnitudes and $R - K^\prime$ colors, although there is a slight
tendency for the redder objects to have brighter $K^\prime$ magnitudes
especially when considering the nuclear quantities (Fig. 1). The more
luminous 1-Jy objects at $R$ and $K^\prime$ show a tendency to have
warm $IRAS$ 25~$\mu$m/60~$\mu$m colors ($f_{25}/f_{60} \ge 0.2$) and
to present Seyfert/AGN characteristics at optical and mid-infrared
wavelengths (Fig. 2). There also appears to be a weak positive
correlation between infrared luminosities and the global $R$ or
$K^\prime$ luminosities (see Figs. 2$g$ and 2$h$). Note, however, that
the 1-Jy sample is flux limited and is therefore subject to
luminosity-dependent distance biases: systems with $L_{\rm IR}$ $>$
10$^{12.3}$ $L_\odot$ are more distant on average ($<z>$ = 0.183 $\pm$
0.053) than ULIGs with $L_{\rm IR}$ $\le$ 10$^{12.3}$ $L_\odot$ ($<z>$
= 0.132 $\pm$ 0.029).  Consequently one should not be surprised to
find trends between global luminosities and infrared luminosities. No
trends are seen between global $R$ or $K^\prime$ luminosities and the
$IRAS$ 60~$\mu$m/100~$\mu$m colors or the strengths (equivalent
widths) of the stellar H$\beta$ and Mg~Ib features.

The ULIGs in the 1-Jy sample are on average less luminous at $R$ than
the quasars in the sample of Dunlop et al. (2002; see also Taylor et
al. 1996; McLure et al. 1999), although there is considerable overlap
between the two samples.  Adopting $H - K^\prime \approx 0.8$ for
typical low-redshift quasars (e.g., PG QSOs; Surace 2002, private
communication), we find that 1-Jy ULIGs are a better match in terms of
total $K^\prime$ luminosities to the set of ``low-luminosity'' quasars
of McLeod \& Rieke (1994a; $-23.2 \le M_B \le - 22.1$ mag. for our
adopted $H_0$) than the ``high-luminosity quasars'' of McLeod \& Rieke
(1994b) and McLeod \& McLeod (2001) with $M_B \le -23.2$ mag. However,
we note in \S 4.2 that the situation is reversed when considering only
the magnitudes of the host galaxies instead of the integrated
magnitudes.

\subsection{Colors}

The (nuclei of the) 1-Jy ULIGs are (significantly) redder than normal
galaxies [median $R - K^\prime$ and $(R - K^\prime)_4$ = 3.25 and 4.27
mag., respectively, versus 2.62 $\pm$ 0.34 mag. for normal galaxies;
see \S 3.3 in Paper I]. These colors are also redder than those of the
PG QSOs ($R - K^\prime$ = 2.7 mag. on average if redshifted to $z
\approx 0.1$ to match the average redshift of the 1-Jy ULIGs; Surace
2002, private communication). The red ULIGs with integrated $R -
K^\prime \ga 4$ mag. or nuclear $(R - K^\prime)_4 \ga 5$ mag. have a
tendency to host Seyfert nuclei and have warm $IRAS$
25~$\mu$m/60~$\mu$m colors (Fig. 3).  There is no significant
correlation between the integrated or nuclear $R - K^\prime$ colors
and the infrared luminosities, $IRAS$ 60~$\mu$m/100~$\mu$m colors, and
strengths of the stellar absorption features.  Ten single-nucleus
objects in the sample have $(R - K^\prime)_4 > 5$ mag. and would
therefore be considered extremely red objects (EROs; Elston, Rieke, \&
Rieke 1988) if it were not for the circumnuclear emission (which
presumably would not be detected at slightly higher redshifts;
neglecting $k$-corrections). Of these ten objects, only two
(F11119+3257 and F13218+0552) have {\em integrated} colors which meet
the ERO criterion; both of these objects are optically classified
Seyfert 1 galaxies.

\subsection{Compactness}

The compactness is defined as the ratio of the luminosity from the
inner 4 kpc to the total luminosity.  This quantity is physically
meaningful only for single-nucleus systems.  In Paper I, we mentioned
that the mean compactness of our galaxies at $K^\prime$
(0.36$\pm$0.17) is significantly higher than at $R$
(0.14$\pm$0.09). We find no significant trend between $K^\prime$
compactness and the infrared luminosity or $IRAS$ $f_{60}/f_{100}$
color within the 1-Jy sample.  However, Figure 4$b$ shows that compact
objects with $(L_{K^\prime})_4/L_{K^\prime} \ge \frac{1}{3}$ host most
of the optically identified Seyfert galaxies in our sample. Given the
well-known correlations between warm $IRAS$ 25 $\mu$m-to-60 $\mu$m
color, the presence of an AGN, and small equivalent widths of Mg~Ib
(especially in Seyfert 1s; e.g., Veilleux et al. 1995, 1999a), it is
not surprising to find a tendency for these compact objects to also
have $f_{25}/f_{60} > 0.2$ and EW(Mg~Ib) $\le$ 1 \AA. The trends with
$f_{25}/f_{60}$ and EW(Mg~Ib) are weaker than the trend with optical
spectral type so it appears that the presence of a Seyfert nucleus is
the key factor determining the value of the $K^\prime$
compactness. Similar trends are found when considering the $R$
compactness. These results may reflect the fact that Seyfert nuclei
contribute significantly to the emission at $R$ and $K^\prime$ or that
the host galaxies of Seyfert nuclei are more compact than those of
other galaxies.  Our data on host galaxy properties appear to rule
out this last possibility (see \S 4.5).

\section{Morphological Evidence for Galaxy Interactions}

\subsection{Frequency of Multiple Mergers}

A careful inspection of the images in the 1-Jy atlas (Fig. 1 in Paper
I) combined with Keck spectroscopy of field galaxies (described in
detail in Appendix A) indicates that nearly 100\% of the objects in
1-Jy sample show signs of a strong tidal interaction/merger in the
form of distorted or double nuclei, tidal tails, bridges, and
overlapping disks. The only possible exception is F23233+2817 (see
Appendix A). Nearly all of these objects are involved in the
interaction/merger of two, and no more than two, galaxies. At most
five systems out of 118 ($\la$ 5\%) are comprised of more than two
interacting systems: F13443+0802, F14394+5332, F15001+1433,
F17068+4027, and F21477+0502. In all five cases, triplets appear to be
involved although the physical connection between the various
components of the triplets have not all been confirmed. The fraction
(number) of multiple ($>$ 2) nuclei, double/pair mergers, and
single-nucleus merger remnants in the 1-Jy sample is 4\% (5), 39\%
(46), and 56\% (66, excluding F23233+2817), respectively.

The general lack of multiple ($>$ 2) systems in the 1-Jy sample is not
consistent with the relatively large fraction ($\sim$ 20\%) of
multiple systems claimed by Borne et al. (2000) and Cui et al. (2001)
from an optical $HST$ dataset. Part of the discrepancy may be in the
sample selection. Very few objects in the $HST$ sample are also part
of the 1-Jy sample. It is possible, though probably not very likely,
that the $HST$ sample studied by both Borne et al. and Cui et al.
just happens to include a larger fraction of multiple systems. Other
sources of uncertainties will affect the measured fraction of multiple
systems. The criteria for interaction differ slightly from one study
to the next due to different observational parameters (e.g., high
spatial resolution will favor the detection of compact nuclei {\em
and} stellar clusters; deep surface brightness limits will favor the
detection of faint tidal features) and the subjective nature of
deciding if an interaction is indeed taking place without accurate
spectroscopy of all galaxies in the system.  In our study, galaxies
are said to be interacting only if they show convincing signs of
large-scale tidal tails and/or present distorted outer isophotes.  In
many cases, Keck long-slit spectra of the galaxies in the systems were
obtained to confirm the interaction and to reject a few stellar
sources or foreground/background galaxies. Some of the systems are
accompanied by one or two galaxies at the same redshift as the $IRAS$
sources but with no convincing signs of interaction. These galaxies
are labeled ``G'' in Figure 1 of Paper I; they may be part of the same
group of galaxies as the $IRAS$ source but they do not appear to be
interacting with it (see Appendix A).

Another source of discrepancy is the definition of a galactic
nucleus. Cui et al. define nuclei as having $M_I < -17$ mag.  The HST
study by Surace et al. (1998) of 12 warm ULIGs has found several
genuine stellar superclusters with $M_I < -17$ mag. These would be
misclassified as nuclei according to the criterion of Cui et al.  For
comparison, all of the galaxies in the present paper have global $M_R
< -20.1$ mag. or $M_I \la -20.7$ mag. if we assume a typical $R - I$
$\approx$ 0.6 mag. (e.g., Fukugita et al. 1995).

Finally, the waveband sampled by the observations is an important
consideration when comparing the results from different studies. The
$HST$ data in Borne et al. and Cui et al. were obtained in the
$I$-band while our observations are taken at $R$ and $K^\prime$. The
near-infrared data are particularly useful to reduce the effects of
obscuration and thus to reveal the true stellar morphology of these
galaxies. This is made obvious when we compare our data with those of
Cui et al. for the few objects that we have in common.  Five of these
objects (F13539+2920, F14060+2919, F14202+2615, F22206--2715,
F22491--1808) are part of the group of 17 multiple systems of Cui et
al. In all cases except F14202+2615, $K^\prime$ images reveal the
presence of only one or two galactic nuclei (in F14202+2615, the third
galaxy shows no signs of interaction with the northern pair; see
Appendix A). In all four of these cases, bright star-forming knots
which were apparent in the $R$-band images disappear when observed at
$K^\prime$, indicating the lack of an obvious old stellar population in
these clusters.

A recent $HST$ $H$-band imaging study by Bushouse et al. (2002; see
also Colina et al. 2001) of a subset of objects from the sample of
Borne et al. (2002) and Cui et al. (2001) confirm our conclusions on
the frequency of multiple systems. Out of 27 ULIGs, Bushouse et
al. find only one candidate for multiple merging (IRAS 18580+6527; see
their Table 1), or $\sim$ 4\% of their sample. This fraction is
virtually identical to that found from our sample. This excellent
agreement emphasizes the importance of imaging galaxies at
near-infrared wavelengths rather than in the optical when trying to
determine the stellar distribution in these objects. We refer the
reader to Appendix A for a discussion of the four 1-Jy ULIGs
(F11095--0238, F13469+5833, F20414--1651, and F22206--2715) which were
also observed by Bushouse et al. (2002).

In the rest of this section, we use the $R$ and $K^\prime$ images of
the 1-Jy sample to look for possible trends in the properties of ULIGs
along the merger sequence. In the discussion below, three
morphological parameters are used to trace the merger sequence: the
observed (projected) nuclear separation, the length of tidal tail(s),
and the overall morphology of the system.

\subsection{Apparent Nuclear Separations}

Numerical simulations of merging disk systems indicate that the
nuclear separation between the two galaxies in the system is not
simply a monotonically decreasing function of time. The system often
goes through a ``pre-merger phase'' (Mihos 1999 calls it the ``hanging
out'' phase) characterized by two distinct galaxies with
well-developed tidal tails and bridges as a result of the first close
encounter (e.g., Barnes \& Hernquist 1992; Mihos \& Hernquist 1996;
Gerritsen 1997; Bekki et al. 1999). This result complicates the
interpretation of the measured (projected) nuclear separations ($NS$)
in the 1-Jy sample of galaxies since, without kinematic information,
we cannot tell if the two galaxies are pre- or post-apgalacticon
(farthest approach). The measurements are listed in Table 1 and shown
in Figure 5.  An excellent agreement is found between our measurements
and those of Surace (1998; adjusted to $q_0 = 0$). The distribution of
nuclear separations is highly peaked at small values but also presents
a significant tail at high values.  The median separation of the
sample is below the seeing limit ($\la$ 1$\arcsec$ $\sim$ 3 kpc).
More than 70\% (60\%) of the objects in the 1-Jy sample have nuclear
separations of less than 5.0 (2.5) kpc, but 16\% (7\%) of the objects
have separations in excess of 10 (20) kpc. These widely separated
pairs may be difficult to explain in the standard ULIGs scenario,
where the main activity is triggered in the late phase of the galaxy
merger (e.g., Mihos \& Hernquist 1996), unless the ULIG phase was
triggered by an earlier merger event which is not evident at the
present epoch.  This isssue is discussed in more detail in \S 5.1.

We have looked for possible correlations between the measured $NS$ and
the optical spectral type, infrared luminosity, $IRAS$
25~$\mu$m/60~$\mu$m and 60~$\mu$m/100~$\mu$m colors within the 1-Jy
sample.  We find a significant tendency for the Seyfert 1 galaxies
and, to a lesser extent, the Seyfert 2 galaxies to have small nuclear
separations. This is shown in Figure 6$a$. 100\% (68\%) of the Seyfert
1s (2s) have $NS \le$ 2.5 (5.0) kpc.  Similar trends are observed with
$IRAS$ $f_{25}/f_{60}$ colors (Fig. 6$b$): 79\% of the 14 warm objects
with $f_{25}/f_{60} \ga 0.2$ have nuclear separations of 2.5 kpc or
less (all of them have a nuclear separation of less than 10 kpc).

The trend between $NS$ and AGN activity is confirmed when we use the
$ISO$ spectral classification of Lutz et al. (1999). Twenty-five
objects are in common with our sample. Of the 10 objects classified as
AGN-dominated according to $ISO$, all of them have $NS \le 2.5$ kpc.
In contrast, starburst-dominated galaxies show a broad range of
nuclear separation.

No obvious trends are observed between the measured nuclear
separations and the equivalent widths of the H$\beta$ and Mg Ib
stellar absorption features, two traditional indicators of the
starburst age (Veilleux et al. 1995, 1999a).

As shown in Figure 7, single-nucleus sources and close pairs are more
common in the highest infrared luminosity bin (83\% of the sources
with log~[$L_{\rm IR}$/$L_\odot$] $\ge$ 12.5 have $NS \le$ 2.5 kpc
versus 58\% at log~[$L_{\rm IR}$/$L_\odot$] $<$ 12.24).  This trend
appears to extend to lower infrared luminosities: In a recent study of
56 galaxies with log~[$L_{\rm IR}$/$L_\odot$] = 11.10 -- 11.99, Ishida
(2002) finds that only about a third of the objects with log~[$L_{\rm
IR}$/$L_\odot$] $\ge$ 11.5 are advanced mergers with a single
nucleus. 

\subsection{Lengths of Tidal Tails}

The strength and length of tidal tails in galaxy mergers are strong
functions of the encounter geometry and the merger phase.  A strong
resonance between the orbital and rotational motions of stars and gas
produces strong tidal tails in prograde disks (e.g., Toomre \& Toomre
1972; Mihos \& Hernquist 1996). The lack of resonance in retrograde
disks inhibits the formation of extended tidal tails, but still
results in the formation of tidally induced spiral arms, central bars,
and diffuse tidal debris (e.g., Mihos \& Hernquist 1996).  Systems in
the pre-merger phase, where the two galaxies are still distinct but
have gone through the first encounter, are expected to have
well-developed tidal features. As the final merger takes place, the
tidal tails gradually disappear as the material in the tails is
accreted back onto the remnant or escapes the system altogether (e.g.,
Mihos, Dubinski, \& Hernquist 1998; Hibbard \& van Gorkom 1996).  

In an attempt to further constrain the phase of the interaction in
1-Jy ULIGs, we have measured the total projected lengths of the tidal
tails from our $R$-band images (the $K^\prime$ images are less
sensitive to these faint features, so they were not used for this
analysis). The measurements were done along each tail down to a
constant $R$ surface brightness of 24 magnitudes per square
arcsecond. The results are listed in Table 1 and shown in Figure
8. Slightly more than half of the objects in the 1-Jy sample have tail
lengths of 10 kpc or less, indicating an advanced merger. There is no
obvious correlation between tail lengths and optical spectral types,
infrared luminosities, and $IRAS$ 25~$\mu$m/60~$\mu$m and
60~$\mu$m/100~$\mu$m colors. Note, however, that 70\% of the objects
with $f_{25}/f_{60} \ge 0.2$ have tail lengths of 10 kpc or less. This
fraction is 60\%, 36\%, 52\%, and 66\% when considering Seyfert 1s,
Seyfert 2s, LINERs, and H~II galaxies, respectively.

\subsection{Interaction Classification}

Using a classification scheme first proposed by Surace (1998) and
based on the results of published numerical simulations (e.g., Barnes
\& Hernquist 1992; Mihos \& Hernquist 1996; Gerritsen 1997), we have
classified each object in the 1-Jy sample (except for the five triplet
candidates and F23233+2817, which is an isolated spiral galaxy)
according to their morphology:

\begin{itemize}
\item[I.] {\em First Approach.} This phase refers to the earliest
stage of the interaction, prior to the first close passage of the
galaxies, when the galaxy disks remain relatively unperturbed and
separate and hence do not show evidence for tidal tails or bridges.

\item[II.] {\em First Contact.} At this stage, the disks overlap but
strong bars and tidal tails have not yet formed. These are the primary
morphological differences between this phase and the early merger
phase (Phase IV below). 

\item[III.] {\em Pre-Merger.} This stage is characterized by two
identifiable nuclei with well-developed tidal tails and
bridges. While our imaging observations cannot prove that the
double-nucleus systems will necessarily merge, the presence of
ultraluminous activity combined with the prevalence of mergers in
ultraluminous galaxies, as well as the presence of tidal tails and
other structure interconnecting the nuclei in every case, would seem
to strongly indicate eventual merger. Within this category the
systems were further divided into two subclasses based on their
projected separations: \\ 
\indent IIIa. {\em Wide Binary.} Systems with apparent separation $>$ 10 kpc.\\
\indent IIIb. {\em Close Binary.} Systems with apparent separation $\le$ 10 kpc.\\ 
Note that these subclasses are an approximation only to the merger
phase since without detailed kinematic information for many of these
systems, we cannot tell if the two galaxies are pre- or
post-apgalacticon (farthest approach).

\item[IV.] {\em Merger.} This stage occurs after the nuclei have
apparently coalesced. These systems have prominent tidal features, but
only one nucleus can be detected at optical and near-infrared
wavelengths. Additionally, the sole galaxy core is often noticeably
extended and tends to be cut by many dust lanes. Following Surace
(1998), we further subdivide this class into two categories: \\
\indent IVa. {\em Diffuse Merger.} Systems with
$(L_{K^\prime})_4/L_{K^\prime} < \frac{1}{3}$.\\
\indent IVb. {\em Compact Merger.} Systems with
$(L_{K^\prime})_4/L_{K^\prime} \ge \frac{1}{3}$.\\
Objects in class IVa consists of systems that have diffuse, extended
central regions typically composed of several smaller extended
emission region which appear to be cut by dust lanes. Objects in class
IVb are dominated, particularly at long wavelengths ($K^\prime$), by a
single point source. As pointed out by Surace (1998), there
is no a priori reason to believe that the stage IVb systems are any
older than the stage IVa systems, other than the fact that stage IVa
resembles stage III, and stage IVb resembles stage V, and that stage V
is older than stage III based on the appearance of tidal structure.

\item[V.] {\em Old Merger.} These are systems which do not show any
direct {\em unmistakable} signs of tidal tails, yet have disturbed
central morphologies similar to those of the merger stage IV systems,
particularly those with knots of star formation. At this stage, the
surface brightness of the tidal features in the system has fallen
below our detection limit, leaving nothing but the high surface
brightness relaxed merger remnant core visible.
\end{itemize}

The results of the classification are given in Table 1. The
distribution is shown in Figure 9. None of the 1-Jy sources appears to
be in the early stage of the interaction (I and II). Systems comprised
of two distinct objects all show prominent tidal features. Most (56\%)
of the ULIGs harbor a single nucleus and are therefore in the later
stages of the merger.  This is a very significant result: the ULIG
phase is triggered only after the first close encounter between the
two galaxies and generally when the two galaxies have coalesced into a
single object. The fraction of singles to doubles in the 1-Jy sample
varies with infrared luminosity, increasing steeply for the most
extreme ULIGs with log [$L_{\rm IR}$/$L_\odot$] $>$ 12.5
(Fig. 10$a$). This fraction seems to decrease when considering lower
luminosity systems with log [$L_{\rm IR}$/$L_\odot$] = 11.00 -- 11.99
(Ishida 2002). This class of objects is generally comprised of two
distinct galaxies often widely separated and sometimes showing no
obvious tidal features. Many of these luminous infrared galaxies (i.e.
LIGs: log[$L_{\rm IR}/L_\odot$] = 11 -- 11.99) therefore fall into
interaction classes I -- III.

Figure 10$b$ shows that the interaction class is a strong function of
the optical spectral type.  We find that all ten Seyfert 1s are
advanced mergers (class IVb or V) and that half (50\%) of the Seyfert
2s are early or advanced mergers (class IVa, IVb, or V). This is
consistent with the results of our analysis of their nuclear
separations (\S 3.2). On the other hand, LINERs and H~II region-like
galaxies show no preference in interaction class (other than to avoid
classes I and II).  A strong correlation is also found when we consider
the warm objects with $f_{25}/f_{60}$ $>$ 0.2 (Fig. 10$c$): Nearly
80\% of these systems (11 objects out of 14) are advanced mergers
(class IVb or V).

We find no obvious trend between the interaction class and $IRAS$
60~$\mu$m/100~$\mu$m color. The strengths of the H$\beta$ or Mg~Ib
stellar features measured in the nuclei of the 1-Jy objects (either
considering all the objects in the sample or only in systems optically
identified as starburst galaxies) show no correlation with interaction
class.  This negative result indicates that traditional nuclear
indicators of the starburst age are {\em not} particularly good
indicators of the merger phase or the epoch of the merger
event. Detailed studies of star clusters in merger-induced starbursts
confirm the existence of a broad range of stellar age which reflects
the complex star formation history of these systems (e.g., Surace et
al. 1998; Whitmore et al. 1999; Gallagher et al. 2001; Zhang, Fall, \&
Whitmore 2001, and references therein).

\section{Properties of the Host Galaxies}

\subsection{Surface Brightness Profiles}

In an attempt to characterize the brightness distribution of the host
galaxies in the ULIGs of the 1-Jy sample, we have performed the
following isophotal analysis. First, a central point source was
removed from the centers of the objects with obvious central excess
emission.  For this procedure, a stellar source within the field of
view of the detector was shifted and scaled appropriately in intensity
to produce a smooth and monotonic profile in the central regions of
the galaxies. Ellipses were then fitted to the residuals using the
standard Fourier expansion (e.g., Binney \& Merrifield 1998)
\begin{equation}
\delta(\phi) = \bar{\delta} + \sum_{n = 1}^{4}[A_n~{\rm cos}(n\phi) + B_n~{\rm sin}(n\phi)], 
\end{equation}
where $\delta(\phi) \equiv R_i(\phi) - R_e(\phi)$ is the distance at
position angle $\phi$ between the radii of the corresponding points on
the ellipse and on the isophote, and $A_n$ and $B_n$ are the
coefficients of the Fourier expansion. The typical error for the
ellipse fitting is estimated to be less than 0.1
mag. arcsec$^{-2}$. Unless otherwise noted, the error bars on the
various quantities derived from these fits are equivalent to 1-sigma
values and were directly derived from the IRAF routines (see the IRAF
help page and in particular Jedrezejewski 1987 and Busko 1996 for a
more detailed discussion of the error analysis).

The $R$ and $K^\prime$ surface brightness profiles derived from the
ellipse fitting (not corrected for cosmological dimming) are presented
in Figure 11 using two different scales on the horizontal axis: a
logarithmic scale on the left and a linear$^{1/4}$ scale on the
right. An exponential disk and an elliptical-like object with a de
Vaucouleurs profile would produce straight lines in the left and right
panels, respectively. A vertical arrow in each panel represents the
size of the seeing disk.  Least-square fits to the profiles were
carried out to determine whether an exponential or de Vaucouleurs
profile best fits the observed profile at $R$ and $K^\prime$. To avoid
complications due to the seeing, the least-squares fit was applied to
the data points only at radii larger than twice the seeing disk. The
value of the reduced chi-square is indicated in the upper right hand
corner of each panel in Figure 11.  The results from the fits are
summarized in Table 2. In the calculation of the reduced chi-square,
the number of degrees of freedom is the number of data points in the
observed radial profile minus the number of free parameters in the fit
(= 2 for both the exponential and de Vaucouleurs profiles). A reduced
chi-square of order unity indicates a good fit. If the chi-square is
much larger than unity, then the fit is poor. If the reduced
chi-square is much less than unity, then there is not enough data
points to carry out the fit and the parameters derived from the fit
are not reliable.

Figure 12$a$ shows the fraction of single-nucleus objects (i.e. merger
remnants with interaction class IV or V) in the 1-Jy sample with pure
disk-like (D: 2\%), pure elliptical-like (E: 35\%),
elliptical/disk-like (E/D: 38\%), and ambiguous (A: 26\%) radial
profiles.  ``E/D'' profiles are defined as being equally well fit by
an exponential or de Vaucouleurs profile, while neither fits galaxies
with ``ambiguous'' profiles. Both E/D and A profiles are quite common
among ULIGs.  The large number of E/D galaxies in our sample probably
reflects the limitations of our data: images with a larger dynamical
range in surface brightness (i.e. higher spatial resolution to probe
the nuclear region and fainter surface brightness limits to map the
outer portions of the galaxies) would help rule out one type of
profile over the other.  The large fraction of ambiguous profiles
probably reflects the residual effects of the galaxy interaction on
the morphology of the merger remnants. If we combine the E and E/D
galaxies, we find that 73\% of the single-nucleus objects are fit
adequately by a de Vaucouleurs profile.

The fraction of ambiguous profiles among double-nucleus systems
(Fig. 12$b$) is significantly larger than among single-nucleus
systems. Several binaries show obvious tidal tails and bridges which
distort the radial profile. Among the other binaries, we find that the
fraction of elliptical-like galaxies is slightly larger than disk-like
hosts, perhaps an indication that the progenitor galaxies generally
have strong spheroidal components (e.g., bulges).

Strong correlations are found between the type of host galaxy profile
and the nuclear and global properties of single-nucleus systems. In
the following analysis, we take the conservative approach to
double-count each E/D object as an E object {\em and} a D object (it
is conservative because it has the effect of smoothing out possibly
stronger trends with host galaxy profiles). Figure 13$a$ shows that
the hosts of most merger remnants with Seyfert 1 (83\%) or Seyfert 2
(69\%) optical characteristics are E-like. This preponderance of
E-like hosts among AGN mergers is also seen when using the {\em ISO}
classification (Fig. 13$b$; 8 out of 10 or 80\% of $ISO$ AGNs are
E-like). A similar trend is seen among warm objects (10 out of the 13
objects with $f_{25}/f_{60} \ge 0.2$ or 77\% are E-like;
Fig. 13$c$). Starburst and LINER systems, on the other hand, show no
preference between disk-like and E-like hosts.

The nature of the host does not seem to depend strongly on the
infrared luminosity, the 60~$\mu$m/100~$\mu$m ratio, or the strengths
of the stellar H$\beta$ and Mg~Ib features. But, as expected in the
merger scenario for elliptical galaxy formation, E- and E/D profiles
are more common in the later phases of interaction [class V (27\%) and
especially class IVb (49\%); Fig. 13$d$], while ambiguous profiles are
more often seen in early mergers (IVa, 59\%), a sign perhaps that many
of these systems are still showing the damages caused by the recent
merger event.

\subsection{Absolute Magnitudes}

The absolute $R$ and $K^\prime$ magnitudes of the host galaxies in the
1-Jy sample are listed in the last two columns of Table 2.  Median and
mean values are listed in Table 3. The distributions for double- and
single-nucleus systems are shown separately in Figure 14. The median
and mean $R$-band luminosities of the individual galaxies involved in
binaries are 0.86 and 0.85 $\pm$ $^{0.86}_{0.43}$ $L^\ast$ (1
$\sigma$), respectively. At $K^\prime$, the median and mean for these
objects are 0.91 and 0.90 $\pm$ $^{1.94}_{0.61}$ $L^\ast$. For
single-nucleus systems, the median and mean $R$-band luminosities are
1.69 and 1.69 $\pm$ $^{2.25}_{0.97}$ $L^\ast$, and the median and mean
$K^\prime$-band luminosities are 1.92 and 2.36 $\pm$
$^{3.24}_{1.38}$). Previous studies have obtained similar host galaxy
magnitudes: $\sim$ 2.5 $L^\ast_B$ (Armus, Heckman, \& Miley 1990),
$\sim$ 3 $L^\ast_r$ (ranging from $\sim$ 1 -- 24 $L^\ast_r$) for the
ULIGs in the 2-Jy sample (Murphy et al. 1996), $\sim$ 2.4 $L^\ast_H$
for warm ULIGs from the $IRAS$ BGS sample (Surace \& Sanders 1999),
and $\sim$ 1 -- 2.5 $L^\ast_H$ for cool ULIGs from the BGS and 1-Jy
samples (Surace, Sanders, \& Evans 2000).  Colina et al. (2001) derive
noticeably fainter, sub-$L^\ast$, host galaxy magnitudes ($<M_H>$ =
--23.5 $\pm$ 0.7 mag. scaled down to our cosmology) for 27 mostly cool
ULIGs. The $H$-band $HST/WFPC2$ data used for this analysis are shown
in Bushouse et al. (2002). Direct comparisons with our data for the
four objects in common with the 1-Jy sample indicate that much of the
diffuse low surface brightness features in our data are not visible in
the snapshot images of Bushouse et al. (2002). This underestimate of
the light contribution from the host galaxies explains the fainter
host galaxy magnitudes derived by Colina et al. (2001).

The large number of galaxies in the 1-Jy sample allows us to look for
trends between host galaxy luminosities and infrared luminosities,
optical spectral types, and $IRAS$ colors. Focussing on single-nucleus
systems, there appears to be a tendency to find luminous hosts in the
most infrared luminous systems of the 1-Jy sample (see Fig. 15$a$ and
Table 3).  This is due to the fact noted in \S 2.1 that systems with
$L_{\rm IR}$ $>$ 10$^{12.3}$ $L_\odot$ are more distant on average
($<z>$ = 0.183 $\pm$ 0.053) than ULIGs with $L_{\rm IR}$ $\le$
10$^{12.3}$ $L_\odot$ ($<z>$ = 0.132 $\pm$ 0.029).

Figure 15 and Table 3 indicate that the most luminous hosts also tend
to be found in systems with warm $IRAS$ 25~$\mu$m/60~$\mu$m colors
($\ge$ 0.2) and AGN characteristics. The hosts of the warm objects are
0.6 -- 0.7 magnitudes brighter on average than the hosts of the cool
objects, although both distributions overlap considerably. In this
case, the difference in mean host luminosity cannot be attributed to a
difference in redshift since both classes of objects have virtually
the same median redshift (= 0.148). The same is true when comparing
the Seyfert and non-Seyfert populations ($<z>$ = 0.143 $\pm$ 0.056 and
0.150 $\pm$ 0.041, respectively), yet the host galaxies of Seyferts
are brighter by about 0.8 -- 1.0 mag. on average at both $R$ and
$K^\prime$ than the hosts of LINERs and H~II region-like galaxies. K-S
tests between the distributions of Seyferts vs non-Seyferts and warm
vs cool ULIGs confirm these differences (P[null] $<$ 10$^{-5}$ and
10$^{-2}$, respectively). These results can be attributed to real
differences in the host luminosities of AGNs vs non-AGNs, or may be
due to a prominent starbursting population in warm Seyferts
contributing significantly at $R$ and $K^\prime$.

One should be cautious when interpreting these results, however. An
object-by-object comparison with the $H$-band data of Surace \&
Sanders (1999) suggests that the host luminosities measured in a few
of the warm Seyferts appear to be slightly higher than those measured
from AO imaging (assuming $H - K^\prime \la 0.5$ for the host; see
below). This may indicate that we are slightly underestimating the
contribution from the central point source in some of these objects
(the possibility that the AO imaging of Surace \& Sanders is
underestimating the host luminosities appears less likely because the
exposure times for these observations were fairly long and therefore
the surface brightness limits of the data were fairly deep). The PSF
subtraction in our data is more difficult in the warm, highly
nucleated systems of the 1-Jy sample than in the cooler, more diffuse
systems; this may introduce a slight bias which may account for the weak
trends between host luminosities, $IRAS$ colors and spectral types.
These issues are discussed in more detail in \S\S 4.5, 5.2, and 5.3.

Next we compare the host magnitudes of ULIGs with the results from
studies of quasar host galaxies. At $R$, the host galaxies in ULIGs
appear slightly less luminous on average than the HST measurements of
Dunlop et al. (2002) on radio-quiet quasars (--22.35 $\pm$ 0.15 mag.,
adjusted to $H_0$ = 75 km s$^{-1}$ Mpc$^{-1}$) and radio-loud quasars
(--22.85 $\pm$ 0.10 mag.).  Note, however, that ULIG hosts span a
broad range in luminosity so there is considerable overlap between
ULIGs and the other classes of objects.  Given the correlations
between host luminosities, infrared luminosities, $IRAS$
25~$\mu$m/60~$\mu$m colors, and optical spectral types discussed in
the previous paragraphs, we find that if we focus our analysis on the
subset of objects with log~[L$_{\rm IR}$/$L_\odot$] $\ge$ 12.3,
$f_{25}/f_{60} \ge$ 0.2, and with Seyfert characteristics at optical
wavelengths, the apparent shift in host galaxy $R$ magnitudes between
these ULIGs and the quasar dataset of Dunlop et al. disappears.

To directly compare our data with the ground-based results of McLeod
\& Rieke (1994a, 1994b) and HST/NICMOS data of McLeod \& McLeod (2001)
on quasars, we need to transform the $H$-band magnitudes listed in
these papers into $K^\prime$ magnitudes. For an old stellar population
at $z = 0$, $H - K^\prime = 0.16$ (Surace 1998), but $H - K^\prime
\approx 0.3 - 0.4$ if $<z> = 0.1 - 0.2$ (e.g., Lilly \& Longair
1984). Given these corrections, the low-luminosity sample of quasars
of McLeod \& Rieke (1994a) have an average $K^\prime$ host magnitude
$M_{K^\prime} \approx -24.3 \pm 0.6$, while the high-luminosity
quasars have $M_{K^\prime} \approx -25.1 \pm 0.7$. The host galaxies
in the 1-Jy ULIGs are therefore comparable in luminosity to those of
the {\em high}-luminosity quasars. So even though we found in \S 2.1
that the 1-Jy ULIGs have lower {\em integrated} $K^\prime$
luminosities than the high-luminosity quasars of McLeod \& Rieke
(1994b), we find that they have host galaxies with similar $K^\prime$
luminosities. This good agreement is also found in the more recent
analysis of the high-luminosity sample by McLeod \& McLeod (2001).

The claim by McLeod \& Rieke (1994a, 1994b; ground-based $H$-band
data) and McLeod, Rieke, \& Storrie-Lombardi (1999; HST/NICMOS data)
of a trend between nuclear and host galaxy luminosities in quasars has
recently come under question. The $R$-band results of Dunlop et
al. (2002) do not seem to confirm this trend.  A recent $K$-band
ground-based study of 14 very high luminosity ($M_V \la -25$)
radio-quiet quasars with 0.26 $\ge$ $z$ $\ge$ 0.46 by the same group
(Percival et al. 2001) find $<M_K> \approx -24.3 \pm 0.04$, a value
closer to the host galaxy magnitudes of the {\em low}-luminosity
quasar sample of McLeod \& McLeod (1994a) and significantly {\em less}
luminous on average than the host galaxies of the 1-Jy ULIGs.
However, as stated for instance by Dunlop et al.  (2002) when
discussing the study of Taylor et al. (1996), the results from
ground-based studies of quasars beyond $z = 0.15$ like that of
Percival et al. should be treated with caution given the technical
difficulties involved in removing the nuclear PSF to characterize the
underlying host galaxy of these objects (see also discussion in \S
4.5).

Adaptive optics imaging of quasars alleviates the problems associated
with PSF subtraction in ground-based data. The recent study by Surace
et al. (2001) of 18 PG quasars with strong far-infrared excesses finds
$H$-band host galaxy luminosities that range from 0.5 to 7.5 $L^\ast$,
with a mean of 2.3 $L^\ast$, consistent with the average $K^\prime$
host galaxy luminosity of 1-Jy ULIGs measured in the present study
and similar to that found by McLeod \& McLeod (2001) among quasars.

\subsection{Luminosity Ratios in Binary Systems}

The mass ratio of a binary system of galaxies is an important
parameter in predicting the amount of ``damage'' incurred in the
interaction. Minor mergers comprised of objects with mass ratios of
order $\sim$ 10 are predicted to have only benign effects on the main
component of the system (e.g., Walker, Mihos, \& Hernquist 1996; Bendo
\& Barnes 2000; Naab \& Burkert 2001). Major mergers with mass ratio
of $\sim$ 3 or less appear to be needed to create a ULIG (e.g., Mihos
\& Hernquist 1996). The kinematic properties of ULIGs also appear
consistent with this scenario (e.g., Genzel et al. 2001). Given our
lack of detailed kinematic data, we have no reliable way to determine
the total mass of each galaxy in our binary systems. We make the
assumptions that the total masses scale linearly with the stellar
masses and that the stellar masses scale with the $R$ or $K^\prime$
luminosities of the host galaxies. Figure 16 shows the distributions
of $\Delta M(R)$ and $\Delta M(K^\prime)$ (equivalent to luminosity
ratios) of the Class III systems in the 1-Jy sample for which we were
able to reliably separate each galaxy (i.e. most of them are Class
IIIa). Most systems have a magnitude difference of less than 1.5 or a
luminosity ratio of less than 4, in general agreement with the
predictions from numerical simulations of galaxy mergers if mass
indeed scales linearly with luminosity. The magnitude difference is
slightly larger at $K^\prime$ than at $R$.  The two objects with the
largest $K^\prime$ magnitude differences are F00456--2904 and
F12127--1412.

\subsection{Colors}

In Figure 17, we find no significant correlation between the nuclear
color and the underlying galaxy color $(R - K^\prime)_{\rm host}$
(P[null] $>$ 0.01 for H~II, LINER, and Seyfert galaxies). In contrast
to the nuclear colors, the median color of the underlying galaxies of
ULIGs is only slightly redder than that of normal galaxies or the host
galaxies of quasars [$(R - K^\prime)_{\rm host}$ = 3.0 for either H~II
galaxies, LINERs, or Seyfert 2 galaxies compared to 2.6 and 2.5 for
normal galaxies and quasars, respectively; de Vaucouleurs \& Longo
1988; Dunlop et al. 2002]. This result confirms that the obscuring gas
and dust is preferentially concentrated within the inner 4 kpc of
ULIGs (e.g., Scoville et al. 2000). The slightly redder colors of
Seyfert 1 hosts [with median $(R - K^\prime)_{\rm host}$ of 3.4] may
be due to contamination at $K^\prime$ by hot dust emission from the
AGN or residuals from the bright central source subtraction
procedure. The colors of ULIG hosts are similar to those found by
Surace et al. (2001) among infrared-excess PG quasars.

\subsection{Half-Light Radii and Surface Brightnesses}

The seeing-deconvolved half-light radii, $r_e$, of the elliptical-like
host galaxies in our sample are listed in Table 2 and plotted in
Figure 18. The median and mean of $r_e$ at $R$ are 4.60 kpc and 4.80
$\pm$ 1.37 kpc (1 $\sigma$), while at $K^\prime$ we find 3.55 kpc and
3.48 $\pm$ 1.39 kpc (1 $\sigma$).  These half-light radii are
virtually indistinguishable from those measured at $H$ by McLeod \&
McLeod (2001) in the eight E-like ``high-luminosity'' quasars of their
sample ($<r_e>$ = 3.39 $\pm$ 1.90 kpc; Fig. 18$b$). On the other hand,
the values measured by Dunlop et al. (2002) among high-luminosity
radio-quiet quasars (7.63 $\pm$ 1.11 kpc adjusted to $H_0$ = 75 km
s$^{-1}$ Mpc$^{-1}$) and radio-loud quasars (7.82 $\pm$ 0.71 kpc)
appear to be significantly larger than those measured in the 1-Jy
ULIGs, although there is some overlap between the two
distributions. We do not try to compare our results with the
half-light radii measured by Taylor et al. (1996) at $K$ because these
data have proven to be unreliable: An object-by-object comparison
between the data of Taylor et al. (1996) and Dunlop et al. (2002)
reveal discrepancies of up to a factor of nearly $\sim$ 25 (e.g.,
1549+203) in $r_e$, which cannot be attributed to color gradients ($R
- K^\prime$) but rather are probably due to incorrect PSF subtraction
in the ground-based data of Taylor et al.  Similar technical
difficulties are likely to affect to some extent the ground-based data
of Percival et al. (2001), given the relatively high redshifts of
their sources.

Combining measurements of the half-light radius and mean $R$-band
surface brightness within half-light radius, $\mu_e$, Dunlop et
al. (2002) have argued that the host galaxies of luminous quasars
follow the same $\mu_e$ -- $r_e$ relation as that of normal (inactive)
ellipticals. The surface brightnesses of ULIGs listed in Table 2 have
been corrected for cosmological dimming [$\propto (1 + z)^4$] and are
compared with those of elliptical galaxies and quasars in Figure
19$a$.  The elliptical-like (E and E/D) host galaxies of the 1-Jy
ULIGs have slightly brighter $R$-band surface brightnesses than the
hosts of the quasars studied by Dunlop et al. Nevertheless, the
elliptical-like host galaxies of ULIGs appear to follow the $\mu_e$ --
$r_e$ relation of normal ellipticals, although with considerable
scatter.  In this figure, we used the fit of Hamabe \& Kormendy (1989)
and shifted it in magnitude by $V - R$ = 0.5 mag (keeping the slope
constant). Seyfert 1 galaxies show a tendency to fall slightly above
the data for non-Seyfert 1s. The source of this tendency is discussed
below.

Interpretation of our data at $K^\prime$ is more difficult because
$r_e$ and $\mu_e$ have been measured reliably in only 18 ULIGs of our
sample (this analysis was carried out only when the $K^\prime$ images
were deep enough to reliably measure the surface brightness profiles
out to a radius comparable to the optical radius). The results are
presented in Figure 19$b$. A weak anti-correlation is found between
apparent surface brightnesses and galaxy sizes. This anti-correlation
is stronger if we exclude Seyfert 1 galaxy hosts from the fit (solid
line in Fig. 19$b$).  The data for the Seyfert 1s should be
considered with caution since they are significantly more sensitive to
our PSF subtraction procedure than for the other objects.  The bright
$\mu_e$ of the three Seyfert 1 host galaxies near the top of Figure
19$b$ may be due to positive residuals from the AGN.

We first compare our $K^\prime$ results with the $V$-band $\mu_e$ --
$r_e$ relation of Hamabe \& Kormendy (1987) for normal ellipticals
and find a generally good agreement with the non-Seyfert 1 data if we
assume a constant color $V - K^\prime$ = 3.7 mag. This color term is
slightly redder than that of normal elliptical galaxies (e.g., Nolan
et al. 2001; Fukugita et al. 1995), possibly because the hosts of
ULIGs are dustier than elliptical galaxies and star formation (red
supergiants) contributes at $K^\prime$.  [We did not use the $B$-band
data of Bender, Burstein, \& Faber (1992) and Faber et al. (1997) for
this comparison to avoid the even larger color correction.]  We also
compare our results with the $K$-band data of Pahre (1999). The slope
of the $\mu_e - r_e$ relation derived from these data (dotted line in
Fig. 19$b$) is shallower than that of Hamabe \& Kormendy (1987).
Unfortunately, a comparison with quasar hosts is not possible because
of the lack of reliable quasar host data in the near-infrared (McLeod
\& McLeod 2001 list $r_e$ but not $\mu_e$ in their paper, and the data
of Taylor et al. are not considered reliable for the reasons discussed
in the first paragraph of this section).

There is no obvious correlation of the $R$-band $r_e$ (or $\mu_e$)
with $IRAS$ 25~$\mu$m/60~$\mu$m and 60~$\mu$m/100~$\mu$m colors,
infrared luminosities, optical and mid-infrared spectral types, and
the strengths of the stellar H$\beta$ and Mg~Ib absorption features.
There are too few measurements of $r_e$ at $K^\prime$ to search for
statistically significant trends with this parameter. Recall, however,
the existence of correlations between the host galaxy $R$ and
$K^\prime$ luminosities ($\propto \mu_e r_e^2$) and infrared
luminosities, $IRAS$ 25~$\mu$m/60~$\mu$m colors, and AGN
characteristics noted in \S 4.2 (Fig. 15 and Table 3).

\subsection{Bars, Axial Ratios, and Boxiness}

The $\sim$ 1$\arcsec$ spatial resolution of our images prevents us from
studying in detail the cores of ULIGs on scales smaller than a few
kpc.  However, a few statements can be made on galactic structures
seen beyond the seeing disk.  Only three objects in our sample,
F14394+5332 W (triplet), F23233+2817 (isolated spiral), and
F23327+2913 S (widely separated pair), show evidence for
bar-like structures which extend 8 kpc (P.A. = 90$\arcdeg$), 12 kpc
(20$\arcdeg$), and 11 kpc (120$\arcdeg$), respectively. None of the
single-nucleus merger remnants shows a strong large-scale bar.

Our ellipse-fitting routine of the surface brightness distribution
provides quantitative information on the axial ratio ($b/a$) of the
host galaxies on scales larger than twice the seeing disk. Figure 20
shows the radial dependence of the axial ratio for the 58
single-nucleus mergers of our sample with ``E'' or ``E/D'' profiles
(F23233+2817 is not included in this subset of objects because it is
an isolated spiral galaxy).  In the majority of cases, the axial ratio
is constant with radius to within $\pm$ 0.15. We plot in Figure 21 the
distribution of $b/a$ measured at $3 \times r_e$.  The median and mean
of this quantity are 0.77 and 0.75 $\pm$ 0.15 (1 $\sigma$),
respectively. This is similar to the average values measured in
normal elliptical galaxies (e.g., Ryden 1992) and in the hosts of
quasars (Dunlop et al. 2002).

Surface photometry of elliptical galaxies often shows isophotes that
deviate slightly from a perfect ellipse. The departure of the
isophotes from a perfect ellipse is measured by the $A_4$ coefficient
of the cos 4$\phi$ term in equation (1). Typical values for
ellipticals normalized to the major-axis length, $a$, range from
--0.02 to +0.03 (e.g., Bender et al. 1989), where the negative values
indicate boxy isophotes and positive values indicate disky
isophotes. Ellipticals with boxy isophotes are generally luminous
(massive) radio-loud ellipticals with X-ray halos, very slow stellar
rotation and a distinct cuspy cores, whereas ellipticals with disky
isophotes are low-to-moderate luminosity (mass) lenticular and
elliptical galaxies which are generally radio-quiet, show little or no
X-ray emission, and exhibit significant rotational support and a
power-law surface brightness profile on small scales (e.g., Kormendy
\& Bender 1996; Faber et al. 1997; Pahre, de Carvalho \& Djorgovski
1998 and references therein).

The radial profiles of the normalized $R$-band boxiness parameter in
the single-nucleus ULIGs of our sample are shown in Figure 22. Several
of these galaxies have large absolute values of $A_4/a$ (generally
from --0.05 to +0.05) compared to normal ellipticals. Significant
$A_4/a$ gradients are seen in many objects. Thus it is hard to
attribute a single boxiness value to each galaxy. The distributions of
$A_4/a$ measured at $2 \times r_e$ and $3 \times r_e$ in
elliptical-like (E and E/D) ULIGs are presented in Figure 23. There is
no clear preference between positive and negative boxiness values. The
median and mean of $A_4/a$ at $2 \times r_e$ are --0.002 and --0.001
$\pm$ 0.036, and +0.005 and +0.014 $\pm$ 0.059 at $3 \times$
$r_e$. The broad range of boxiness values in ULIG hosts probably
reflects the residual effects of the violent merger that took place in
these galaxies.

\section{Discussion}

In this section, the results from our imaging survey are combined with
those from other published studies on ULIGs, quasars, and normal
elliptical galaxies to attempt to answer the following questions: (1)
What is the nature of ultraluminous infrared galaxies?; (2) Are
ultraluminous infrared mergers elliptical galaxies in formation?  (3)
Are ultraluminous infrared galaxies quasars in formation?

\subsection{What is the Nature of Ultraluminous Infrared Galaxies? }

What types of galaxies make up ultraluminous infrared systems? What
conditions are needed to trigger their quasar-like infrared
luminosities?  Answers to these two questions would go a long way in
clarifying the nature of ULIGs. There is now a consensus among
researchers in this field concerning the frequency of interactions in
ULIGs and the number of galaxies involved in these interactions. The
results from the present survey (\S 3.1) show convincingly that the
great majority of ULIGs ($\sim$ 95\%) are involved in the interaction
of two, and no more than two, galaxies. The recent HST results of
Bushouse et al. (2002) also support this conclusion. The higher
frequency of multiple mergers claimed earlier by Borne et
al. (2001) and Cui et al. (2001) has now been shown to be due
primarily to the presence of several star-forming super star clusters
which mimic low-luminosity nuclei at optical
wavelengths. Near-infrared observations reveal the true nature of
these clusters,  ruling out genuine galactic nuclei. Follow-up
spectroscopy of apparent neighbors to ULIG systems also reveals in
some cases that they are not physically associated with the ULIGs (see
Appendix A).

Our detailed morphological analysis of the 1-Jy sample provides new
statistical evidence that most (56\%) ULIGs are involved in the late
phase of a merger where the two nuclei have merged into a single
object.  This frequency does not seem to depend strongly on spatial
resolution since the $HST$ data of Scoville et al. (2000), Colina et
al. (2001), and Bushouse et al. (2002) reveal very few double-nucleus
systems that were not already known from arcsecond-resolution
ground-based data like ours. As described in \S 3.2, the frequency of
advanced mergers is a function of infrared luminosity, increasing
dramatically above $\sim$ 10$^{12.5}$ $L_\odot$ and decreasing among
luminous infrared galaxies with log~[$L_{\rm IR}$/$L_\odot$] = 11.50
-- 11.99.  So not only can we say that the ultraluminous infrared
phase is generally triggered when the individual galaxies in the
merger have lost their individual identity and the nuclei have fused
into a single entity, but also that the extreme environment during the
final phases of a merger appears to be a {\em necessary} condition to
produce the most extreme examples of ULIGs.

Another $\sim$ 39\% of the ULIGs in the 1-Jy sample harbor two
distinct nuclei but always show tidal tails at $R$, evidence that the
ultraluminous infrared phase here was triggered {\em after} the first
close passage of the two interacting galaxies. These results are at
least qualitatively consistent with the simulations of Mihos \&
Hernquist (1996) where two merging galaxies go through two distinct
episodes of enhanced star formation activity (and therefore large
infrared luminosity; note that Mihos \& Hernquist do not try to model
AGN activity) with the first peak taking place shortly ($\sim$ 1 -- 2
$\times$ 10$^8$ yr) after the first close encounter, when the two
galaxies are still widely separated, and the second peak taking place
during the final merger of the galaxies, $\sim$ 10$^9$ yr after the
first encounter. Our results indicate that in quite a few cases the
star formation activity triggered during the first passage is
sufficient to reach the ULIG threshold (although we have no
high-resolution maps in the infrared to determine whether the
FIR-emitting region in these widely spaced ULIG systems is a kpc-scale
ring, as predicted by the simulations of Mihos \& Hernquist).  The
detailed kinematic study of four ULIGs (including F10190+1322 from the
1-Jy sample) by Murphy et al. (2001a) confirms that the ULIG phase may
sometimes take place well before the final merger.

In the simulations of Mihos \& Hernquist, the relative amplitude of
the peaks of enhanced star formation activity is only weakly dependent
on the initial orientation of the merging galaxies but depends
strongly on the morphological properties of these galaxies,
particularly their bulge-to-disk mass ratio. Unfortunately, reliable
profile decomposition to determine this ratio in our sample of binary
galaxies requires sub-kpc resolution and therefore cannot be done with
the present data. Nevertheless, our data analysis provides some
constraints on the type of galaxies involved in the interactions. Our
crude profile classification in \S 4.1 suggests that an important
fraction of these binary galaxies have a prominent spheroidal
component (possibly a bulge). As discussed in \S 4.2 (and in more
detail in \S 5.2 and \S 5.3 below), the host galaxies of
single-nucleus ULIGs span a broad range in luminosities with average
of order $\sim$ 2 $L^\ast$.  Binary systems involve two $\sim$ 0.9
$L^\ast$ (on average) galaxies with luminosity ratio generally less
than 4. This is evidence that mergers between small galaxies or minor
mergers involving the accretion of small satellite galaxies by larger
ones do not provide the conditions necessary to trigger an
ultraluminous infrared phase. Large torques appear needed to produce
enough ``damage'' to the galaxies involved in the interaction so that
gas transfer to the inner portions of the galaxies is sufficient to
trigger a strong episode of star formation and/or AGN activity and
reach the ULIG phase. These results are qualitatively consistent with
numerical simulations of unequal mass mergers (e.g., Mihos \&
Hernquist 1996; Walker, Mihos, \& Hernquist 1996; Bendo \& Barnes
2000; Naab \& Burkert 2001) and the recent kinematic data of Genzel et
al. (2001).

There are interesting exceptions to the merger scenario for ULIGs.
Among these is F23233+2817, the only spiral galaxy in the 1-Jy sample
which does not seem to be involved in any galactic interaction. The
infrared luminosity of this object lies on the boundary between
luminous and ultraluminous infrared galaxies (log~[L$_{\rm
IR}$/$L_\odot$] = 12.00), so it may be that the Seyfert 2 nucleus in
this object (or unsuspected circumnuclear starburst) is sufficiently
powerful to heat enough dust to reach the ultraluminous threshold
without the need for refueling from a major merger.  Eight ULIGs in
the 1-Jy sample appear to involve interacting galaxies that are widely
($>$ 20 kpc) separated (see Table 1 and Fig. 5; one of these objects,
F11223--1244, hosts a Seyfert 2 nucleus). All of these objects show
tidal features indicative of a pre-merger phase (i.e. after first
contact; this seems to rule out the possibility that these objects are
observed {\em before} first contact, as suggested by Dinh-V-Trung et
al. 2001). It would therefore appear that an ultraluminous infrared
phase can also be triggered when two galaxies in a pair are near
apgalacticon (farthest approach after first passage). This is
difficult to reconcile with the predictions of the merger scenario. A
way out of this uncomfortable situation is to suppose that the ULIG
phase was triggered by an earlier merger event which is not evident at
the present epoch. One should also be cautious when interpreting the
data on these eight ULIGs since the physical connection between the
galaxies in most of these pairs has not yet been confirmed through
detailed spectroscopy (see Appendix A).

\subsection{Are Ultraluminous Infrared Mergers Elliptical Galaxies in 
Formation?}

Among the 66 single-nucleus mergers in the 1-Jy sample, 48
(73\%) of them are fit adequately by a de Vaucouleurs profile at $R$
and $K^\prime$ (\S 4.1).  This fraction is uncertain, however,
since the relatively modest dynamical range of our data (especially
at $K^\prime$) prevents us from deciding if a de Vaucouleurs profile
is a better fit than an exponential disk in 25 (38\%) of these
objects (these are the so-called E/D galaxies in Table 2 and Fig. 12).
Nevertheless, we can state with confidence that elliptical-like hosts
are observed in the majority ($>$ 50\%) of late mergers of the 1-Jy
sample. This result supports the conclusions from previous studies
that were based on smaller samples of ULIGs (e.g., Wright et al. 1990;
Stanford \& Bushouse 1991; Doyon et al. 1994; Scoville et al. 2000;
Cui et al. 2001), and is at least qualitatively consistent with the
scenario that ULIGs evolve into ellipticals through merger-induced,
dissipative collapse (Kormendy \& Sanders 1992).  In the rest of this
section, we further explore the predictions of this scenario by
comparing the properties of the elliptical-like hosts found in the
1-Jy sample with those of nearby spiral bulges, lenticular (S0)
galaxies, and ellipticals, while taking into account the various
sources of errors in our analysis of the ULIGs.

In \S 4.2, the elliptical-like (E and E/D) hosts in the 1-Jy sample
were found to have a median and average luminosities at $R$ of 1.77
and 1.66 $\pm$ $^{1.45}_{0.77}$ $L^\ast$, and 2.00 and 1.98 $\pm$
$^{1.52}_{0.86}$ $L^\ast$ at $K^\prime$.  Residuals from AGNs and
compact circumnuclear starbursts may affect (overestimate) slightly
the luminosities of some of these hosts -- this effect appears to be
more important at $K^\prime$ than at $R$ and more frequent among
Seyfert 1s than among Seyfert 2s (\S 4.2 and \S 4.5). The results at
$R$ are thus considered more reliable, even though the effects
associated with dust extinction may be more significant here than at
near-infrared wavelengths.  Assuming $V - R
\approx 0.6$ mag for typical ellipticals (Fukujita et al. 1995), the
average $V$-band absolute magnitude of the elliptical-like hosts in
the 1-Jy sample is $\sim$ --21.1 $\pm$ 0.7 mag. The elliptical-like
hosts of ULIGs therefore span the intermediate luminosity range
between the low-luminosity ($\la$ 1.0 $L^\ast$) ``power-law
galaxies'' and high-luminosity ($\ga$ 3.5 $L^\ast$) ``core galaxies'' of
Faber et al. (1997).

The distribution of axial ratios among elliptical-like hosts in the
1-Jy sample resembles that of elliptical galaxies (\S 4.6). The
surface brightness profiles of ULIG hosts also follow the $\mu_e$ --
$r_e$ projection of the fundamental plane of nearby ellipticals, but
considerable scatter is present (\S 4.5; Fig. 19). The average $r_e$
of the 1-Jy ULIGs (4.80 $\pm$ 1.37 kpc at $R$ and 3.48 $\pm$ 1.39 kpc
at $K^\prime$) are fairly typical of intermediate-luminosity
ellipticals (e.g., Kormendy \& Djorgovski 1989; Bender et al. 1992;
Faber et al. 1997). We mentioned in \S 4.5 that residuals from the PSF
subtraction may artificially increase the values of the average
surface brightness within half-light radius (the half-light radius may
also be slightly {\em under}estimated). This effect is likely to be at
the origin of some of the scatter in Figure 19 and may also explain
why some Seyfert 1 hosts in our sample lie significantly above the
$\mu_e - r_e$ relation of ellipticals. Dust extinction and emission
and circumnuclear star formation are other likely sources of scatter
in this diagram.

Given the recent violent merger events in ULIGs, it is perhaps
surprising to find that {\em any} of these galaxies follow the $\mu_e
- r_e$ of ellipticals. This result brings support to numerical
simulations which show that violent relaxation in dissipative gas-rich
merger systems is very efficient.  A general rule of thumb pointed out
by Mihos (1999) is that ``violent relaxation and/or dynamical mixing
occurs on a few rotation periods {\em at the radius in question}.''
This time scale is only $\sim$ 10$^8$ yrs in the central ($\la$ kpc)
regions of the mergers, but exceeds 10$^9$ yrs in the outer ($\ga$ a
few $r_e$) regions. Indeed, while the central regions of many ULIG
hosts appear to mimic those of elliptical galaxies, the effects of the
mergers still appear to be visible beyond a few $r_e$ as illustrated
by the broad range of boxiness values measured in ULIG elliptical-like
hosts of the 1-Jy sample (\S 4.6; Fig. 23). The large fraction (26\%)
of ULIG hosts which are not fit adequately by either an exponential or
de Vaucouleurs profile is another indication that the light
distribution in several objects may still be showing the effects of
the recent mergers (patchy dust distribution could also explain these
results).

It will be important in the future to obtain deep high-resolution
($\la$ 100 pc) images of the 1-Jy ULIGs in both the visible and
near-infrared domains to improve on our AGN/PSF subtraction and
determine if the ULIG host galaxies are power-law or core galaxies
with properties that are consistent with the photometric portion
($\mu_e - r_e$) of the core and global fundamental planes of
early-type galaxies (e.g., Faber et al. 1997). Multicolor maps
combined with detailed radiative transfer models will be particularly
useful in these objects to help disentangle the effects of dust
extinction and emission and circumnuclear star formation on this scale
(see, for instance, Regan, Vogel, \& Teuben 1995). ULIGs are known to
have sub-kpc nuclear concentrations of molecular gas with mean surface
densities similar to the stellar densities measured in the cores of
ellipticals (a few x 10$^4$ M$_\odot$ pc$^{-2}$; e.g., Bryant \&
Scoville 1999; Downes \& Solomon 1998). If {\em post}-ULIG hosts are
indeed genuine ellipticals, one might therefore expect the hosts of
current star-forming ULIGs to have {\em lower} stellar densities than
the core densities of classical ellipticals. This effect could be
significant since the star formation rates of ULIGs are $\sim$ few x
100 M$_\odot$ yr$^{-1}$ and the ULIG phase may last 10$^7$ -- 10$^8$
years.  The effect at large radii should be less important because
most of the star formation (and AGN) activity in ULIGs is concentrated
well within the central kpc (e.g., Soifer et al. 2000, 2001).

Given that merger systems are expected to reach their equilibrium
values of light distribution and kinematics on a short time scale
(e.g., Mihos 1999; Bendo \& Barnes 2000), the kinematic properties of
ULIGs should also follow the fundamental plane of
ellipticals. Considerable efforts have been invested in recent years
to constrain the stellar kinematics in the cores of luminous infrared
mergers (e.g., Doyon et al. 1994; Shier \& Fisher 1998; James et
al. 1999; Genzel et al. 2001). The advent of powerful new
near-infrared spectrographs on 8-meter class telescopes now allows the
radial profiles of the velocity dispersion and rotation velocity to be
measured in the brightest systems and compared with those of
elliptical galaxies.  Most relevant to our study of the 1-Jy ULIGs are
the results from Genzel et al. (2001) since they refer to ULIGs while
those of Shier \& Fisher (1998) and James et al. (1999) refer to
systems of lower luminosities. Genzel et al. (2001) find that random
motions dominate the stellar dynamics in the cores of their (12)
systems, but that significant rotation is also common. The properties
of these galaxies closely resemble those of intermediate-mass ($\sim$
$L^\ast$), disky ellipticals. These results are consistent with ours
in the fact that they rule out giant ($\ga$ 3.5 $L^\ast$) ellipticals
with large cores and little rotation as the hosts of ULIGs. A more
quantitative comparison is difficult due the broad range of
photometric properties found in the 1-Jy sample and the relatively
small size of the kinematic sample of Genzel et al. Recently
commissioned three-dimensional spectrographs on large telescopes
should provide the sensitivity and two-dimensional spatial coverage
needed to derive detailed kinematic maps for several more ULIG hosts,
and therefore allow more robust comparisons with our results.

\subsection{Are Ultraluminous Infrared Galaxies Quasars in Formation?}

The results presented in \S 2 and \S 3 indicate clear trends between
merger phase, $IRAS$ colors, and the importance of AGN activity in
ULIGs (in addition to the trend between infrared luminosity and merger
phase, which has already been discussed in \S 5.1). Objects with warm
quasar-like infrared colors (as measured by the $IRAS$ 25 $\mu$m/60
$\mu$m flux ratio) present strong signs of AGN-dominated activity at
both optical and mid-infrared ($ISO$) wavelengths and are generally
found in advanced mergers (based not only on the apparent nuclear
separation but also on the overall morphology of the system including
the shape and compactness of the nuclear core and the strengths and
lengths of the tidal tails). These trends are consistent with an
evolutionary sequence in which cool starburst-dominated ULIGs
transform into warm AGN-dominated ULIGs as the merger of the two
$\sim$ 0.9 $L^\ast$ galaxies takes place.

It is clear that exceptions to these general trends with merger phase
abound in the 1-Jy sample. For instance, 46\% of the 41 advanced
mergers (classes IVb and V in \S 3.4) in our sample show no obvious
signs of Seyfert activity (see Fig. 10). There are also seven
Seyfert 2 nuclei (but no Seyfert 1s) among the 45 pre-mergers (classes
IIIa and IIIb). The cases of an isolated AGN with a spiral galaxy host
and a widely separated pair with an AGN were mentioned in \S 5.1.
Although merger phase appears to be an important factor in determining
the relative importance of star formation and AGN activity in ULIGs,
it is not the only one (see the discussion in Farrah et
al. 2001). Star formation activity depends on the molecular gas supply
and the efficiency to transform this gas into stars (which is a
complicated function of density, metallicity, magnetic filed strength,
local gas kinematics, etc.). On the other hand, the level of AGN
activity in ULIGs depends linearly on the local mass accretion rate
onto the supermassive black holes that are presumed to exist in these
objects, and the efficiency of the system to transform the
gravitational energy into radiation that can escape the system and be
detected by our instruments. One only needs an accretion rate of
$\sim$ 1 M$_\odot$ yr$^{-1}$ to produce the infrared (bolometric)
luminosity of a ULIG (assuming a radiative efficiency of 10\% in
rest-mass units). Given the complexity of the star formation and AGN
phenomena it is not surprising to see large scatter in the
correlations between merger phase and AGN activity. Note, however,
that a late merger phase appears to be a necessary condition to
trigger powerful Seyfert 1 (= quasar) activity.  Our results emphasize
the importance of studying a large and homogeneous set of ULIGs like
the 1-Jy sample to draw statistically meaningful conclusions on these
objects. The problems of small sample sizes and/or inhomogeneous
selection criteria have plagued many studies of luminous infrared
galaxies in the past.

The natural end-result of the merger sequence described in the first
paragraph is an object with quasar-like colors and luminosities that
shows only slight residual effects from the merger. This object has
several properties in common with nearby quasars. A summary of the
properties of quasar host galaxies is beyond the scope of this paper.
Suffice it to say that evidence for galaxy interactions and mergers is
seen in some but not all nearby quasars (see, e.g., Hutchings \& Neff
1992; Hutchings et al. 1994; Disney et al. 1995; Bahcall et al. 1997; 
Stockton 1999; Dunlop et al. 2002 and references therein).  The
present discussion focusses on the so-called infrared-excess quasars
since they are particularly good candidates for transition objects
which have just gone through a ULIG phase and are now in the process
of settling in to become optical quasars. These objects show
morphological evidence for recent mergers (e.g., Surace et al. 2001
and references therein), contain large quantities of molecular gas
(e.g., Evans et al. 2001 and references therein), and sometimes
present spectroscopic evidence for substantial gas outflows (e.g., BAL
phenomenon, Lipari, Colina, \& Macchetto 1994). The host galaxies of
these infrared-excess quasars have luminosities and colors which are
similar to those of the 1-Jy ULIGs (\S 4.2 and \S 4.3).  In a series
of papers, Canalizo \& Stockton (2000a, 2000b, 2001) have shown that
these ``transition'' quasars show strong recent star-forming activity,
which in most case is directly related to the tidal interaction. The
ages they derive for the starburst populations range from currently
active star formation in some objects to poststarburst ages $\sim$ 300
Myr in others. They argue for a direct connection between
interactions, starbursts, and QSO activity. 

The issue of the hosts in ULIGs and quasars is important and deserves
further attention. If these two classes of objects are indeed related,
then they should have similar host galaxy properties, modulo possible
evolutionary effects. In \S 4.2, we found good agreement between the
host luminosities of the 1-Jy ULIGs and those of infrared-excess
quasars (Surace et al. 2001) and optical quasars from the sample of
McLeod \& Rieke (1994b; McLeod \& McLeod 2001), but less so when
comparing with the sample of radio-quiet and radio-loud quasars of
Dunlop et al. (2002). Morphological classification of the hosts in
ULIGs indicates a preponderance of elliptical-like hosts among
single-nucleus systems (\S 4.1 and \S 5.2), a result which is at least
qualitatively consistent with the conclusions of Dunlop et al. (2002;
note that the conclusions of McLeod \& McLeod 2001 are ambiguous in
this respect). We also found that the half-light radii of the 1-Jy
ULIG hosts are similar to those found by McLeod \& McLeod (2001), but
significantly smaller than those measured by Dunlop et al. (2002). The
half-light surface brightnesses of ULIG hosts at $R$ also appear to be
slightly brighter than those measured by Dunlop et al. (McLeod \&
McLeod 2001 do not tabulate this quantity in their paper).

It is not clear at present whether the differences found between our
results and those of Dunlop et al. are significant. Many effects may
complicate the interpretation of our data.  Excess $R$-band emission
from circumnuclear starbursts, known to be common in ULIGs (e.g.,
Veilleux et al. 1995) but less so among classical quasars, could
explain the brighter $\mu_e$ in ULIG hosts but may have difficulties
explaining the $\sim$ 50\% difference in $r_e$ between the ULIG hosts
and the quasar hosts of Dunlop et al. (2002).  As discussed in \S 5.2,
residuals from the AGN subtraction may also artificially brighten the
half-light surface brightnesses and decrease the half-light radii in
our sample but this effect is not thought to be sufficient to
reconcile our results with those of Dunlop et al. (2002). The apparent
lack of agreement between McLeod \& McLeod (2001) and Dunlop et
al. (2002) in terms of the half-light radii suggests that the quasar
results may depend on sample selection (although these two samples
have similar mean redshifts, $<z>$ = 0.2, and luminosities) or the
methods of observations ($R$-band versus $H$-band $HST$ imaging).
On-going adaptive optics imaging studies of quasar hosts (e.g., Guyon
2002, in prep.) will increase the sample size and will span a broader
range in quasar properties; the results from these studies should
settle this issue within the next few years.

In \S 4.2, we mentioned the existence in our data of slight trends
between host luminosities, $IRAS$ 25 $\mu$m/60~$\mu$m colors, and
nuclear optical spectral types.  The most straightforward explanation
for these trends is that quasar activity generally requires the
mergers of massive ($\ga$ $L^\ast$ + $L^\ast$) progenitors. However,
this scenario has difficulties explaining the fact that the trends
appear to be stronger at $K^\prime$ than at $R$.  Given the presence
of circumnuclear starbursts in many of these objects (e.g., Veilleux
et al.  1995; Kim et al. 1998), it is possible that these starbursts
contribute significantly to the host luminosities, perhaps more so in
warm Seyfert galaxies than in the cooler objects. If the effect is to
be more significant at $K^\prime$ than at $R$, then the starburst age
has to be $\ga$ a few $\times$ 10$^7$ yrs at which point the
$K^\prime$ luminosity is dominated by young, red supergiants (Bruzual
\& Charlot 1993; Leitherer et al. 1999).  Another possible
explanation, also discussed in \S 4.2, is that the trends are due to
errors in the PSF subtraction, which are more likely to affect the
warm highly nucleated Seyferts than the cooler more diffuse LINERs and
H~II region galaxies, and are more severe at $K^\prime$ than at
$R$. An object-by-object comparison with the $H$-band adaptive-optics
results of Surace \& Sanders (1999) was carried out in \S 4.2 and
seems to indicate a slight luminosity offset for the few warm Seyferts
in common between the two samples. Needless to say, it will be
important to expand the sample of warm ULIGs observed with adaptive
optics or with $HST$ to refine the measurements of the host
luminosities in these objects and verify the trends detected in our
data.

\section{Conclusions}

An $R$- and $K^\prime$-band atlas of the {\em IRAS} 1-Jy sample of 118
ULIGs was presented in a companion paper (Kim, Veilleux, \& Sanders
2002). The present paper discusses the results from the analysis of
these images and combines them with the results from published
spectroscopic studies of ULIGs at optical, near-infrared, and
mid-infrared wavelengths as well as new Keck spectroscopy. The results
on the 1-Jy sample are compared with those from optical and
near-infrared studies of quasars and normal ellipticals.  The main
conclusions are as follows:

\begin{itemize}

\item[1.] All but one object in the 1-Jy sample show signs of a strong
tidal interaction/merger in the form of distorted or double nuclei,
tidal tails, bridges, and overlapping disks.  Interactions involving
more than two galaxies are seen in only 5 (4\%) of the 118
systems. These results confirm those of previous studies.

\item[2.] Objects with red $R - K^\prime$ colors and large
nuclear-to-total luminosity ratios show a tendency to host Seyfert
nuclei, have warm $IRAS$ 25 $\mu$m/60~$\mu$m colors and be
AGN-dominated according to the $ISO$ classification scheme.

\item[3.] Using a classification scheme first proposed by Surace
(1998) and based on the results of published numerical simulations of
the mergers of two galaxies, we classified the 1-Jy sources according
to their overall morphology and apparent nuclear separations.  None of
the 1-Jy sources appears to be in the early stages (first approach or
first contact) of a merger. Most (56\%) of them harbor a single
disturbed nucleus with and without tidal tails; they are therefore in
the late stages of a merger.  The fraction of advanced mergers with a
single nucleus increases above infrared luminosities of 10$^{12.5}$
$L_\odot$ and decreases below 10$^{12}$ $L_\odot$.

\item[4.]  The strengths of the H$\beta$ and Mg Ib stellar features
measured in the nuclei of ULIGs are not particularly good indicators
of the merger phase or epoch of the merger event.

\item[5.] All Seyfert 1s and most of the Seyfert 2s are advanced
mergers, either based on their overall morphology or their small ($<$
5 kpc) nuclear separations.  A similar result is found when we
consider the warm objects with $f_{20}/f_{60}$ $>$ 0.2 or the
AGN-dominated objects based on $ISO$ classification. LINERs and H~II
region-like galaxies show no preference between pre-merger and advanced
merger phases.

\item[6.]  The individual galaxies making up the binary systems of the
1-Jy sample show a broad distribution in host absolute magnitudes
(luminosities) with a mean of --21.02 $\pm$ 0.76 mag. (0.85 $\pm$
$^{0.86}_{0.43}$ $L^\ast$) at $R$ and --23.98 $\pm$ 1.25 mag. (0.90
$\pm$ $^{1.94}_{0.61}$ $L^\ast$) at $K^\prime$, and a luminosity ratio
at $R$ or $K^\prime$ generally less than $\sim$ 4.  The hosts of
single-nucleus ULIGs have mean absolute magnitudes (luminosities) of
--21.77 $\pm$ 0.92 mag. (1.69 $\pm$ $^{2.25}_{0.97}$ $L^\ast$) at $R$
and --25.03 $\pm$ 0.94 mag. (2.36 $\pm$ $^{3.24}_{1.38}$) at
$K^\prime$.  These magnitudes are similar to those found in previous
studies of ULIGs, except those derived from shallow $HST$ snapshots
which underestimate the contribution from low surface brightness
features. Correlations are observed between host galaxy luminosity and
infrared luminosity, $IRAS$ 25 $\mu$m/60~$\mu$m color, and optical
spectral type. The trend with infrared luminosity is due to a redshift
bias. There is considerable overlap between the host galaxy luminosity
distribution of single-nucleus ULIGs and that of quasars, although the
hosts of the quasars studied by Dunlop et al.  (2002) are slightly
more luminous on average than the 1-Jy ULIG hosts.  This luminosity
shift is not observed when comparing with the results of McLeod \&
McLeod (2001) and Surace et al. (2001). The $R - K^\prime$ colors of
ULIG hosts are similar to those of quasars.

\item[7.]  An analysis of the surface brightness profiles of the host
galaxies in single-nucleus ULIGs reveals that about 35\% and 2\% of
the $R$ and $K^\prime$ surface brightness profiles are fit adequately
by an elliptical-like $R^{1/4}$-law and an exponential disk,
respectively. Another 38\% are equally well fit by either an
exponential or an elliptical-like profile. The remainder (26\%) of the
single-nucleus sources cannot be fit with either one of these
profiles.  Combining these results, we find that a de Vaucouleurs
profile is an adequate fit to $\sim$ 73\% of the single-nucleus ULIGs
in the 1-Jy sample. These elliptical-like hosts are most common in merger
remnants with Seyfert 1 nuclei (83\%), Seyfert 2 characteristics
(60\%) or mid-infrared ($ISO$) AGN signatures (80\%).

\item[8.] The hosts of ULIGs have half-light radii ($<r_e> = 4.80 \pm
1.37$ kpc at $R$ and $<r_e>$ = 3.48 $\pm$ 1.39 kpc at $K^\prime$)
which are similar to those measured by McLeod \& McLeod (2001) in
quasar hosts, but are significantly smaller than the quasar hosts
studied by Dunlop et al. (2002). The origin of this apparent
discrepancy between the two quasar datasets is not clear at present.
The hosts of 1-Jy systems follow with some scatter the $\mu_e - r_e$
relation of normal ellipticals (especially if Seyfert 1s are excluded
from the analysis due to possible AGN residuals). The distributions of
luminosities, $R$-band axial ratios, and half-light radii in
single-nucleus ULIGs are also similar to those of normal ellipticals
of intermediate luminosities. The results at $K^\prime$ are more
uncertain because of the smaller sample size and possible PSF
subtraction residuals. Elliptical-like hosts in the 1-Jy sample show a
broader range of boxiness values than normal ellipticals.

\item[9.] Results \#7 and \#8 provide strong support to the idea that
some of the ultraluminous infrared mergers in the 1-Jy sample may
eventually become intermediate-luminosity elliptical galaxies under
the condition that they get rid of their excess gas or transform this
gas into stars. The significant fraction of single-nucleus ULIGs with
ambiguous surface brightness profiles or with large boxiness
parameters indicate that these objects are still feeling the effects
of the recent mergers.  These results confirm the predictions of
numerical simulations that violent relaxation is very efficient at the
center of the merger but less so in the outer regions.

\item[10.] The results from this study are generally consistent with
the evolutionary scenario in which ULIGs are the results of a merger
of two gas-rich galaxies which first goes through a
starburst-dominated pre-merger phase when the system is seen as a
binary, next reaches a dust-enshrouded AGN-dominated merger phase once
the two nuclei have merged into one, and then finally ends up as an
optical (post-ULIG) AGN where the host is elliptical-like and shows
only limited signs of the merger.  However, our results also indicate
that many ULIGs in the 1-Jy sample may not follow this exact
scenario. For instance, approximately 46\% of the 41 advanced mergers
in the 1-Jy sample show no obvious signs of Seyfert activity, while
seven of the 45 pre-mergers already show Seyfert 2 (but not Seyfert 1)
activity. The possible correlations between host luminosities, $IRAS$
25 $\mu$m/60~$\mu$m color, and optical spectral types may add another
twist to the merger scenario, suggesting that merger-induced quasar
activity may require the merger of massive ($\ga$ $L^\ast$ + $L^\ast$)
galaxies.  These trends may also be due to a brightening of the
circumnuclear starburst in the hosts of the warm Seyfert
ULIGs. However, one needs to be cautious when interpreting these
apparent trends because of the difficulty associated with removing the
contribution of the central AGN in some of the warm Seyfert ULIGs.
\end{itemize}

Deep near-infrared $HST$ and adaptive-optics imaging of ULIGs and
quasars will help further clarify the nature of their host galaxies.
These observing techniques should be applied to a large sample of warm
ULIGs to refine the PSF subtraction in these objects and verify the
possible trends with host luminosities reported in the present paper.
The spatial resolution of the data presented here is not sufficient to
determine the core properties of the elliptical-like ULIG hosts and
the bulge-to-disk ratio of the galaxies in the pre-merger phase.  Both
high spatial resolution and large dynamic range will be needed to
fully characterize the host galaxies of these ULIGs. A similar program
should be carried out on the classical quasars to help us understand
the apparent inconsistencies between the various quasar
datasets. Finally, three-dimensional spectrographs on large telescopes
will provide the sensitivity and two-dimensional spatial coverage
needed to derive detailed kinematic maps in several elliptical-like
ULIG hosts. Comparisons with data on normal old (low-$z$) and young
(high-$z$) ellipticals will help quantify the similarities and
differences between these two classes of objects.

\clearpage

\acknowledgments

We acknowledge the help from Jim Deane, Aaron Evans, Cathy Ishida, and
Joe Jensen in acquiring some of the images presented in the companion
paper.  The authors thank Jason Surace for many useful discussions,
comments, and suggestions which have greatly improved this paper.  We
also thank Dave Rupke who carried out the reduction of the Keck/ESI
spectra. Helpful conversations with Reinhard Genzel, Richard Green,
John Kormendy, and Tod Lauer are also acknowledged. We thank the
anonymous referee for a prompt and thorough review. S.V.  is grateful
for partial support of this research by NASA/LTSA grant NAG 56547.
D.C.K.  acknowledges financial support from the Academia Sinica in
Taipei and the BK21 project of the Korean
Government. D.B.S. gratefully acknowledges the hospitalitiy of the
Max-Planck Institut for Extraterrestriche Physik and is grateful for
support from a senior award from the Alexander von Humboldt-Stiftung
and from NASA JPL contract 961566.  This work has made use of NASA's
Astrophysics Data System Abstract Service and the NASA/IPAC
Extragalactic Database (NED), which is operated by the Jet Propulsion
Laboratory, California Institute of Technology, under contract with
the National Aeoronautics and Space Administration.

\clearpage

\centerline{Appendix A: Notes on Individual Objects}

In this Appendix, we briefly discuss the rationale behind the
interaction classification of the 1-Jy sample presented in Table 1 and
discussed in \S 3.4. For this task, we used the $R$ and $K^\prime$
images presented in Figure 1 of Paper I and new long-slit spectra
obtained on 2000 March 30 -- 31 and 2001 January 22 -- 23 using Keck
II with the Echellette Spectrograph and Imager (ESI) in the prism
mode.  This configuration produces 4000 -- 9000 \AA\ spectra with a
dispersion that ranges from 50 to 300 km s$^{-1}$ pix$^{-1}$. Accurate
wavelength calibration was derived by fitting a 9th-order Chebychev
polynomial to the line positions of xenon and mercury-neon lamps
observed with the same setup as that of the objects. The redshifts
derived from these data are generally accurate to within $\pm$ 0.001
(1 $\sigma$).  This is not accurate enough to allow us to make any
statement about the relative motion of galaxies in a system but it
allows us to determine companionship and group membership.  The
exposure times were typically 10 minutes for each slit position.
Systems for which we obtained long-slit ESI spectra are indicated by
an asterisk.

\par{\em $\bullet$ F00091$-$0738} (IIIb): This system is a close pair
with apparent nuclear separation of only 2.1 kpc. The two nuclei are
more easily visible in the $K^\prime$ image.

\par{\em $\bullet$ F00188$-$0856$^\ast$} (V): This is a single-nucleus
system with slight distortions in the outer isophotes. The bright
source south of the nucleus is a star.

\par{\em $\bullet$ F00397$-$1312} (V): This is a single-nucleus system
with elongated nuclear isophotes and distorted outer isophotes.

\par{\em $\bullet$ F00456$-$2904} (IIIa): The redshift of the
north-east component of this pair is unknown. A broad tidal tail is
apparent on the north-east galaxy and a shorter one extending to the
north-west is apparent on the main galaxy.

\par{\em $\bullet$ F00482$-$2721} (IIIb): This is a relatively close
pair with apparent nuclear separation of 6.7 kpc. The ``wisp'' to the
south-west is barely apparent in the $K^\prime$ image and appears to
be unrelated to the pair. The redshift of this feature is unknown.

\par{\em $\bullet$ F01004$-$2237} (V): This is a highly nucleated
source with slight distortions in the outer isophotes (see $HST$ image
of Surace et al. 1998).

\par{\em $\bullet$ F01166$-$0844} (IIIb): This is a classic looking
pair with a moderately large nuclear separation of 9.8 kpc and a
prominent tidal tail extending to the north of the north-west
component.

\par{\em $\bullet$ F01199$-$2307$^\ast$} (IIIa): This is a pair of
emission-line galaxies separated by 20.3 kpc. The north-east companion
has the same redshift as the main component (0.155 versus 0.156,
respectively). A faint tidal tail is visible on the west side of the
main component, and material also appears to bridge the binary.

\par{\em $\bullet$ F01298$-$0744$^\ast$} (IVb): The $IRAS$ source is
the distorted object to the north-east. This object harbors two
prominent tidal tails, signs of a recent merger.  The galaxy to the
south-west shows no sign of interaction and has an absorption-line
spectrum with a redshift of 0.188 instead of 0.136 for the $IRAS$
source.

\par{\em $\bullet$ F01355$-$1814$^\ast$} (IIIb): This object is
classified as a close pair based on the separation (5.8 kpc) between
the centroid of the bright south-east (SE) component and the fan-like
north-west (NW) component. The feature to the south-east of this pair
appears to be a star-forming tidal feature based on the fact that it
is barely visible at $K^\prime$. All three features have emission-line
redshifts of 0.191 to within $\pm$ 0.001. Deep multi-band imaging or
spectroscopy will be needed to confirm the nature of the faint
south-east feature; the existence of an old underlying stellar
population in this object would require changing the classification of
this system to a triplet.

\par{\em $\bullet$ F01494$-$1845$^\ast$} (IVa): This diffuse
single-nucleus source shows a broad tidal feature to the
north-west. The faint object to the south-east is a star while the
elongated source to the north is an emission-line galaxy at the same
redshift as the $IRAS$ source.  The compact source to the east of the
elongated galaxy is a broad-lined quasar with redshift of $\sim$ 1.62
based on the identifications of the C~IV $\lambda$1549, C~III]
$\lambda$1909, and Mg~II $\lambda$2798 features.

\par{\em $\bullet$ F01569$-$2939} (IVa): This highly distorted source
harbors a narrow tidal tail stretching over 55 kpc. A shorter
counter-tail is also probably present on the west side of the object.

\par{\em $\bullet$ F01572$+$0009 = Mrk 1014} (IVb): This
single-nucleus source presents two curved tidal features extending 46
kpc (also see Surace et al. 1998).

\par{\em $\bullet$ F02021$-$2103} (IVa): This highly distorted object
presents two symmetric tidal features.

\par{\em $\bullet$ F02411$+$0353$^\ast$} (IIIb): This apparently
complex system at $R$ is a close (7.1 kpc) pair at $K^\prime$. The
north-east (NE) component presents a fan-like tidal feature extending
to the east. Keck spectroscopy has confirmed the physical interaction
between the NE and SW components ($z$ = 0.144 and 0.143,
respectively). The compact source near the western edge of the
$R$-band field appears to display a very faint feature extending to
the south and therefore may not be a star.

\par{\em $\bullet$ F02480$-$3745} (IVa): This single-nucleus object
presents two symmetric tidal features extending over 27 kpc.

\par{\em $\bullet$ F03209$-$0806} (IVb): This compact object harbors
spectacular tidal tails extending over 61 kpc.

\par{\em $\bullet$ F03250$+$1606$^\ast$} (IVb): This object shows weak
tidal features to the north and south-east of the nucleus. The faint
compact object near the eastern edge of the field of view is a star,
and the same also appears to be true for the other compact object west
of the star. The compact object (labelled ``G'') to the south-west has
the same redshift (both emission and absorption lines) as the $IRAS$
source, but shows no sign of interaction.

\par{\em $\bullet$ FZ03521$+$0028} (IIIb): This is a close pair
separated by only 3.5 kpc.

\par{\em $\bullet$ F04074$-$2801$^\ast$} (IVa): This source is highly
distorted. A tidal feature is seen extending $\sim$ 28 kpc north of the
nucleus. The two galaxies to the north-west and north-east of the
$IRAS$ source show no signs of interaction but are at the same
emission-line redshift as the $IRAS$ source to within the measurement
uncertainties.

\par{\em $\bullet$ F04103$-$2838} (IVb): This compact source presents
two short curved tidal features extending 8 kpc.

\par{\em $\bullet$ F04313$-$1649} (IVa): This is a classic merger
remnant with two large tidal tails extending 59 kpc.

\par{\em $\bullet$ F05020$-$2941$^\ast$} (IVa): The $IRAS$ source is
the distorted source with short tidal tails to the south and
north-west. The bright source to the east is a star. No obvious
emission or absorption feature was detected in the faint diffuse
object between the star and the $IRAS$ source.

\par{\em $\bullet$ F05024$-$1941$^\ast$} (IVa): The $IRAS$ source is
the distorted source near the center of the image. A tidal feature is
clearly visible on the west side of this object. The east source is a
bright star.

\par{\em $\bullet$ F05156$-$3024} (IVb): This single-nucleus object
presents a short tidal feature to the south.

\par{\em $\bullet$ F05189$-$2524} (IVb): This compact galaxy has a
faint curved tidal feature to the north-west (refer to Surace et
al. 1998 for a high-resolution $HST$ image of this object).

\par{\em $\bullet$ F07599$+$6508} (IVb): The $IRAS$ source lies to the
south-east in this field, and presents faint tidal features on the west
and south sides (more easily visible in the images of Surace
1998). The other bright source in the field is a star.

\par{\em $\bullet$ F08201$+$2801$^\ast$} (IVa): This object harbors a
broad tidal tail that extends over 25 kpc. The spectrum of the faint
source to the north-west presents faint [O~II] $\lambda$3727 emission
at the same redshift as the $IRAS$ source, but it is not clear whether
the emission really originates in the north-west source; it may be
contamination from the diffuse tidal feature surrounding the $IRAS$
source.  No obvious emission or absorption feature was detected in the
southern object.

\par{\em $\bullet$ F08474$+$1813$^\ast$} (V): This apparently complex
system comprises the $IRAS$ source to the west and two stars to the
south-east and north-east of this source. No obvious tidal tail is
apparent in the $IRAS$ source.

\par{\em $\bullet$ F08559$+$1053$^\ast$} (IVb): A diffuse tidal tail
seems to emerge from the south-west side of the bright galaxy. The two
galaxies to the west of the brighter source have the same
emission-line redshifts as that of the main source to within $\pm$
0.001, but show no sign of interaction with the main object. This
system is therefore classified as a single-nucleus source, and the
other two galaxies are considered (non-interacting) group members.

\par{\em $\bullet$ F08572$+$3915} (IIIb): This is a well-known
interacting pair separated by 5.7 kpc (e.g., Sanders et al. 1988a,b;
Armus et al. 1990; Surace et al. 1998).

\par{\em $\bullet$ F08591$+$5248$^\ast$} (V): The $IRAS$ source to the
east is a compact object with slightly distorted isophotes but no
obvious tidal tails. The source to the west is a faint emission-line
galaxy at the same redshift as the $IRAS$ source ($z$ = 0.157), but it
shows no sign of interaction. 

\par{\em $\bullet$ F09039$+$0503} (IVa): This is a single-nucleus
system with a prominent curved tidal tail extending 24 kpc to the east.

\par{\em $\bullet$ F09116$+$0334$^\ast$} (IIIa): The compact source
located east of the bright Seyfert 2 nucleus is a small
absorption-line galaxy at the same redshift as the Seyfert
galaxy. Tidal features are seen to the east and west of the Seyfert
galaxy.

\par{\em $\bullet$ F09463$+$8141} (IVa): This source appears to be a
pair at $R$ but the $K^\prime$ image shows that it is in fact a single
nucleus with a prominent tidal tail extending out to 76 kpc to the west.

\par{\em $\bullet$ F09539$+$0857} (V): This compact object has
distorted outer isophotes but shows no obvious tidal tails.

\par{\em $\bullet$ F10035$+$2740} (IVa): This object appears to be
double at $R$ but the $K^\prime$ image shows that it is in fact a
single nucleus with a prominent tail extending out to 54 kpc to the
north. The redshift of the bright galaxy to the south-west is unknown;
it shows no obvious sign of interaction with the main object.

\par{\em $\bullet$ F10091$+$4704} (IVa): This highly distorted source
presents two complex tidal features to the south and north-west of the
nucleus.  The southern feature presents a luminous knot at $R$ which is not
apparent at $K^\prime$.

\par{\em $\bullet$ F10190$+$1322} (IIIb): This is a close (5.5 kpc)
pair with weak tidal features. The compact source to the south is
star. 

\par{\em $\bullet$ F10378$+$1108} (IVb): This single-nucleus source
presents broad tidal features extending to the north and south.

\par{\em $\bullet$ F10485$-$1447$^\ast$} (IIIa): Stars complicates the
classification of this system. The two brightest stars in the field
have been confirmed spectroscopically. The identification of the third
one located next to the galaxy labelled ``E'' is based on its
unresolved point spread function. The eastern (absorption-line) and
western (emission-line) components have similar redshifts ($z$ =
0.133). Tidal features seem to emanate from the western component and
one of them possibly creates a bridge with the eastern component.
This interpretation is not unique, however. It is possible that the
eastern component is not interacting with the western component and
that the apparent ``bridge'' is instead a second tidal feature from
the western component as a result of an earlier merger.

\par{\em $\bullet$ F10494$+$4424} (IVb): This object presents two
peculiar ``spikes'' to the north which are presumably of tidal origin
and a fainter diffuse feature to the south-west.

\par{\em $\bullet$ F10594$+$3818} (IIIb): This close (4.4 kpc) pair
shows prominent tidal tails extending out to 90 kpc to the west. The
bright source to the west is a star.

\par{\em $\bullet$ F11028$+$3130} (IVa): This single-nucleus source
shows a broad tidal feature to the west.

\par{\em $\bullet$ F11095$-$0238} (IVb): A prominent curved tidal tail
is visible to the north of this compact source. The high-resolution
$H$-band image of Bushouse et al. (2002) reveals an interesting double
core in this object. However, as mentioned in their paper, it is not
clear whether the double core represents two galactic nuclei or a
single nucleus bisected by a dust lane. This is the reason why this
object is classified as a late merger (IVb) rather than a close binary
(IIIb).

\par{\em $\bullet$ F11119$+$3257} (IVb): This compact source shows an
interesting bootlegged tidal feature to the east.

\par{\em $\bullet$ F11130$-$2659} (IVa): This single-nucleus system
presents a complex tidal feature to the north-west and a shorter
feature to the south-east.

\par{\em $\bullet$ F11180$+$1623$^\ast$} (IIIa): The $IRAS$ source is
the wide (21.8 kpc) pair of galaxies to the south-east in this field.
The emission lines in the spectrum of the western component in
this pair are quite faint. No spectrum was obtained of the two
galaxies labelled ``G''.

\par{\em $\bullet$ F11223$-$1244} (IIIa): This source is
conservatively classified as a very widely separated (88 kpc)
pair. The redshift of the eastern component is unknown. The signs of
interaction in the eastern component are very slight and it is
possible that the tidal features of the western component were caused
instead by an earlier galaxy merger.

\par{\em $\bullet$ F11387$+$4116} (V): This compact source shows
slight outer isophotal distortions. The redshifts of the two faint
sources north-west of the nucleus are not available. 

\par{\em $\bullet$ F11506$+$1331} (IVb): This is a single-nucleus
object with two symmetric tidal tails to the east and west of the
nucleus. The nature of the object to the south is unknown, but does
not appear to participate in an interaction.

\par{\em $\bullet$ F11582$+$3020} (V): The $IRAS$ source is the
distorted object on the east in this field.  The compact object to the
west is a star, and the other two diffuse sources (labelled ``G'') are
galaxies with redshifts similar to that of the $IRAS$ source but they
show no sign of interaction with it.

\par{\em $\bullet$ FZ11598$-$0112} (IVb): The $IRAS$ source is the
object in the middle of the field with two prominent tidal tails
extending out to 12 kpc. The redshift of the galaxy to the north is
unknown. 

\par{\em $\bullet$ F12018$+$1941} (IVb): The $IRAS$ source presents
two symmetric tidal tails.  

\par{\em $\bullet$ F12032$+$1707} (IIIa): The $K^\prime$ image of this
object reveals the presence of two interacting galaxies separated by
12.0 kpc.

\par{\em $\bullet$ F12072$-$0444} (IVb): This single-nucleus source
presents a curved tidal tail to the south and a shorter feature to the
north-west (see high-resolution image of Surace et al. 1998). 

\par{\em $\bullet$ F12112$+$0305} (IIIb): This peculiar-looking object
is a close pair separated by 3.8 kpc. A bright tidal tail is present
in the $R$-band image.

\par{\em $\bullet$ F12127$-$1412$^\ast$} (IIIa): The members of this
widely separated (21.9 kpc) pair have similar emission-line redshifts
($z$ = 0.133) although the emission lines in the north-east component
are considerably stronger than those in the south-west component

\par{\em $\bullet$ F12265$+$0219 = 3C 273} (IVb): This object is
highly nucleated, but shows a thick tidal feature towards the
north-east and a very faint feature to the north-west (e.g., Tyson,
Baum, \& Kreidl 1982). The faint linear feature emerging to the
south-west is the well-known optical jet in this system (e.g., 
Bahcall et al. 1995).

\par{\em $\bullet$ F12359$-$0725$^\ast$} (IIIa): This apparent triplet
is made of a widely separated pair (labeled ``N'' and ``S''; separated
by 22.9 kpc) and of a background galaxy at $z$ = 0.396 to the west.

\par{\em $\bullet$ F12447$+$3721} (IVa): This diffuse object shows
broad tidal features to the east and north-west.

\par{\em $\bullet$ F12540$+$5708 = Mrk 231} (IVb): This object is made
of a compact nucleus surrounded by a diffuse tidal complex (e.g.,
Surace et al. 1998).

\par{\em $\bullet$ F13106$-$0922$^\ast$} (IVa): The $IRAS$ source is
the highly distorted source to the north-east in this field with a
very long (82 kpc) tidal tail. The source north of the $IRAS$ source
is a star and so is the source to the west.

\par{\em $\bullet$ F13218$+$0552$^\ast$} (V): This is a highly
nucleated source with no obvious tidal tails. The compact object to
the south-west of the $IRAS$ source appears to be a background
absorption-line galaxy at $z$ = 0.23 although additional data are
needed to confirm this redshift. 

\par{\em $\bullet$ F13305$-$1739$^\ast$} (V): This highly nucleated
galaxy shows slight isophotal distortions. Our spectrum of the compact
source to the north-west is not of sufficient quality to determine the
nature and redshift of this object.

\par{\em $\bullet$ F13335$-$2612} (IIIb): The two nuclei in this close
(3.2 kpc) pair are more easily visible in the $K^\prime$ image.

\par{\em $\bullet$ F13342$+$3932$^\ast$} (IVb): This object is
misclassified as an interacting group by Borne et al. (2000). The
$IRAS$ source is in the middle of the field.  The galaxy to the
north-east (labelled ``G'') has an emission-line redshift of 0.180,
very similar to that of the $IRAS$ source. However, it shows no sign
of interaction with the $IRAS$ source. The elongated galaxy to the
north-west lies in the foreground at $z$ = 0.165. No redshift is
available for the other galaxy to the north-west.

\par{\em $\bullet$ F13428$+$5608} (IVb): This merger remnant harbors a
spectacular tidal tail which extends out 31 kpc to the south.

\par{\em $\bullet$ F13443$+$0802$^\ast$} (Tpl) All three objects in
the field are emission-line galaxies with similar redshifts ($z$ =
0.135). A bridge of material may be present between the eastern binary
and the south-west component.

\par{\em $\bullet$ F13451$+$1232} (IIIb): This is a well-known pair
separated by 4.0 kpc (e.g., Surace et al. 1998; Evans et al. 1999).

\par{\em $\bullet$ F13454$-$2956} (IIIa): A tidal tail appears to
emerge south-west of the southern component in this widely separated
(15.8 kpc) pair.

\par{\em $\bullet$ F13469$+$5833} (IIIb): This close (4.3 kpc) pair
shows prominent curved tidal tails to the north-east and south-west.

\par{\em $\bullet$ F13509$+$0442$^\ast$} (IVb): The $IRAS$ source is
at the center of the field and shows a prominent tidal tail which
extends 15 kpc to the south. The galaxy near the northern edge of the
field of view is a background emission-line galaxy at $z$ = 0.188.
The object to the south-west of the $IRAS$ source is a star.

\par{\em $\bullet$ F13539$+$2920} (IIIb): This relatively close (7.0
kpc) pair presents a strong tidal feature to the north which gives the
appearance of a third nucleus in the $R$-band image (or $I$-band $HST$
image of Cui et al. 2001). This feature is barely apparent in the
$K^\prime$ image.

\par{\em $\bullet$ F14053$-$1958} (IIIb): This very close (1.5 kpc)
pair is barely resolved in the $K^\prime$ image and presents two tidal
features to the north and south.

\par{\em $\bullet$ F14060$+$2919} (IVa): This is a highly distorted
source with a strong linear tidal feature to the north-east surrounded
by a diffuse envelope of material. A single nucleus is observed in
this object at $K^\prime$.  This object is misclassified as multiple
($>$ 2) by Cui et al. (2001).

\par{\em $\bullet$ F14070$+$0525} (V): This compact object shows no
obvious tidal features but has disturbed isophotes.

\par{\em $\bullet$ F14121$-$0126} (IIIb): This moderately close (9.1
kpc) pair shows a prominent curved tidal tail to the north-west and a
fainter feature which seems to emerge from the southern component. The
compact source to the south-west appears to be a star based on its
unresolved point spread function.

\par{\em $\bullet$ F14197$+$0813} (V): A star to the north is
superposed on the body of this object.  There is no evidence for tidal
tails in this object.

\par{\em $\bullet$ F14202$+$2615} (IIIa): The $IRAS$ source is the
pair of galaxies to the north. The redshift of the southern galaxy is
unknown but since it shows no obvious sign of interaction with the
pair, we classify this system as a pair instead of a triplet (Cui et
al. 2001).

\par{\em $\bullet$ F14252$-$1550} (IIIb): This appears to be a
moderately close (8.3 kpc) pair with a fairly large luminosity ratio
of $\sim$ 8 at $K^\prime$.

\par{\em $\bullet$ F14348$-$1447} (IIIb): This is a close (4.8 kpc)
binary with distinct tidal features to the north and south of the
galaxies. The compact source near the eastern edge of the field is a
star based on its unresolved point spread function.

\par{\em $\bullet$ F14394$+$5332} (Tpl) This system is classified as a
triplet because the eastern component has two close (2.6 kpc) nuclei
and a prominent tidal feature appears to bridge the eastern and
western components.  However, we have no confirmation that the two
components lie at the same distance.

\par{\em $\bullet$ F14485$-$2434} (IVb): The $IRAS$ source lies west
of the field center. The redshift of the galaxy to the north-east is
unknown; it shows no sign of tidal disturbance.

\par{\em $\bullet$ F15001$+$1433$^\ast$} (Tpl) This object is
classified as a triplet because the tidal feature west of the eastern
component appears to point towards the other two western sources. One
of these (labelled W in Fig. 1 of Paper I) has the same redshift as
the main component ($z$ = 0.162). The low signal-to-noise of the
spectrum for the westernmost (WW) component does not allow us to
confirm its physical connection with the other galaxies in the system.

\par{\em $\bullet$ F15043$+$5754} (IIIb): This is a close (5.1 kpc)
double with short and stubby tidal features.

\par{\em $\bullet$ F15130$-$1958} (IVb): This single-nucleus source
has two tidal features extending to the north and south over 14 kpc.

\par{\em $\bullet$ F15206$+$3342} (IVb): This is a relatively compact
source with a distinct tidal feature to the west. $HST$ images of this
image reveal a complex morphology (Surace et al. 1998). A recent
detailed kinematic study of this object by Arribas \& Colina (2002)
confirms the existence of a single nucleus in this system. The source
to the north is a star based on its unresolved point spread function.

\par{\em $\bullet$ F15225$+$2350} (IVa): This merger remnant presents a
prominent tidal tail to the south-west and a fainter one to the
south-east.

\par{\em $\bullet$ F15327$+$2340 = Arp 220} (IIIb): This system is
well known to harbor two near-infrared nuclei separated by 0$\farcs$9
or 0.4 kpc (e.g., Scoville et al. 2000), but these two nuclei are not
distinct at the resolution of our images.

\par{\em $\bullet$ F15462$-$0450$^\ast$} (IVb): The $IRAS$ source
presents tidal tails extending over 8 kpc. The compact source to the
south is a spectroscopically confirmed star.

\par{\em $\bullet$ F16090$-$0139} (IVa): This single-nucleus object
presents a diffuse tidal feature to the north-east.

\par{\em $\bullet$ F16156$+$0146$^\ast$} (IIIb): The two objects in
this pair have emission-line spectra with similar redshifts ($z$ =
0.133). The system presents a faint stubby tidal feature to the south
and a very faint tail to the north-west.  A tidal bridge may also be
connecting the two objects in the pair. The compact source near the
galaxy pair appears to be a star based on its unresolved point spread
function, although underlying continuum emission from the north-west
component makes this statement uncertain.

\par{\em $\bullet$ F16300$+$1558} (V): This is a compact object with
distorted outer isophotes, but no obvious signs of tidal tails.

\par{\em $\bullet$ F16333$+$4630} (IIIa): This widely separated (13.1
kpc) binary appears to be linked by a tidal bridge. The object to the
north-east of the pair is a star based on its unresolved point spread
function.

\par{\em $\bullet$ F16468$+$5200} (IIIb): This moderately close (7.8
kpc) pair presents a tidal tail which seems to emerge from the western
component and extends 14 kpc to the north.

\par{\em $\bullet$ F16474$+$3430} (IIIb): This object is similar to
F14252$-$1550 in that it is a moderately close (6.5. kpc) pair with a
fairly large {\em nuclear} luminosity ratio of $\sim$ 5.5 at
$K^\prime$ (the components are too close to each other to determine
reliably the global luminosity ratio). A prominent tidal tail is
visible on the west side.

\par{\em $\bullet$ F16487$+$5447} (IIIb): This close (5.4 kpc) pair
appears to be linked by a tidal bridge.

\par{\em $\bullet$ F17028$+$5817$^\ast$} (IIIa): The western component
of this widely separated (23.3 kpc) pair presents diffuse tidal
features and the isophotes of the eastern component are extended in
the direction of the western companion. Both components have the same
redshift within the uncertainties ($z$ = 0.106 versus 0.107,
respectively)

\par{\em $\bullet$ F17044$+$6720} (IVb): The brighest source in the
field is a star.  The $IRAS$ source presents a peculiar tidal feature
to the north-west of the nucleus.

\par{\em $\bullet$ F17068$+$4027} (Tpl): This source is tentatively
classified as a triplet because of its resemblance with the other
triplet candidate F15001+1433. The main source (labelled ``E'') in the
field has a single nucleus but is elongated along the east-west
direction, i.e. in the same direction as the two fainter western (W
and WW) galaxies.  Note that the redshifts of these two galaxies are
unknown and that no obvious signs of interaction are visible in these
objects. If future observations do not confirm the interaction, the
$IRAS$ source should be re-classified as a IVa.

\par{\em $\bullet$ F17179$+$5444$^\ast$} (IVb): This single-nucleus
source is surrounded by what seems to be tidal material extending
asymmetrically to the south-west. The compact source to the south-west
is a star based on its unresolved point spread function and its
spectrum.

\par{\em $\bullet$ F20414$-$1651$^\ast$} (IVb): The isophotes of this
single-nucleus object are highly elongated. The spectra of the two
compact and diffuse sources to the south of the main galaxy are not of
sufficient quality to allow us determine the nature of these objects
(a faint H$\alpha$ feature at the same redshift as the main galaxy may
be present in the spectrum of the compact object, but this feature is
very faint and may be due to contaminating emission from the main
object). These southern sources may be galaxies in interaction with
the main source but the evidence is not sufficiently convincing to
classify this system as an interacting pair.

\par{\em $\bullet$ F21208$-$0519} (IIIa): This wide binary is
separated by 14.1 kpc and shows distinct tidal features which extend
out to 14 kpc.

\par{\em $\bullet$ F21219$-$1757} (V): The isophotes of this compact
object are only slightly distorted. 

\par{\em $\bullet$ F21329$-$2346} (IVa): This single-nucleus object
presents a faint tidal feature to the south and a peculiar bootlegged
feature to the north which is also presumably of tidal origin.

\par{\em $\bullet$ F21477$+$0502} (Tpl) This is arguably the best case
for a triple system in the 1-Jy sample. A bridge of material appears
to connect the western component to the eastern pair. 

\par{\em $\bullet$ F22088$-$1831} (IIIb): The members of this pair are
separated by 4.3 kpc. The compact source to the north-east is a star
based on its unresolved point spread function.

\par{\em $\bullet$ F22206$-$2715} (IIIb): This moderately close (7.7
kpc) binary shows two prominent tidal features to the south-east and
north-west.  It is misclassified as a multiple ($>$ 2) system by Cui et
al. (2001).

\par{\em $\bullet$ F22491$-$1808} (IIIb): This close (2.2. kpc) pair
shows two classic tidal tails to the east and north-west. It is
misclassified as a multiple ($>$ 2) system by Cui et al. (2001).

\par{\em $\bullet$ F22541$+$0833} (IIIa): This system is tentatively
classified a widely separated (18.5 kpc) binary. The outer isophotes
of the south-east galaxy are highly elongated in the direction of the
north-west companion.  Note, however, that the redshift of the
south-east galaxy is unknown.

\par{\em $\bullet$ F23060$+$0505} (IVb): This single-nucleus system
presents a diffuse tidal feature to the south-west.

\par{\em $\bullet$ F23129$+$2548} (IVa): The $IRAS$ source lies north
in this field and present tidal features to the south and west.  The
redshift of the galaxy to the south is unknown. This galaxy shows no
sign of interaction with the $IRAS$ object.

\par{\em $\bullet$ F23233$+$2817} (Isolated): This source shows a
spiral-like pattern surrounding a very compact nucleus. No obvious
tail is visible in this object.  The point sources directly south and
north-east of the nucleus are presumed to be bright H~II regions based
on their luminosities and colors. This is the only object in the 1-Jy
sample with no obvious sign of interaction.

\par{\em $\bullet$ F23234$+$0946} (IIIb): This moderately close (7.4
kpc) pair shows a hook-like tidal feature to the south-east. The
nature of the two faint point sources to the south-east and north-west
of the pair are unknown. Neither one shows signs of interaction with
the $IRAS$ source.

\par{\em $\bullet$ F23327$+$2913} (IIIa): Each member of this wide
(22.7 kpc) binary presents tidal features and appears to be linked to
each other by a bridge of material.

\par{\em $\bullet$ F23389$+$0300} (IIIb): This close (4.9 kpc) binary
presents a short tidal feature to the south.

\par{\em $\bullet$ F23498$+$2423} (IIIa): This wide (12.7 kpc) binary
source presents prominent tidal tails at $R$. The compact source west
of the binary is a star based on its unresolved point spread function.

\clearpage

\setcounter{figure}{0}
\begin{figure}
\caption{Color -- magnitude diagrams of the 1-Jy sample. ($a$)
Integrated $R$-band absolute magnitudes versus $R - K^\prime$. ($b$)
Integrated $K^\prime$ absolute magnitudes versus $R - K^\prime$. ($c$)
Nuclear (4-kpc diameter) $R$-band absolute magnitudes versus nuclear
$R - K^\prime$. ($d$) Nuclear (4-kpc diameter) $K^\prime$ absolute
magnitudes versus nuclear $R - K^\prime$. Nuclear quantities are
measured within a diameter of 4 kpc.}
\end{figure}

\setcounter{figure}{1}
\begin{figure}
\caption{Trends between the integrated $R$ (left) and $K^\prime$ (right) 
absolute magnitudes and ($a$) -- ($b$) the $IRAS$ 25~$\mu$m/60 $\mu$m
colors; ($c$) -- ($d$) the optical spectral types; ($e$) -- ($f$) the
$ISO$ spectral types; ($g$) -- ($h$) the infrared luminosities. }
\end{figure}

\setcounter{figure}{2}
\begin{figure}
\caption{Trends between the integrated (left) and nuclear (4-kpc; right) 
$R - K^\prime$ colors and ($a$) -- ($b$) the optical spectral types;
($c$) -- ($d$) the $IRAS$ 25~$\mu$m/60 $\mu$m colors. No trends are
observed with the infrared luminosities, $IRAS$ 60~$\mu$m/100~$\mu$m
colors and the strengths of the H$\beta$ and Mg~Ib stellar features
(not shown here).}
\end{figure}

\setcounter{figure}{3}
\begin{figure}
\caption{Trends between the $R$ (left) and $K^\prime$ (right) 
compactness values and ($a$) -- ($b$) the optical spectral types;
($c$) -- ($d$) the $IRAS$ 25~$\mu$m/60 $\mu$m colors; ($e$) -- ($f$)
the equivalent widths of the stellar Mg Ib feature. }
\end{figure}

\begin{figure}
\caption{Apparent nuclear separations in the 1-Jy sample of
galaxies. The distribution is highly peaked at small values but also
presents a significant tail at high values. The very uncertain
separation measured in F11223--1244 (87.9 kpc) is not shown in this
figure.}
\end{figure}

\setcounter{figure}{5}
\begin{figure}
\caption{Trends between the apparent nuclear separations and ($a$) the
optical spectral types; ($b$) the $IRAS$ 25~$\mu$m/60 $\mu$m colors.
No trends are observed with the nuclear values of the equivalent
widths of the H$\beta$ and Mg Ib stellar absorption features, two
traditional indicators of the starburst age (not shown here). }
\end{figure}

\setcounter{figure}{6}
\begin{figure}
\caption{Apparent nuclear separations as a function of infrared
luminosities. The great majority of the extreme ULIGs with log [L$_{\rm
IR}$/$L_\odot$] $>$ 10$^{12.5}$ $L_\odot$ are single-nucleus systems and
therefore have $NS <$ 2.5 kpc. }
\end{figure}

\setcounter{figure}{7}
\begin{figure}
\caption{Total projected tail lengths as a function of ($a$) the
optical spectral types; ($b$) the infrared luminosities; ($c$) the
$IRAS$ 25~$\mu$m/60~$\mu$m colors; ($d$) the $IRAS$
60~$\mu$m/100~$\mu$m colors; ($e$) the equivalent widths of the
H$\beta$ stellar feature; ($f$) the equivalent widths of the Mg Ib
stellar feature.  No obvious correlations are observed with any of
these parameters. }
\end{figure}

\setcounter{figure}{8}
\begin{figure}
\caption{Interaction classes for the objects 
in the 1-Jy sample. See text in \S 3.4 for a description of each
class. The five triplets in the sample are plotted separately on the
left. The merger of two galaxies is expected to follow the sequence I
$\rightarrow$ V. Contrary to systems of lower infrared luminosities
which often comprise two distinct galaxies (classes I, II and III),
most (56\%) of the objects in 1-Jy sample are single-nucleus systems
and are therefore in the later stage of a merger (IV or V). }
\end{figure}

\setcounter{figure}{9}
\begin{figure}
\caption{Positive trends between the interaction class and ($a$) the infrared
luminosity, ($b$) the optical spectral type, and ($c$) the $IRAS$
25~$\mu$m/60 $\mu$m color.  No trends are observed with the $IRAS$
60~$\mu$m/100~$\mu$m colors and the strengths of the H$\beta$ and Mg
Ib stellar features (not shown here).  }
\end{figure}

\setcounter{figure}{10}
\begin{figure}
\caption{Observed surface brightness profiles of (left) $R$-band
images and (right) $K^\prime$ images. These profiles have not been
corrected for cosmological dimming. The profile of each image is
plotted on two different radial scales: a linear scale and a $R^{1/4}$
scale.  Solid lines in each of the panels are least-square exponential
or de Vaucouleurs fits to the data.  To avoid seeing effects, only the
data outside of twice the radius of the seeing disk are considered in
the fits. Arrows indicate the size of the seeing disk.  The value of
the reduced chi-square is indicated on the upper right hand corner of
each panel. Values much larger than unity indicate poor fits, while
values much less than unity indicate fits which are poorly
constrained.}
\end{figure}

\setcounter{figure}{11}
\begin{figure}
\caption{Fraction of ($a$) single-nucleus and ($b$) double-nucleus
1-Jy ULIGs with disk-like (D), elliptical-like (E),
elliptical/disk-like (E/D) and ambiguous (A) radial profiles. E/D
profiles are equally well fitted by an exponential or de Vaucouleurs
function, while neither fits galaxies with ``ambiguous'' profiles. As
expected, the fraction of ambiguous profiles is considerably larger among
binary systems.}
\end{figure}

\setcounter{figure}{12}
\begin{figure}
\caption{Trends between host galaxy profiles and ($a$) optical
spectral types, ($b$) $ISO$ spectral types, ($c$) $IRAS$ 25~$\mu$m/60
$\mu$m colors, and ($d$) interaction classes. Elliptical-like systems
show a tendency to host warm AGN and to lie in advanced mergers. The
nature of the host does not seem to depend strongly on the infrared
luminosity, the 60~$\mu$m/100~$\mu$m ratio, or the strengths of the
stellar H$\beta$ and Mg Ib features (not shown here).}
\end{figure}

\setcounter{figure}{13}
\begin{figure}
\caption{Absolute $R$ (left) and $K^\prime$ (right) 
magnitudes of the host galaxies for individual objects in ($a$) --
($b$) double-nucleus systems and ($c$) -- ($d$) single-nucleus
systems.  For comparison, the absolute magnitudes at $R$ and
$K^\prime$ of a $L^\ast$ galaxy are --21.2 and --24.1, respectively.}
\end{figure}

\setcounter{figure}{14}
\begin{figure}
\caption{Trends between the $R$-band absolute magnitude of the host
galaxies and ($a$) the infrared luminosities; ($b$) the $IRAS$
25~$\mu$m/60 $\mu$m colors; ($c$) the optical spectral types. Luminous
host galaxies show a tendency to host IR-luminous and warm AGNs. The
tendency with infrared luminosity is due to distance-dependent
effects.}
\end{figure}

\setcounter{figure}{15}
\begin{figure}
\caption{Distribution of the host magnitude differences at (left) $R$
and (right) $K^\prime$ among the binary systems (class III) of the
1-Jy sample. Notice the difference in the magnitude scale between the
two figures. The magnitude difference is slightly larger at $K^\prime$
than at $R$, although most systems have a magnitude difference in both
bands of less than 1.5 or, equivalently, a luminosity ratio of less
than 4. }
\end{figure}

\clearpage

\setcounter{figure}{16}
\begin{figure}
\caption{The nuclear (4-kpc diameter) 
$R - K^\prime$ colors plotted against the colors of the host
galaxies. HII galaxies, LINERs, Seyfert 2s and Seyfert 1s are
represented as stars, open circles, filled circles, and filled
squares, respectively.  No significant correlation is observed between
these two quantities, especially when considering only the objects
with $(R - K^\prime)_{\rm host} < 4$.}
\end{figure}

\setcounter{figure}{17}
\begin{figure}
\caption{Seeing-deconvolved half-light radii, $r_e$, of the
elliptical-like host galaxies in the 1-Jy sample measured at ($a$) $R$
and ($b$) $K^\prime$.  The mean $r_e$ of ULIGs at $R$ (4.80 $\pm$ 1.37
kpc) is significantly smaller than the mean values measured by Dunlop
et al. (2002) for the hosts of radio-quiet quasars (7.63 $\pm$ 1.11)
and radio-loud quasars (7.82 $\pm$ 0.71 kpc), although there is some
overlap. In contrast, the mean $r_e$ of ULIGs at $K^\prime$ (3.48
$\pm$ 1.39 kpc) is nearly identical to that of the quasars measured in
the $H$-band by McLeod \& McLeod (2001; 3.39 $\pm$ 1.90). The origin
of the discrepancy between the two quasar datasets is not known. }
\end{figure}

\setcounter{figure}{18}
\begin{figure}
\caption{Surface brightnesses, $\mu_e$, as a function of the
half-light radii, $r_e$, measured at ($a$) $R$ and ($b$) $K^\prime$
for the elliptical-like host galaxies in the 1-Jy sample.  The surface
brightnesses have been corrected for cosmological dimming and the
half-light radii for seeing effects. The hosts of these ULIGs appear
to be smaller and slightly brighter at $R$ than the hosts of the
radio-loud and radio-quiet quasars from the sample of Dunlop et
al. (2002), but still follow with some scatter the $\mu_e$ -- $r_e$
relation of normal (inactive) ellipticals [represented by the dotted
line in panel ($a$); Hamabe \& Kormendy (1987) assuming $V - R$ = 0.5
mag]. At $K^\prime$, the interpretation is more difficult because data
exist for a fewer number of ULIGs and no quasars. The dash line
represents the fit of Hamabe \& Kormendy (1987) assuming $V -
K^\prime$ = 3.7 mag, while the dotted line is the best fit to the
$K$-band data of Pahre (1999) for normal ellipticals.  The hosts of
Seyfert 1 ULIGs are noticeably brighter (especially at $K^\prime$)
than normal galaxies; this may be due to positive residuals from the
PSF subtraction. The solid line in Fig. 19$b$ is the best fit through
the ULIG data, excluding the Seyfert 1 galaxies.}
\end{figure}

\setcounter{figure}{19}
\begin{figure}
\caption{Radial profiles of the axial ratio in the hosts of single-nucleus
ULIGs. The arrow in each panel indicates $ r = 3 \times r_e$.}
\end{figure}

\setcounter{figure}{20}
\begin{figure}
\caption{Axial ratios measured at $3 \times r_e$ in single-nucleus
ULIGs. The median and mean of this distribution are 0.77 and 0.75
$\pm$ 0.15 (1 $\sigma$), respectively. These numbers are similar to those
measured in normal ellipticals and in quasar hosts.}
\end{figure}

\setcounter{figure}{21}
\begin{figure}
\caption{Radial profiles of the normalized $R$-band boxiness
parameter, $A_4/a$, in the host galaxies of single-nucleus ULIGs.
Note the large variations of $A_4/a$ with radius in many of these
objects. The arrows in each panel indicate $r = 2 \times r_e$ and $3
\times r_e$.}
\end{figure}

\setcounter{figure}{22}
\begin{figure}
\caption{$R$-band boxiness values normalized to the semi-major axis at
($a$) $2 \times r_e$ and ($b$) $3 \times r_e$ for ULIGs with
elliptical-like host galaxies. Both distributions are broader than
that of normal ellipticals and peak near zero.}
\end{figure}

\clearpage

\setcounter{figure}{0}
\begin{figure}
\epsscale{1.0}
\plotone{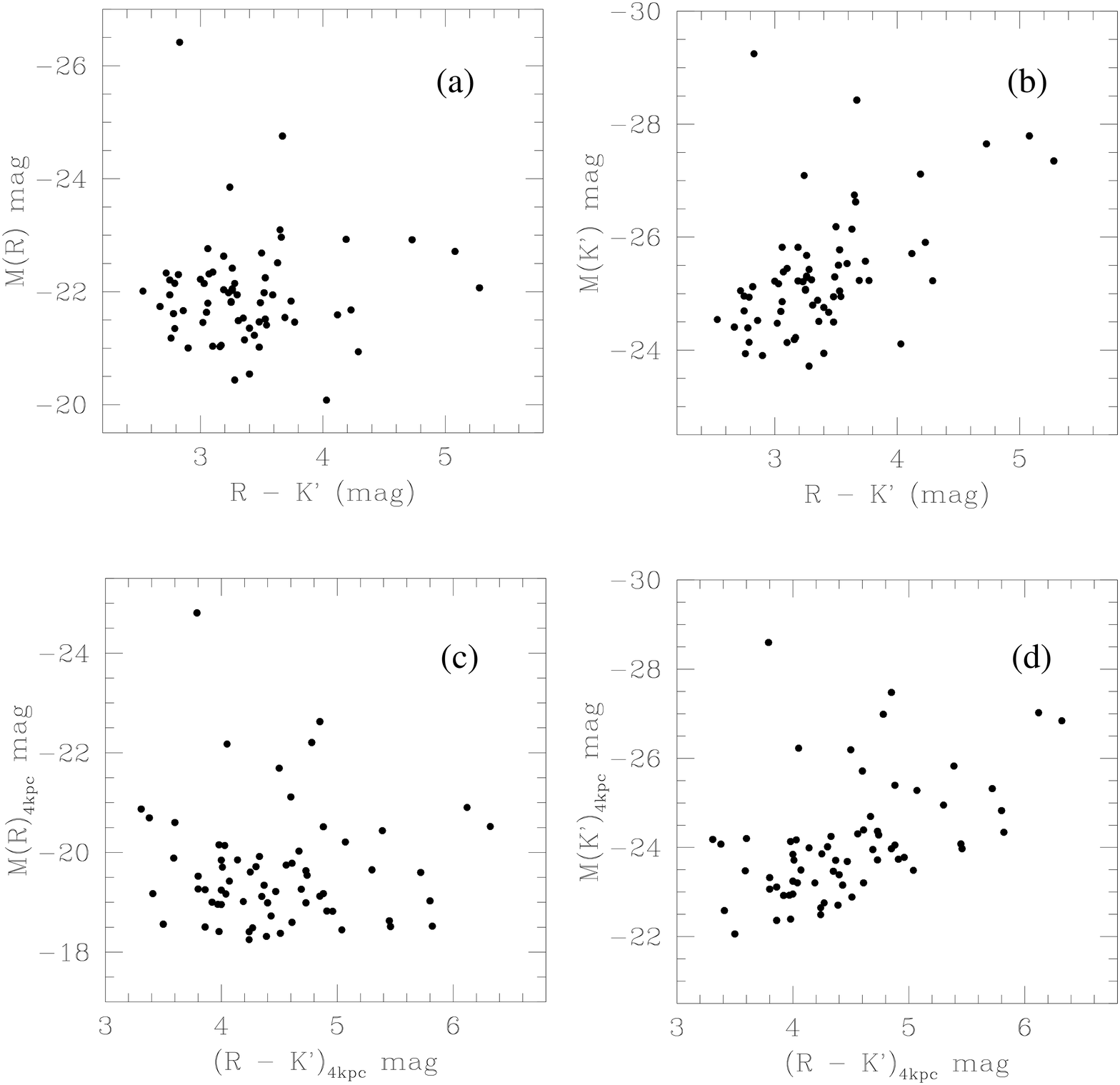}
\caption{}
\end{figure}

\clearpage

\setcounter{figure}{1}
\begin{figure}
\epsscale{1.0}
\plotone{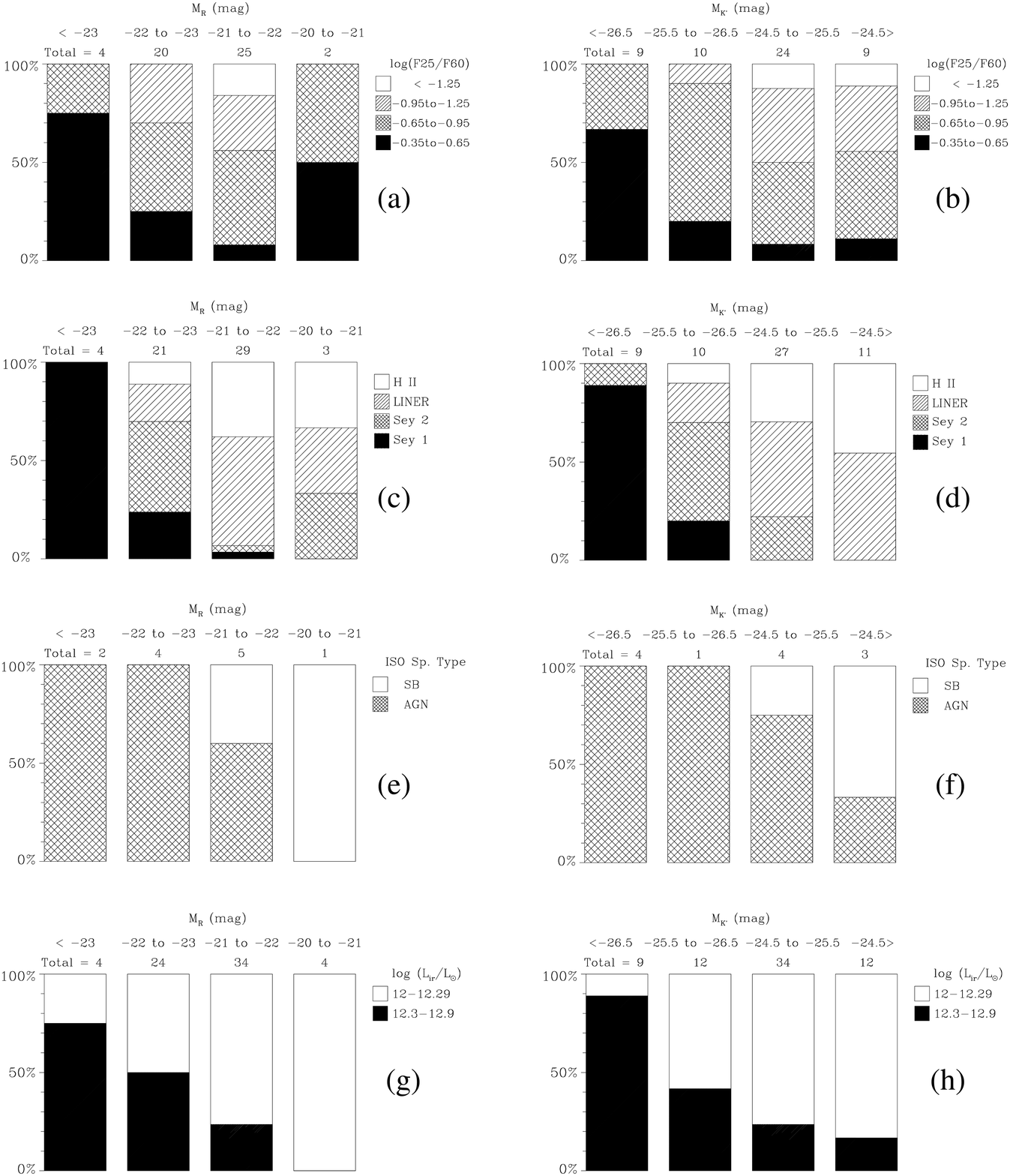}
\caption{}
\end{figure}

\clearpage

\setcounter{figure}{2}
\begin{figure}
\epsscale{1.}
\plotone{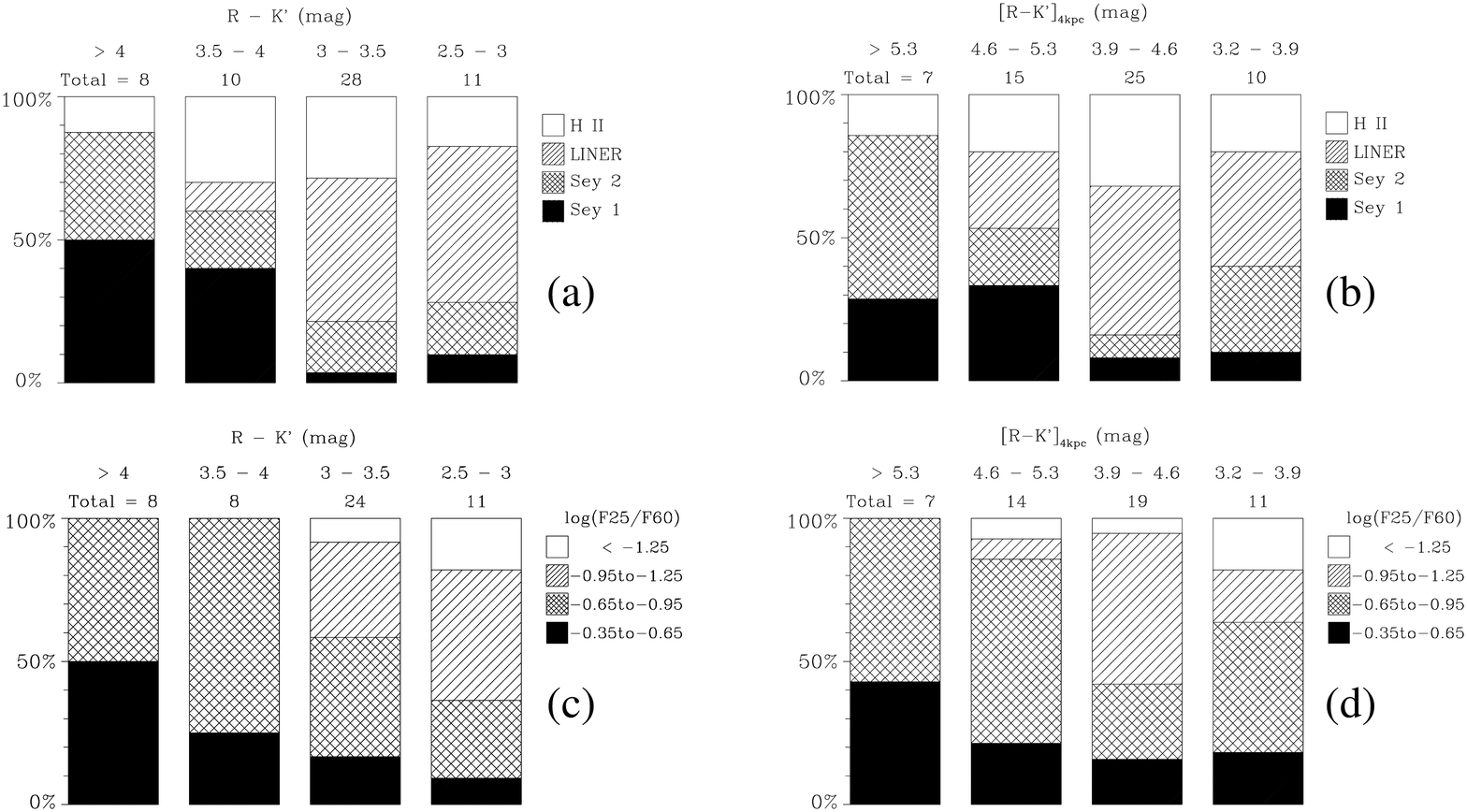}
\caption{}
\end{figure}

\clearpage

\setcounter{figure}{3}
\begin{figure}
\epsscale{1.}
\plotone{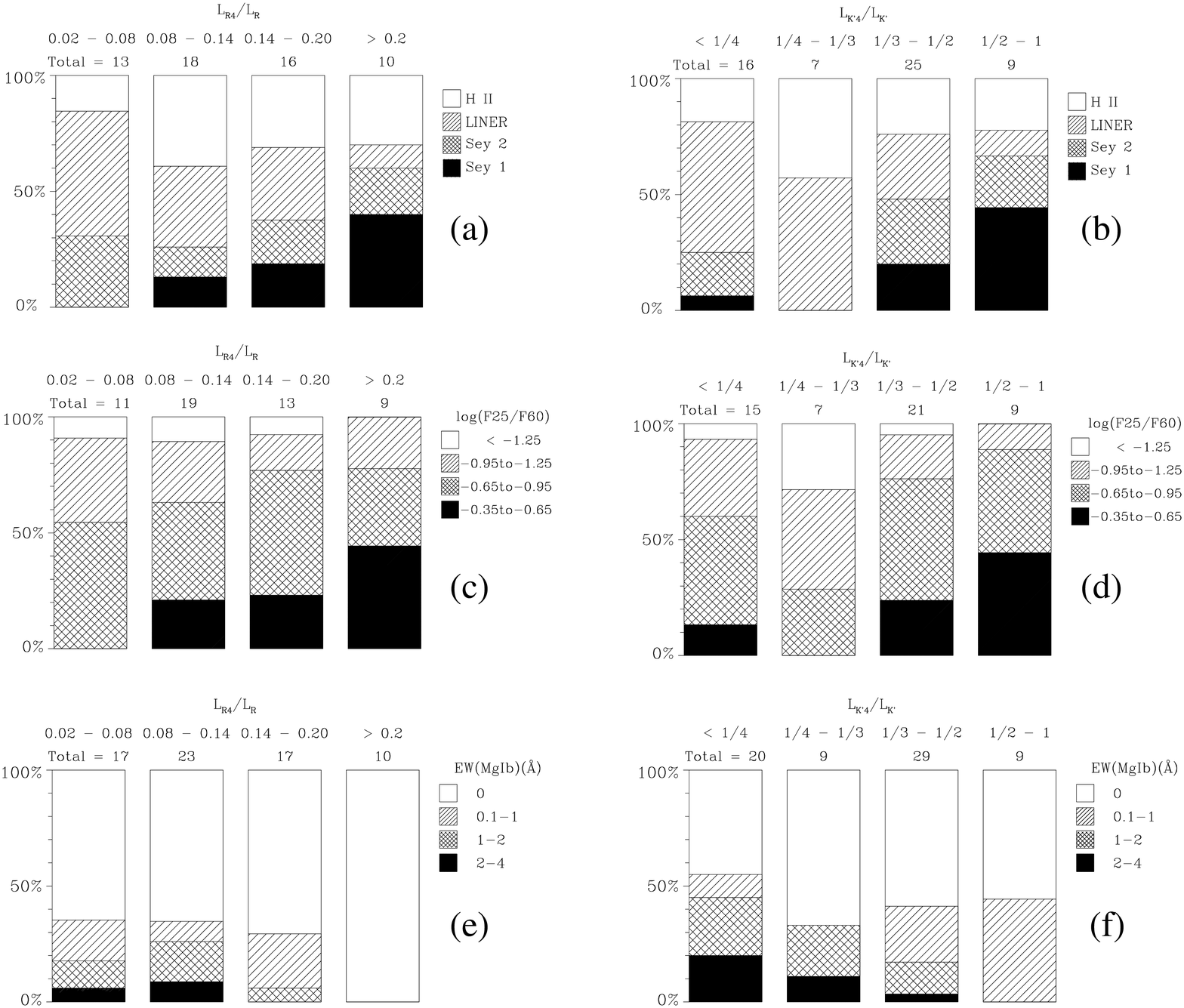}
\caption{}
\end{figure}

\clearpage

\setcounter{figure}{4}
\begin{figure}
\plotone{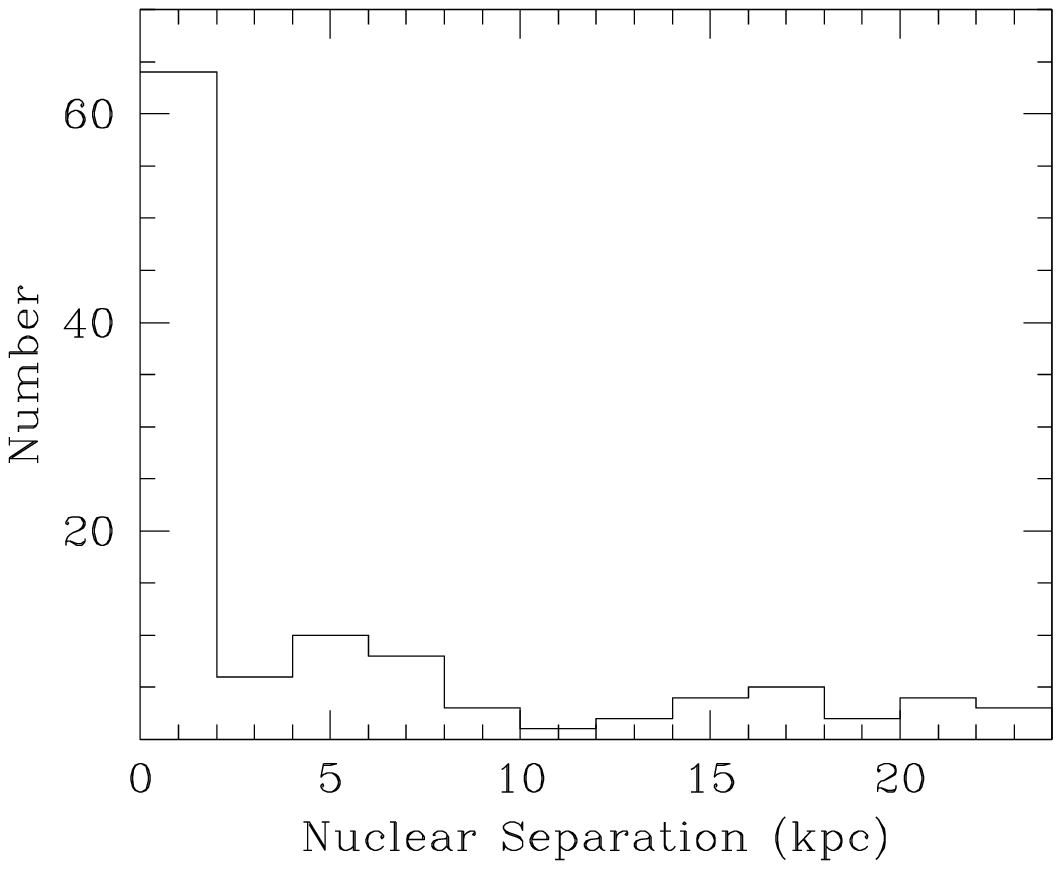}
\caption{}
\end{figure}

\clearpage

\setcounter{figure}{5}
\begin{figure}
\epsscale{0.85}
\plotone{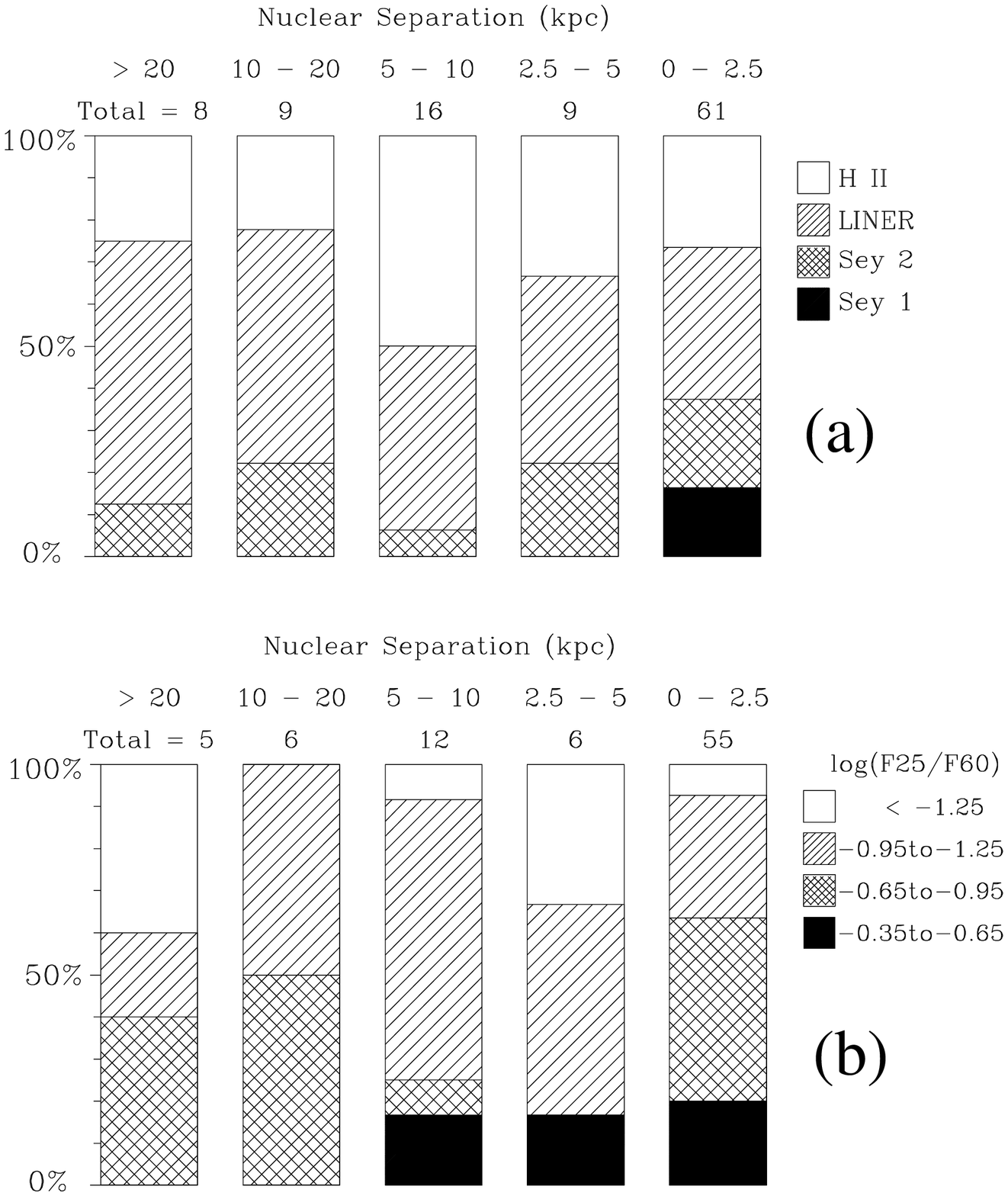}
\caption{}
\end{figure}

\clearpage

\setcounter{figure}{6}
\begin{figure}
\epsscale{1.0}
\plotone{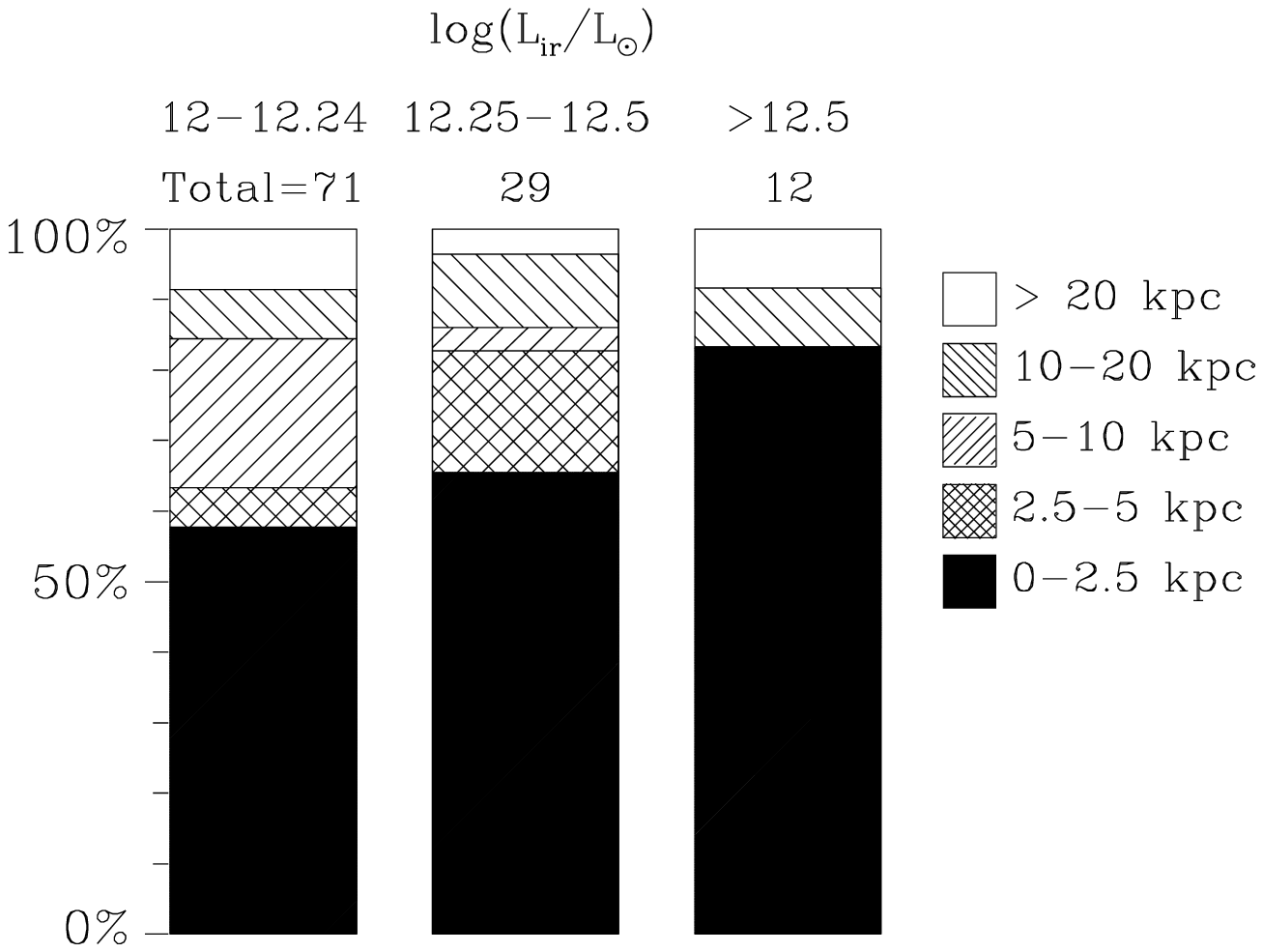}
\caption{}
\end{figure}

\clearpage

\setcounter{figure}{7}
\epsscale{1.}
\begin{figure}
\plotone{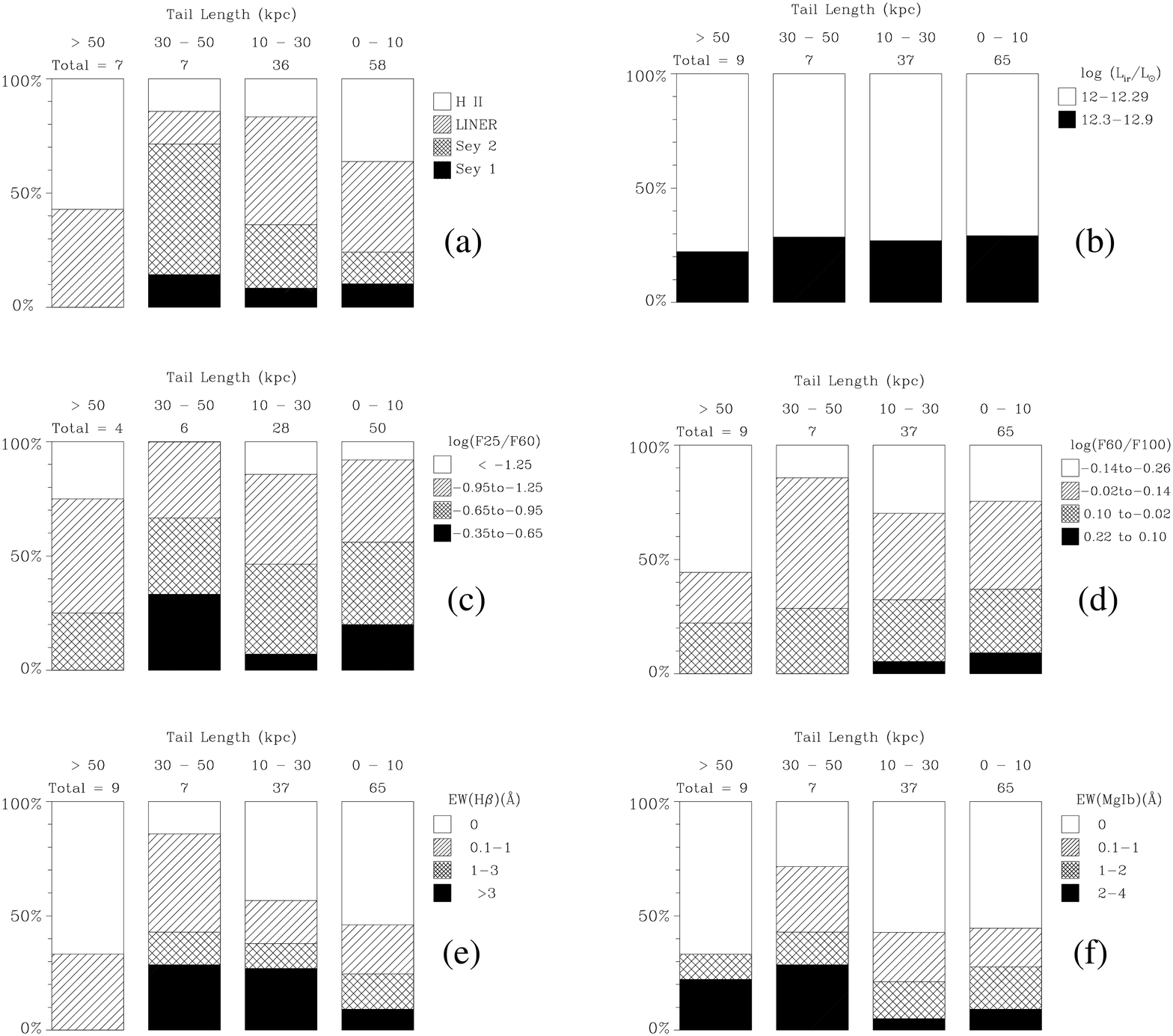}
\caption{}
\end{figure}

\clearpage

\setcounter{figure}{8}
\begin{figure}
\epsscale{1.0}
\plotone{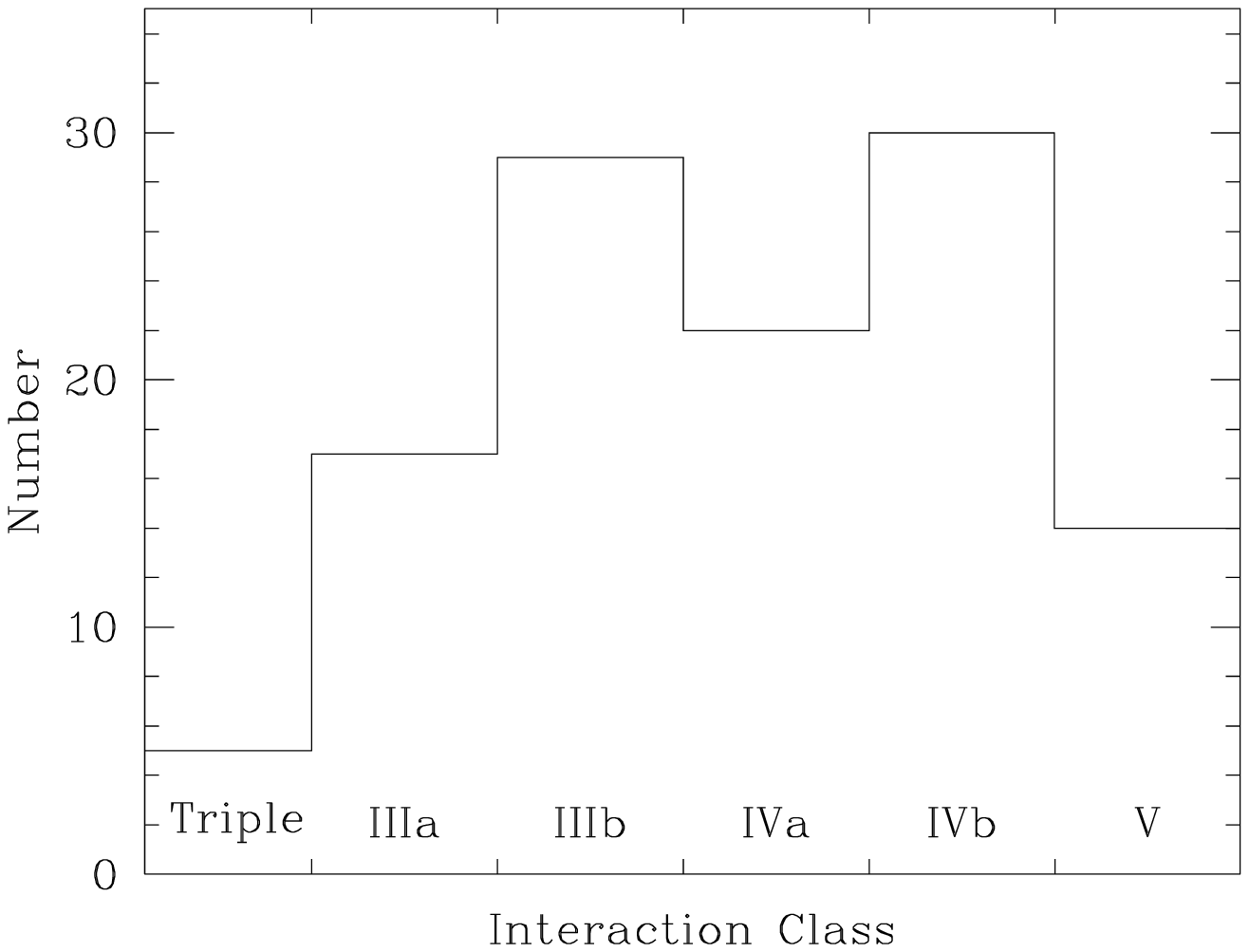}
\caption{}
\end{figure}

\clearpage

\setcounter{figure}{9}
\begin{figure}
\epsscale{0.65}
\plotone{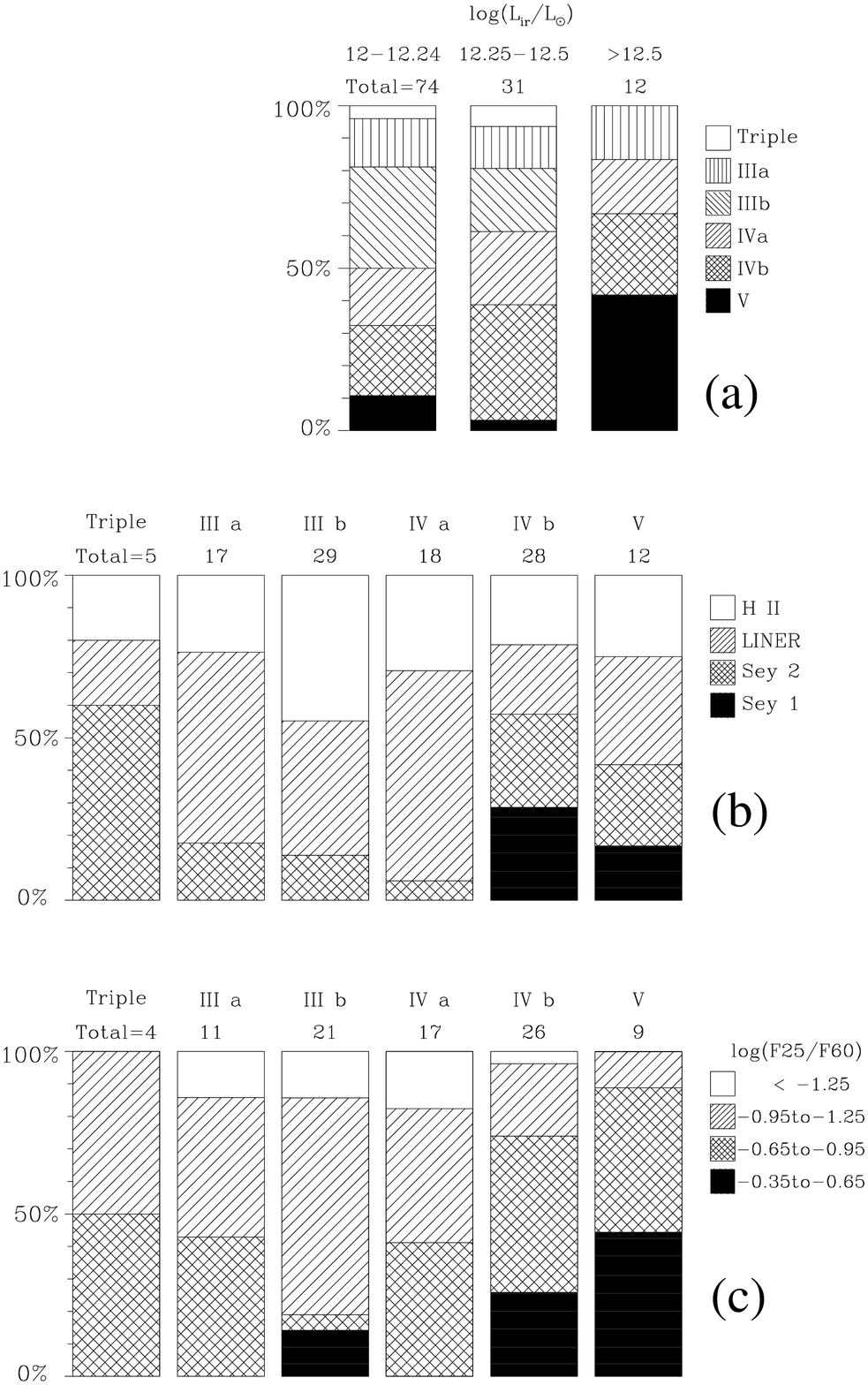}
\caption{}
\end{figure}

\clearpage

\setcounter{figure}{11}
\epsscale{0.6}
\begin{figure}
\plotone{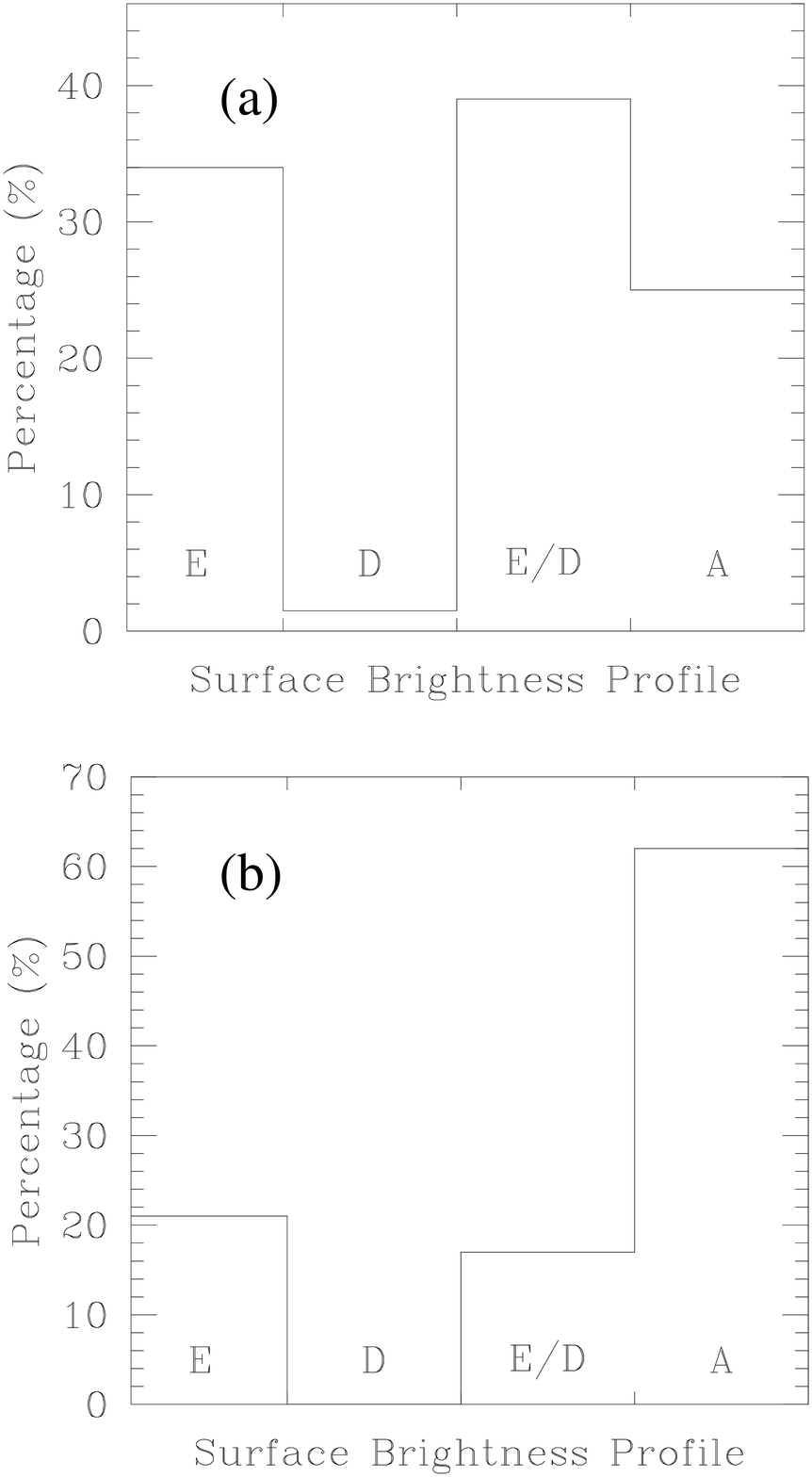}
\caption{}
\end{figure}

\clearpage

\setcounter{figure}{12}
\epsscale{1.}
\begin{figure}
\plotone{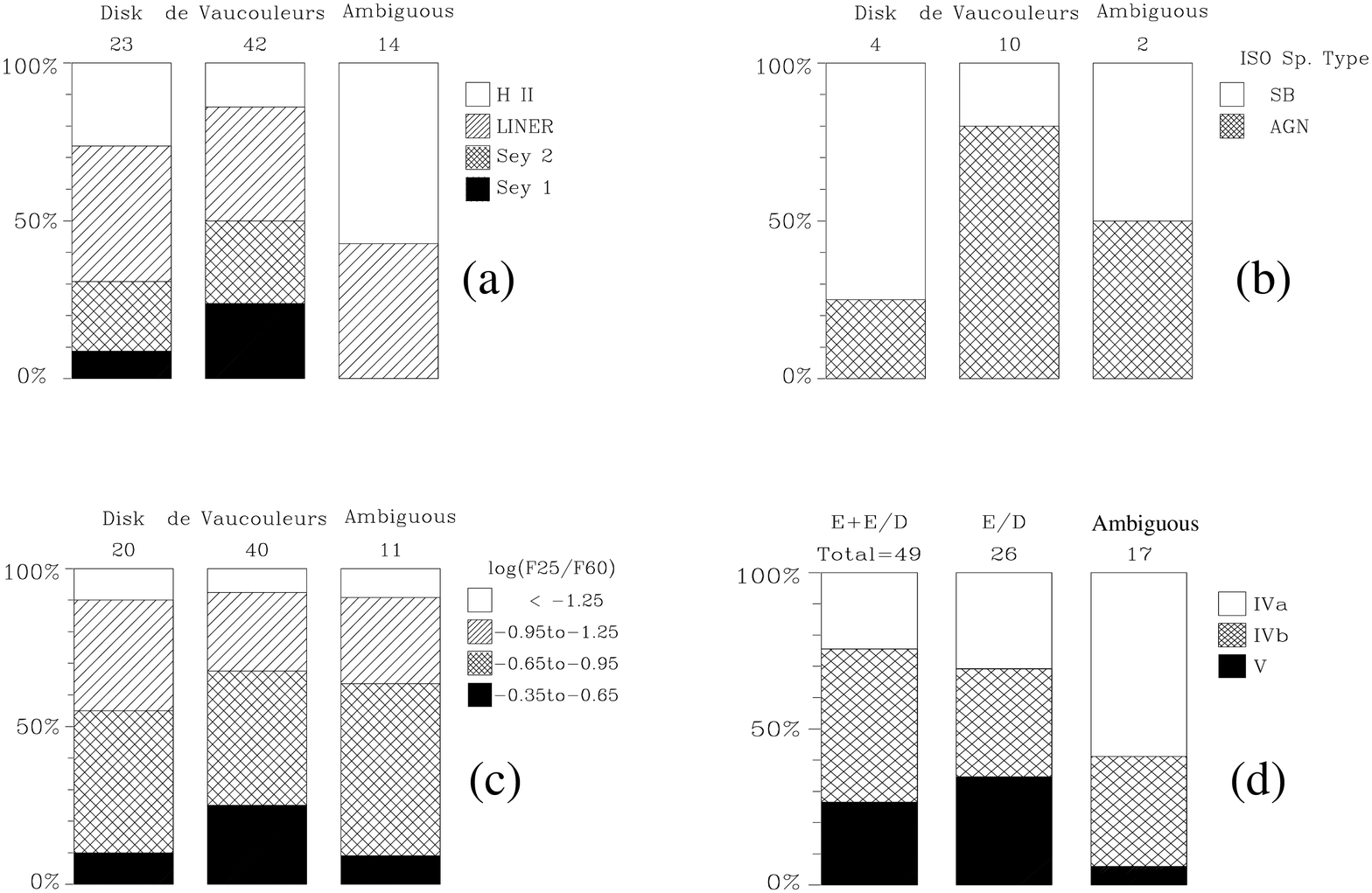}
\caption{}
\end{figure}

\clearpage

\setcounter{figure}{13}
\epsscale{1.0}
\begin{figure}
\plotone{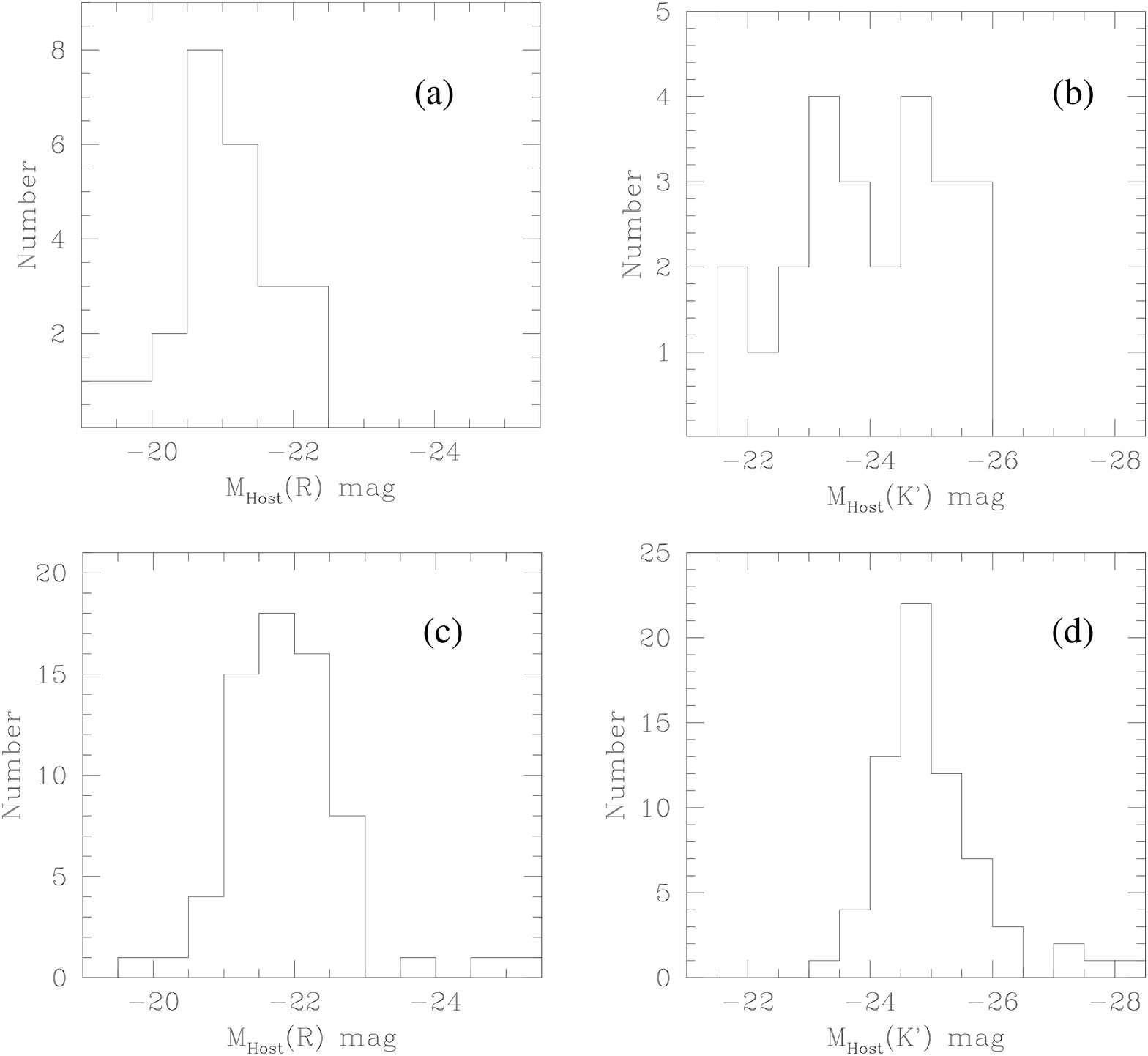}
\caption{}
\end{figure}

\clearpage

\setcounter{figure}{14}
\epsscale{0.6}
\begin{figure}
\plotone{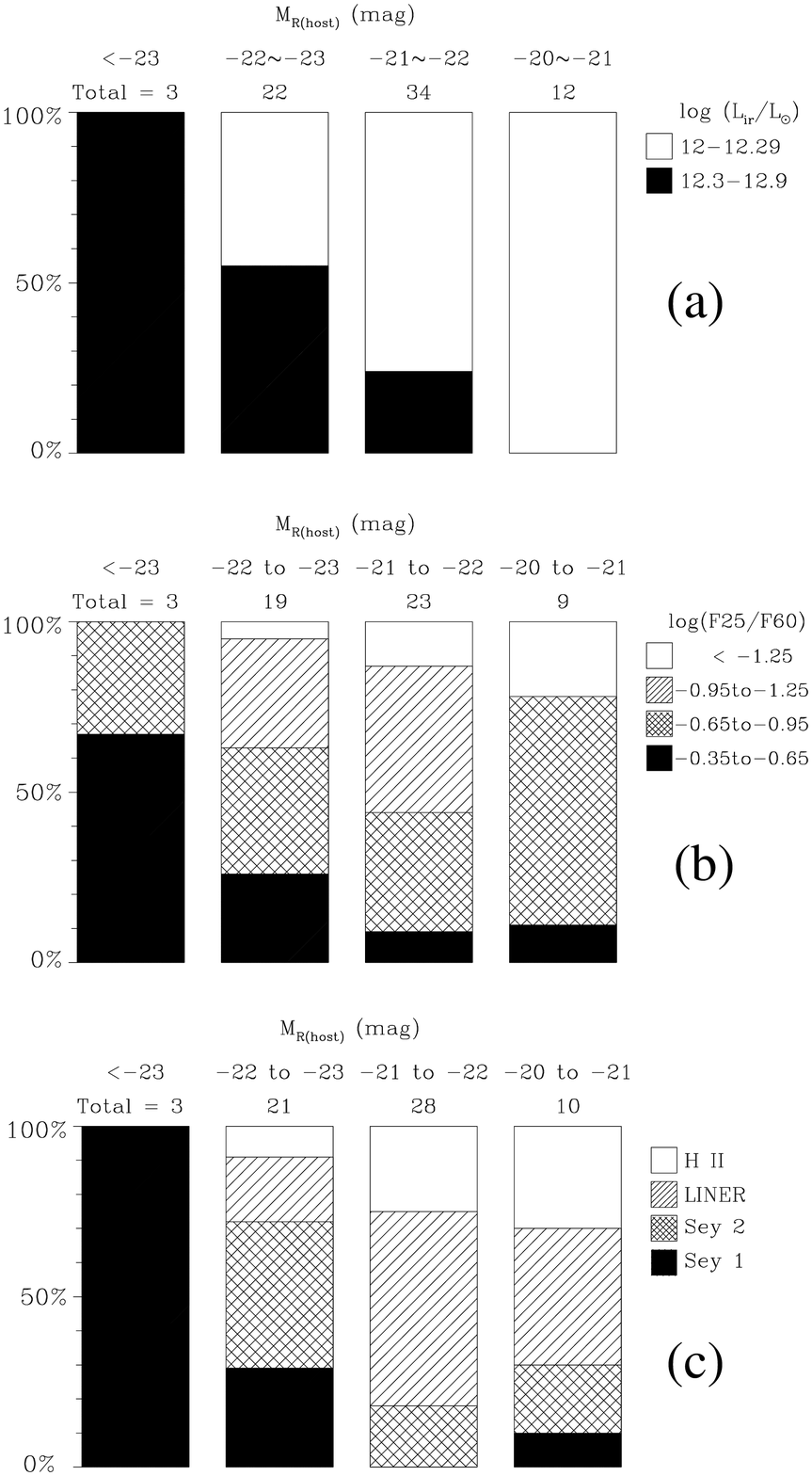}
\caption{}
\end{figure}

\clearpage

\setcounter{figure}{15}
\epsscale{1.0}
\begin{figure}
\plotone{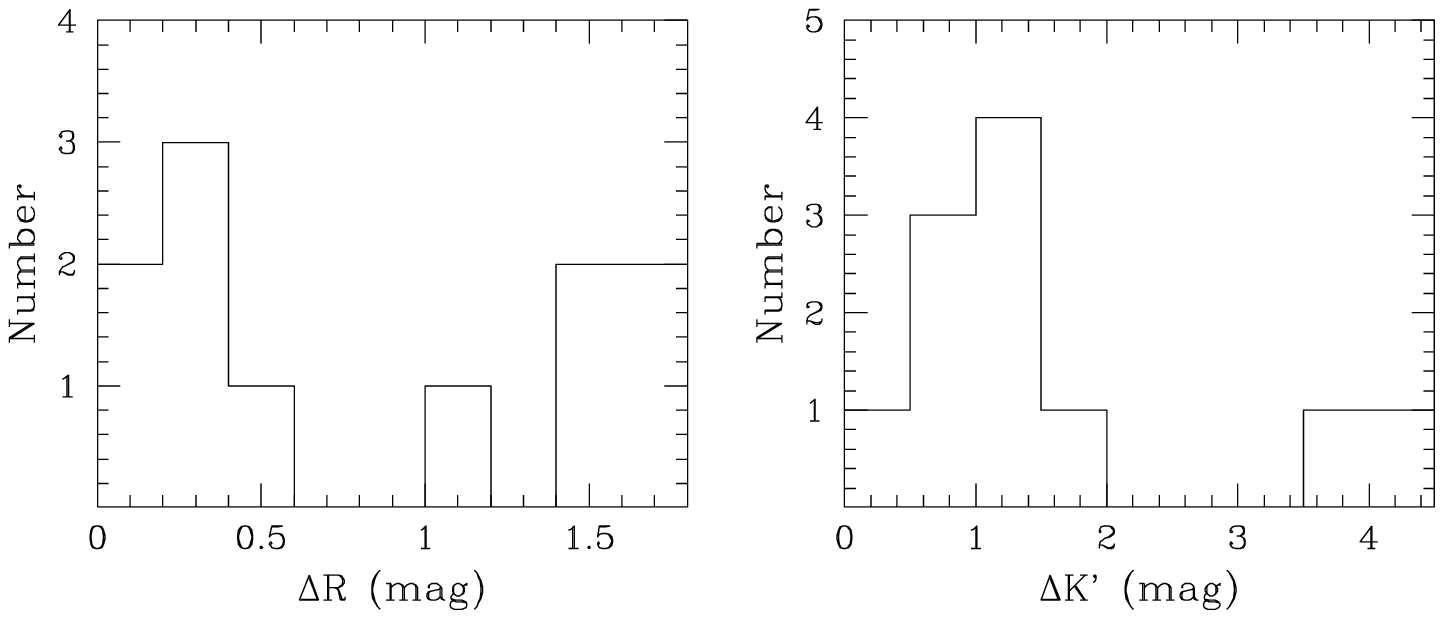}
\caption{}
\end{figure}

\clearpage

\setcounter{figure}{16}
\epsscale{1.0}
\begin{figure}
\plotone{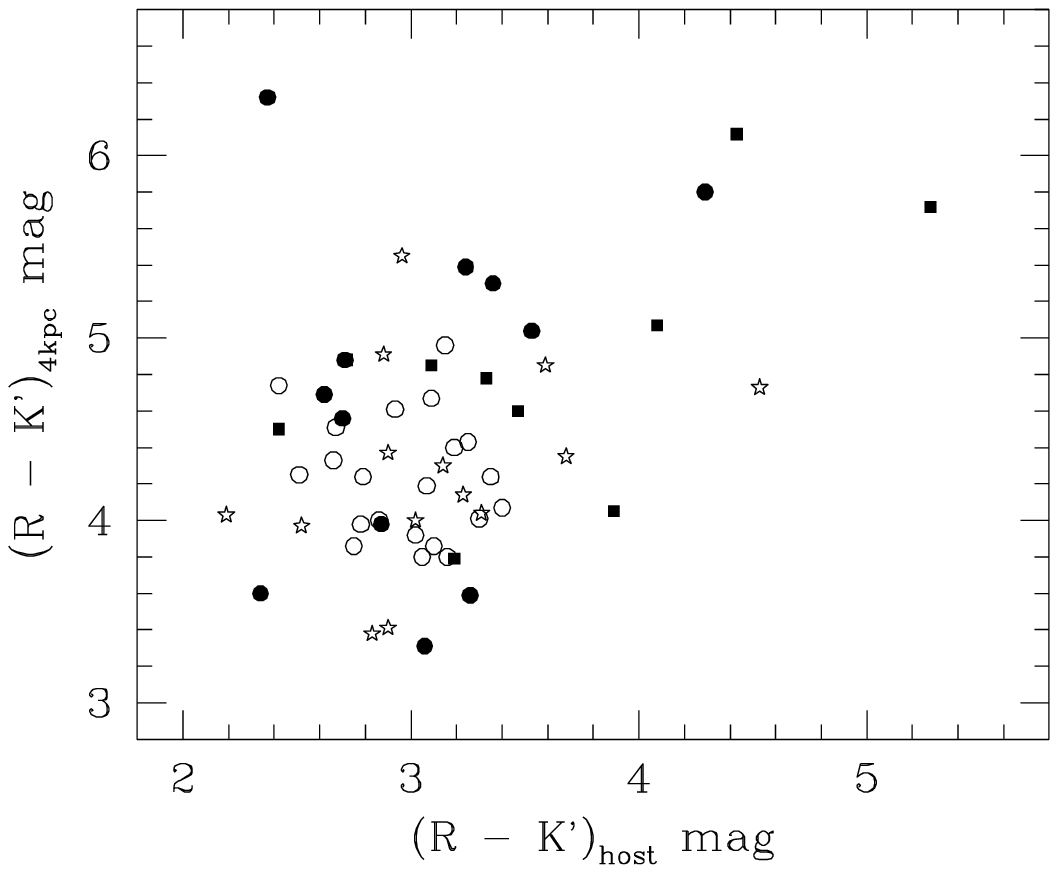}
\caption{}
\end{figure}

\clearpage

\setcounter{figure}{17}
\epsscale{0.6}
\begin{figure}
\plotone{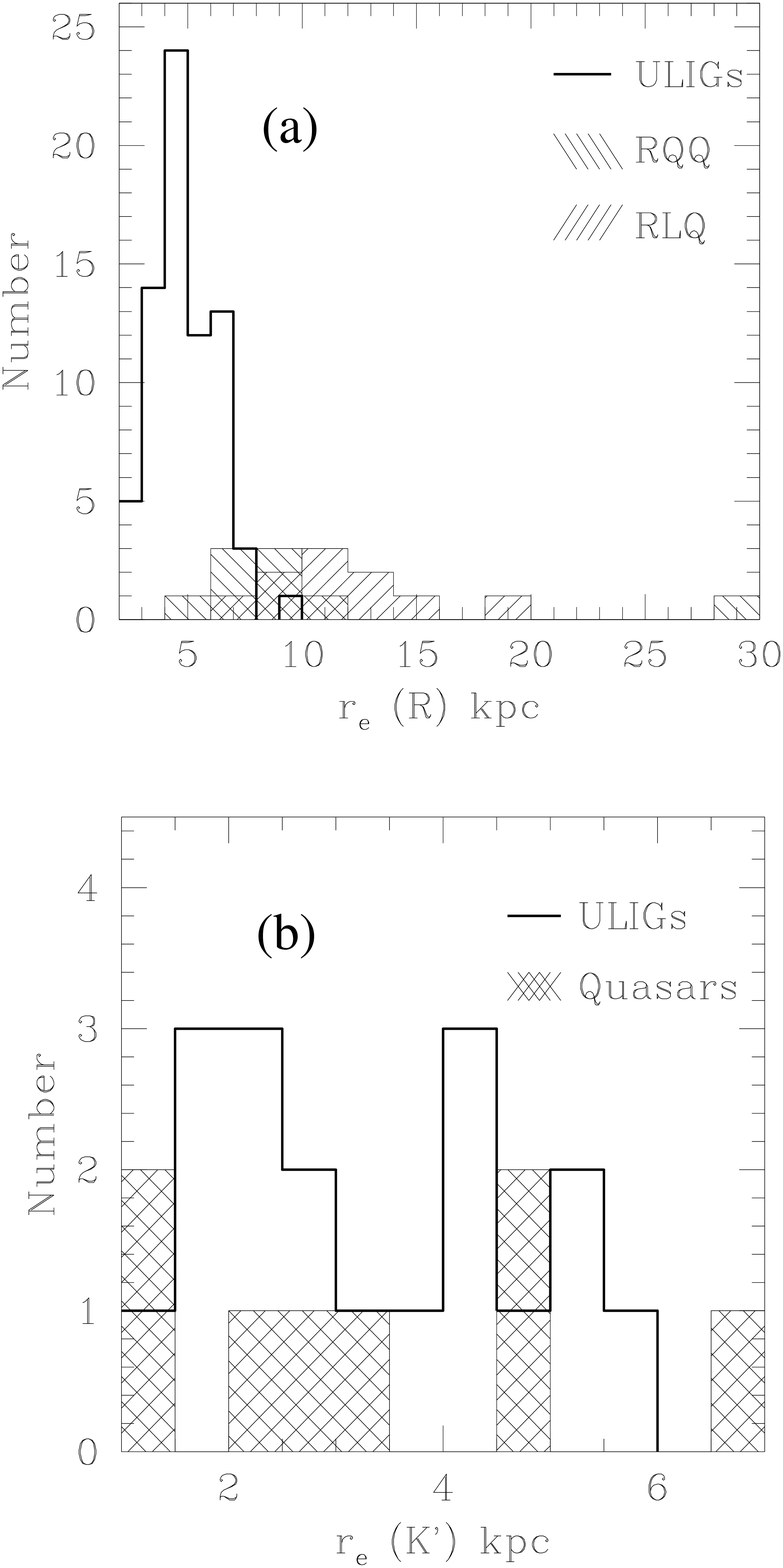}
\caption{}
\end{figure}

\clearpage

\setcounter{figure}{18}
\epsscale{0.6}
\begin{figure}
\plotone{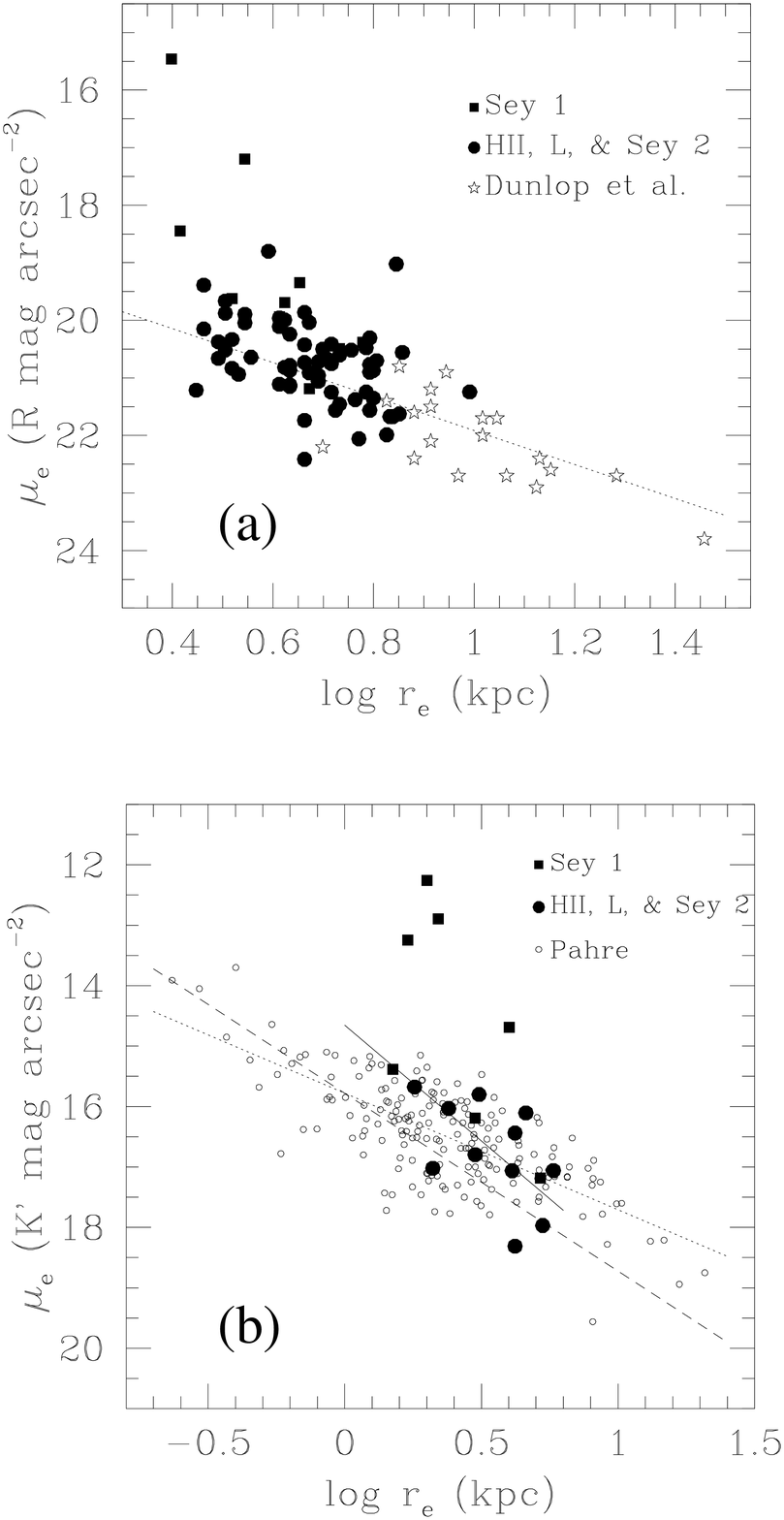}
\caption{}
\end{figure}

\setcounter{figure}{19}
\epsscale{1.0}
\begin{figure}
\caption{}
\end{figure}

%
%
%
%

\setcounter{figure}{20}
\epsscale{0.8}
\begin{figure}
\plotone{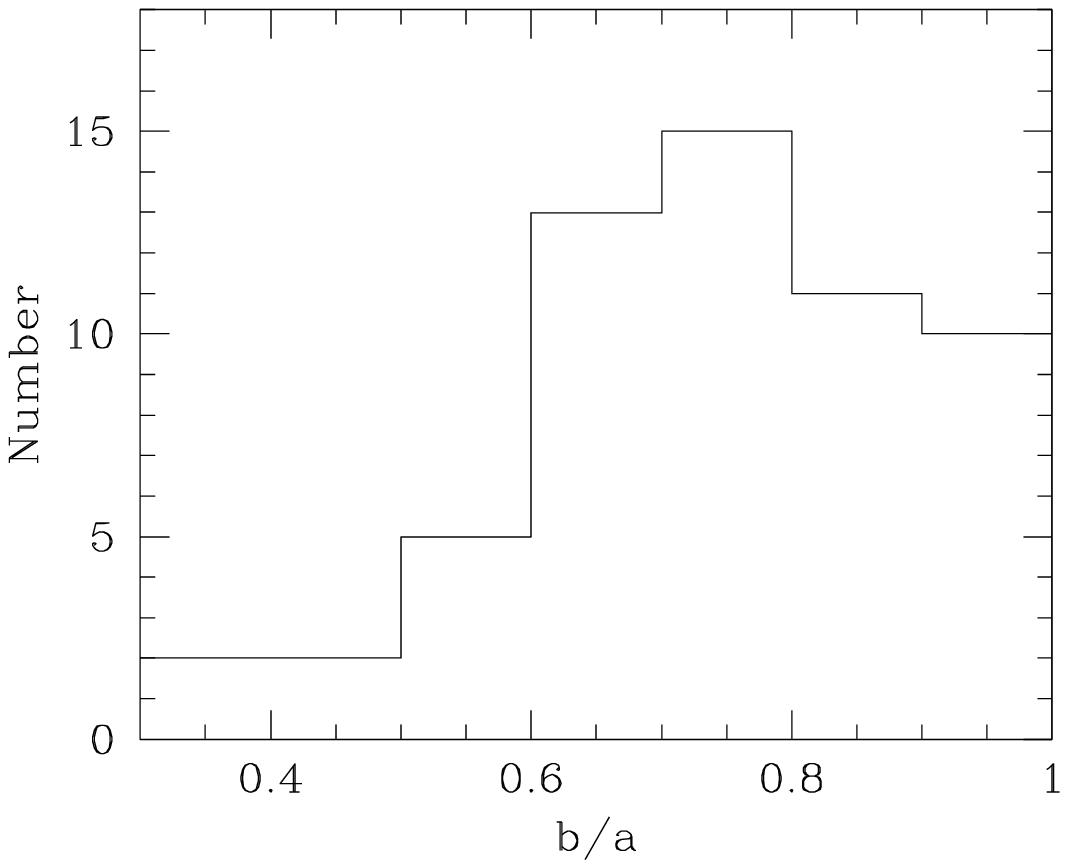}
\caption{}
\end{figure}

\clearpage

\setcounter{figure}{21}
\epsscale{1.0}
\begin{figure}
\caption{}
\end{figure}

%
%
%

\setcounter{figure}{22}
\epsscale{0.6}
\begin{figure}
\plotone{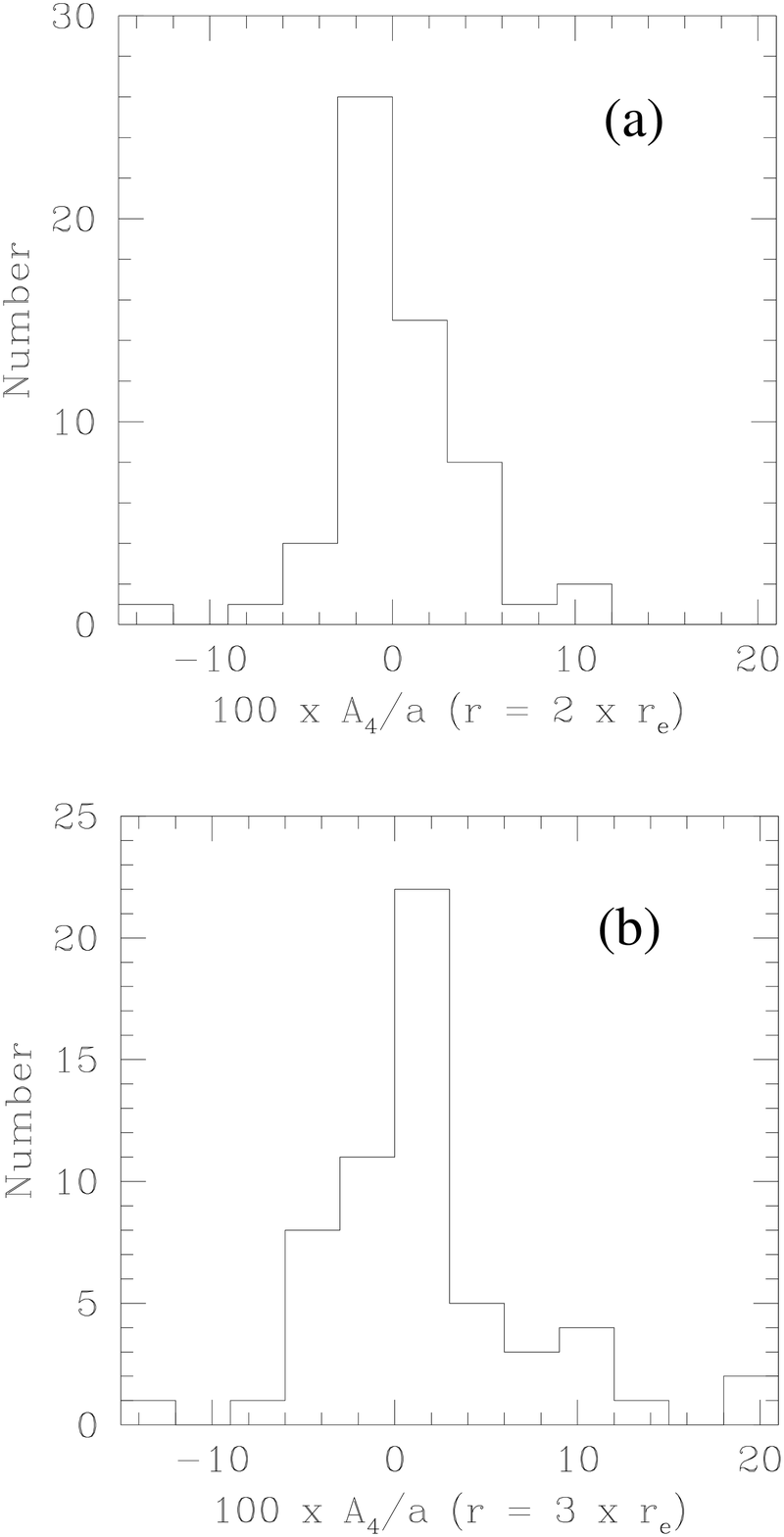}
\caption{}
\end{figure}

\end{document}


\newcommand{\ts}{\thinspace} 
\begin{deluxetable}{lrrcclrrcc}
\footnotesize
\tablenum{1}
\tablewidth{0pt}
\tablecaption{Morphological Parameters}
\tablehead{
\multicolumn{1}{c}{IRAS} &
\multicolumn{1}{c}{NS} &
\multicolumn{1}{c}{TL} &
\multicolumn{1}{c}{IC} &
\multicolumn{1}{c}{ST} &
\multicolumn{1}{c}{IRAS} &
\multicolumn{1}{c}{NS} &
\multicolumn{1}{c}{TL} &
\multicolumn{1}{c}{IC} &
\multicolumn{1}{c}{ST} \nl
\multicolumn{1}{c}{$FSC$} &
\multicolumn{1}{c}{kpc} &
\multicolumn{1}{c}{kpc} &
\multicolumn{1}{c}{} &
\multicolumn{1}{c}{} &
\multicolumn{1}{c}{$FSC$} &
\multicolumn{1}{c}{kpc} &
\multicolumn{1}{c}{kpc} &
\multicolumn{1}{c}{} &
\multicolumn{1}{c}{}\nl
\multicolumn{1}{c}{(1)} &
\multicolumn{1}{c}{(2)} &
\multicolumn{1}{c}{(3)} &
\multicolumn{1}{c}{(4)} &
\multicolumn{1}{c}{(5)} &
\multicolumn{1}{c}{(1)} &
\multicolumn{1}{c}{(2)} &
\multicolumn{1}{c}{(3)} &
\multicolumn{1}{c}{(4)} &
\multicolumn{1}{c}{(5)} 
}
\startdata
00091$-$0738  & 2.1&12     &IIIb&HII &  12359$-$0725  &22.9&16     &IIIa&  L \nl
00188$-$0856  &    &       &   V&  L &  12447$+$3721  &    &       & IVa&HII \nl
00397$-$1312  &    &       &   V&HII &  12540$+$5708  &    &14     & IVb& S1 \nl
00456$-$2904  &20.7&9      &IIIa&HII &  13106$-$0922  &    &82     & IVa&  L \nl
00482$-$2721  & 6.7&15     &IIIb&  L &  13218$+$0552  &    &       &   V& S1 \nl
01004$-$2237  &    &       &   V&HII &  13305$-$1739  &    &       &   V& S2 \nl
01166$-$0844  & 9.8&7      &IIIb&HII &  13335$-$2612  & 3.2&14     &IIIb&  L \nl
01199$-$2307  &20.3&       &IIIa&HII &  13342$+$3932  &    &8      & IVb& S1 \nl
01298$-$0744  &    &9      & IVb&HII &  13428$+$5608  &    &31     & IVb& S2 \nl
01355$-$1814  & 5.8&       &IIIb&HII &  13443$+$0802  &    &       & Tpl& S2 \nl
01494$-$1845  &    &       & IVa&    &  13451$+$1232  & 4.0&35     &IIIb& S2 \nl
01569$-$2939  &    &55     & IVa&HII &  13454$-$2956  &15.8&10     &IIIa& S2 \nl
01572$+$0009  &    &46     & IVb& S1 &  13469$+$5833  & 4.3&56     &IIIb&HII \nl
02021$-$2103  &    &       & IVa&    &  13509$+$0442  &    &15     & IVb&HII \nl
02411$+$0353  & 7.1&       &IIIb&HII &  13539$+$2920  & 7.0&       &IIIb&HII \nl
02480$-$3745  &    &27     & IVa&    &  14053$-$1958  & 1.5&11     &IIIb& S2 \nl
03209$-$0806  &    &61     & IVb&HII &  14060$+$2919  &    &9      & IVa&HII \nl
03250$+$1606  &    &       & IVb&  L &  14070$+$0525  &    &       &   V& S2 \nl
03521$+$0028  & 3.5&       &IIIb&  L &  14121$-$0126  & 9.1&22     &IIIb&  L \nl
04074$-$2801  &    &28     & IVa&  L &  14197$+$0813  &    &       &   V&    \nl
04103$-$2838  &    &8      & IVb&  L &  14202$+$2615  &14.8&19     &IIIa&HII \nl
04313$-$1649  &    &59     & IVa&    &  14252$-$1550  & 8.3&       &IIIb&  L \nl
05020$-$2941  &    &8      & IVa&  L &  14348$-$1447  & 4.8&12     &IIIb&  L \nl
05024$-$1941  &    &12     & IVa& S2 &  14394$+$5332  &    &35     & Tpl& S2 \nl
05156$-$3024  &    &10     & IVb& S2 &  14485$-$2434  &    &       & IVb&    \nl
05189$-$2524  &    &20     & IVb& S2 &  15001$+$1433  &    &       & Tpl& S2 \nl
07599$+$6508  &    &       & IVb& S1 &  15043$+$5754  & 5.1&       &IIIb&HII \nl
08201$+$2801  &    &25     & IVa&HII &  15130$-$1958  &    &14     & IVb& S2 \nl
08474$+$1813  &    &       &   V&    &  15206$+$3342  &    &       & IVb&HII \nl
08559$+$1053  &    &18     & IVb& S2 &  15225$+$2350  &    &31     & IVa&HII \nl
08572$+$3915  & 5.7&5      &IIIb&  L &  15327$+$2340  & 0.4&11     &IIIb&  L \nl
08591$+$5248  &    &       &   V&    &  15462$-$0450  &    &8      & IVb& S1 \nl
09039$+$0503  &    &24     & IVa&  L &  16090$-$0139  &    &       & IVa&  L \nl
09116$+$0334  &15.8&12     &IIIa&  L &  16156$+$0146  & 7.5&       &IIIb& S2 \nl
09463$+$8141  &    &76     & IVa&  L &  16300$+$1558  &    &       &   V&  L \nl
09539$+$0857  &    &       &   V&  L &  16333$+$4630  &13.1&       &IIIa&  L \nl
10035$+$2740  &    &54     & IVa&    &  16468$+$5200  & 7.8&14     &IIIb&  L \nl
10091$+$4704  &    &29     & IVa&  L &  16474$+$3430  & 6.5&11     &IIIb&HII \nl
10190$+$1322  & 5.5&       &IIIb&HII &  16487$+$5447  & 5.4&       &IIIb&  L \nl
10378$+$1108  &    &       & IVb&  L &  17028$+$5817  &23.3&       &IIIa&  L \nl
10485$-$1447  &20.0&56     &IIIa&  L &  17044$+$6720  &    &8      & IVb&  L \nl
10494$+$4424  &    &12     & IVb&  L &  17068$+$4027  &    &       & Tpl&HII \nl
10594$+$3818  & 4.4&90     &IIIb&HII &  17179$+$5444  &    &7      & IVb& S2 \nl
11028$+$3130  &    &5      & IVa&  L &  20414$-$1651  &    &       & IVb&HII \nl
11095$-$0238  &    &14     & IVb&  L &  21208$-$0519  &14.1&14     &IIIa&HII \nl
11119$+$3257  &    &25     & IVb& S1 &  21219$-$1757  &    &       &   V& S1 \nl
11130$-$2659  &    &21     & IVa&  L &  21329$-$2346  &    &49     & IVa&  L \nl
11180$+$1623  &21.8&22     &IIIa&  L &  21477$+$0502  &    &       & Tpl&  L \nl
11223$-$1244  &87.9&35     &IIIa& S2 &  22088$-$1831  & 4.3&       &IIIb&HII \nl
11387$+$4116  &    &       &   V&HII &  22206$-$2715  & 7.7&       &IIIb&HII \nl
11506$+$1331  &    &       & IVb&HII &  22491$-$1808  & 2.2&       &IIIb&HII \nl
11582$+$3020  &    &       &   V&  L &  22541$+$0833  &18.5&       &IIIa& S2 \nl
11598$-$0112  &    &12     & IVb& S1 &  23060$+$0505  &    &22     & IVb& S2 \nl
12018$+$1941  &    &       & IVb&    &  23129$+$2548  &    &11     & IVa&  L \nl
12032$+$1707  &12.0&       &IIIa&  L &  23233$+$2817  &    &       &Iso.& S2 \nl
12072$-$0444  &    &28     & IVb& S2 &  23234$+$0946  & 7.4&18     &IIIb&  L \nl
12112$+$0305  & 3.8&8      &IIIb&  L &  23327$+$2913  &22.7&7      &IIIa&  L \nl
12127$-$1412  &21.9&       &IIIa&  L &  23389$+$0300  & 4.9&       &IIIb& S2 \nl
12265$+$0219  &    &       & IVb& S1 &  23498$+$2423  &12.7&17     &IIIa& S2 \nl
\tablenotetext{}
{{\it Col 1:}\ $IRAS$ name}
\tablenotetext{}
{{\it Col 2:}\ Nuclear Separation in kpc unit}
\tablenotetext{}
{{\it Col 3:}\ Tail length in kpc unit}
\tablenotetext{}
{{\it Col 4:}\ Morphological class}
\tablenotetext{}
{{\it Col 5:}\ Spectral type: HII = H II region galaxy, L = LINER, S2 = Seyfert~2, S1 = Seyfert~1}
\enddata
\end{deluxetable}


\newcommand{\ts}{\thinspace} 
\begin{deluxetable}{lccccccccccccc}
\footnotesize
\tablenum{2}
\tablewidth{0pt}
\tablecaption{Parameters of Surface Brightness Profiles}
\tablehead{
\multicolumn{1}{l}{IRAS FSC} &
\multicolumn{2}{c}{PSF (\%)} &
\multicolumn{4}{c}{$\chi^2_r$ value} &
\multicolumn{1}{c}{Best Fit} &
\multicolumn{2}{c}{$r_e$(kpc)} &
\multicolumn{2}{c}{$\mu_e$(mag/as$^2$)} &
\multicolumn{2}{c}{Host Galaxy(mag)} \nl
\cline{2-7}
\cline{9-14}
\multicolumn{1}{c}{} &
\multicolumn{1}{c}{R} &
\multicolumn{1}{c}{$K^\prime$} &
\multicolumn{1}{c}{R(DP)} &
\multicolumn{1}{c}{R(EP)} &
\multicolumn{1}{c}{$K^\prime$(DP)} &
\multicolumn{1}{c}{$K^\prime$(EP)} &
\multicolumn{1}{c}{} &
\multicolumn{1}{c}{R} &
\multicolumn{1}{c}{$K^\prime$} &
\multicolumn{1}{c}{R} &
\multicolumn{1}{c}{$K^\prime$} &
\multicolumn{1}{c}{M$_{R}$} &
\multicolumn{1}{c}{M$_{K^\prime}$} \nl
\multicolumn{1}{c}{(1)} &
\multicolumn{1}{c}{(2)} &
\multicolumn{1}{c}{(3)} &
\multicolumn{1}{c}{(4)} &
\multicolumn{1}{c}{(5)} &
\multicolumn{1}{c}{(6)} &
\multicolumn{1}{c}{(7)} &
\multicolumn{1}{c}{(8)} &
\multicolumn{1}{c}{(9)} &
\multicolumn{1}{c}{(10)} &
\multicolumn{1}{c}{(11)} &
\multicolumn{1}{c}{(12)} &
\multicolumn{1}{c}{(13)} &
\multicolumn{1}{c}{(14)} 
}
\startdata
 00091$-$0738    &\nodata & \nodata &     5.2 &    2.2 &   27.2 &   11.0 &   E  &  3.2   &\nodata & 21.0    &\nodata & -21.37  & -24.37 \nl
 00188$-$0856    &\nodata & 53      &     0.8 &    0.4 &    8.3 &    0.7 &   E  &  3.2   &1.8     & 20.4    &16.2    & -21.81  & -24.90 \nl
 00397$-$1312    &\nodata & \nodata &     0.3 &    1.6 &    0.5 &    1.0 &  E/D &  4.2   &\nodata & 21.0    &\nodata & -21.94  & -25.53 \nl
 00456$-$2904    &\nodata & \nodata &     1.2 &    1.5 &    4.2 &    0.5 &   E  &  3.5   &\nodata & 20.5    &\nodata & -22.08  & -25.17 \nl
 00482$-$2721    &\nodata & \nodata &    22.9 &   16.2 &    7.1 &    7.8 &   A  &\nodata &\nodata & \nodata &\nodata & -21.17  & -24.08 \nl
 01004$-$2237    &     71 &      50 &     0.8 &    0.2 &    0.4 &    1.0 &  E/D &  2.8   &\nodata & 21.7    &\nodata & -19.52  & -24.05 \nl
 01166$-$0844    &\nodata & \nodata &    21.7 &   63.3 &   37.3 &   52.5 &   A  &\nodata &\nodata & \nodata &\nodata & -21.60  & -24.49 \nl
 01199$-$2307    &\nodata & \nodata &     0.4 &    1.8 &    0.4 &    0.4 &  E/D &  4.3   &\nodata & 21.5    &\nodata & -21.99  & -24.14 \nl
 01298$-$0744    &\nodata & 76      &    23.1 &   52.1 &    1.5 &    1.9 &   A  &\nodata &\nodata & \nodata &\nodata & -21.18  & -23.70 \nl
 01355$-$1814    &\nodata & \nodata &    51.2 &   58.2 &   36.0 &   24.9 &   A  &\nodata &\nodata & \nodata &\nodata & -21.43  & -24.21 \nl
 01494$-$1845    &\nodata & \nodata &     5.6 &    1.9 &    9.2 &    6.7 &   E  &  6.2   &\nodata & 22.2    &\nodata & -21.54  & -25.23 \nl
 01569$-$2939    &\nodata & \nodata &     5.5 &   30.1 &    5.9 &    2.7 &   A  &\nodata &\nodata & \nodata &\nodata & -21.46  & -24.48 \nl
 01572$+$0009    &\nodata & 66      &     8.7 &    1.0 &   17.6 &    3.1 &   E  &  4.5   &\nodata & 20.0    &\nodata & -23.85  & -26.27 \nl
 02021$-$2103    &\nodata & \nodata &     7.1 &    1.9 &   13.5 &    1.4 &   E  &  7.0   &\nodata & 19.5    &\nodata &\nodata  & -25.67 \nl
 02411$+$0353    &\nodata & \nodata &    18.7 &   61.1 &   18.4 &   28.6 &   A  &\nodata &\nodata & \nodata &\nodata & -22.62  & -25.35 \nl
 02480$-$3745    &\nodata & \nodata &    16.7 &    8.6 &   10.6 &   11.6 &   A  &\nodata &\nodata & \nodata &\nodata & -21.15  & -24.51 \nl
 03209$-$0806    &\nodata & 62      &    29.7 &   15.9 &    0.7 &    1.1 &   A  &\nodata &\nodata & \nodata &\nodata & -22.22  & -24.41 \nl
 03250$+$1606    &\nodata & 64      &     2.3 &    0.1 &    7.8 &    2.9 &   E  &  5.1   &\nodata & 21.2    &\nodata & -21.98  & -24.91 \nl
 03521$+$0028    &\nodata & \nodata &    11.5 &   15.1 &    5.9 &   12.6 &   A  &\nodata &\nodata & \nodata &\nodata & -20.62  & -25.02 \nl
 04074$-$2801    &\nodata & \nodata &    45.0 &   23.9 &    0.9 &    3.4 &   A  &\nodata &\nodata & \nodata &\nodata & -21.94  & -24.69 \nl
 04103$-$2838    &\nodata & 64      &     9.3 &   19.8 &   25.7 &   10.3 &   A  &\nodata &\nodata & \nodata &\nodata & -22.05  & -24.71 \nl
 04313$-$1649    &\nodata & \nodata &     8.6 &    6.9 &    4.2 &    3.9 &  E/D &  6.2   &5.3     & 21.8    &19.0    & -22.01  & -24.54 \nl
 05020$-$2941    &\nodata & \nodata &     2.7 &    1.7 &    0.0 &    0.0 &  E/D &  5.8   &\nodata & 22.0    &\nodata & -21.35  & -24.14 \nl
 05024$-$1941    &\nodata & \nodata &     2.3 &    0.8 &    8.8 &    3.5 &  E/D &  4.7   &4.2     & 20.8    &17.2    & -22.42  & -25.68 \nl
 05156$-$3024    &\nodata & 61      &     2.4 &    6.5 &    7.8 &    3.2 &  E/D &  5.2   &\nodata & 21.1    &\nodata & -22.04  & -24.75 \nl
 05189$-$2524    &     47 &      74 &    25.9 &    5.9 &   51.0 &   13.6 &   E  &  4.3   &\nodata & 21.3    &\nodata & -21.48  & -24.72 \nl
 07599$+$6508    &\nodata & 46      &    28.2 &    4.2 &  142.1 &   18.4 &   E  &  3.5   &2.2     & 17.8    &13.5    & -24.76  & -27.85 \nl
 08201$+$2801    &\nodata & \nodata &     9.3 &    9.3 &    1.4 &    0.3 &   A  &\nodata &\nodata & \nodata &\nodata & -21.49  & -24.80 \nl
 08474$+$1813    &\nodata & 54      &    14.6 &   28.8 &    0.7 &    1.6 &  E/D &  4.1   &\nodata & 21.7    &\nodata & -20.44  & -23.30 \nl
 08559$+$1053    &\nodata & 55      &    10.3 &    2.8 &   20.3 &    2.7 &   E  &  6.2   &3.1     & 21.5    &16.4    & -22.51  & -25.87 \nl
 08572$+$3915    &\nodata & \nodata &     4.6 &   20.0 &   33.7 &   38.2 &   A  &\nodata &\nodata & \nodata &\nodata & -21.68  & -23.93 \nl
 08591$+$5248    &\nodata & 70      &     3.1 &    5.4 &    4.5 &    1.9 &  E/D &  4.1   &4.1     & 20.6    &17.7    & -21.83  & -25.05 \nl
 09039$+$0503    &\nodata & \nodata &    16.7 &   14.7 &   13.8 &    9.9 &   A  &\nodata &\nodata & \nodata &\nodata & -21.35  & -24.75 \nl
 09116$+$0334    &     64 &      64 &     4.5 &    2.3 &   13.9 &    2.9 &   E  &  4.1   &\nodata & 20.7    &\nodata & -22.11  & -26.16 \nl
 09463$+$8141    &\nodata & \nodata &     9.2 &   11.2 &    3.8 &    6.9 &   A  &\nodata &\nodata & \nodata &\nodata & -21.83  & -25.08 \nl
 09539$+$0857    &     26 &      61 &     1.1 &    3.1 &    0.7 &    1.3 &  E/D &  3.1   &\nodata & 20.9    &\nodata & -20.54  & -23.56 \nl
 10035$+$2740    &\nodata & \nodata &    12.4 &   12.1 &    5.4 &   17.2 &   A  &\nodata &\nodata & \nodata &\nodata & -22.15  & -25.43 \nl
 10091$+$4704    &\nodata & \nodata &    19.6 &    6.7 &    3.6 &    0.6 &   E  &  9.8   &\nodata & 22.2    &\nodata & -22.31  & -25.38 \nl
 10190$+$1322    &\nodata & \nodata &    11.2 &   21.1 &   13.0 &   45.7 &   A  &\nodata &\nodata & \nodata &\nodata & -21.79  & -25.45 \nl
 10378$+$1108    &\nodata & 76      &     2.8 &    5.5 &   21.9 &    7.9 &  E/D &  6.1   &\nodata & 21.8    &\nodata & -22.15  & -24.66 \nl
 10485$-$1447 E  &\nodata & \nodata &     3.9 &   11.2 &    2.0 &    1.4 &   A  &\nodata &\nodata & \nodata &\nodata & -20.53  & -22.72 \nl
 10485$-$1447 W  &\nodata & \nodata &     2.7 &    2.8 &    1.9 &    1.1 &  E/D &  5.4   &\nodata & 22.0    &\nodata & -20.91  & -23.90 \nl
 10494$+$4424    &\nodata & 33      &    11.3 &    1.6 &   11.3 &    7.6 &   E  &  4.2   &\nodata & 21.2    &\nodata & -21.23  & -24.38 \nl
 10594$+$3818    &\nodata & \nodata &    21.3 &    2.9 &    4.9 &    8.9 &   E  &  4.6   &\nodata & 20.5    &\nodata & -22.27  & -25.17 \nl
 11028$+$3130    &\nodata & \nodata &     0.8 &    2.1 &    3.0 &    2.7 &  E/D &  4.7   &4.2     & 21.7    &19.1    & -21.03  & -24.13 \nl
 11095$-$0238    &\nodata & \nodata &    32.7 &   11.3 &    2.2 &    6.3 &   A  &\nodata &\nodata & \nodata &\nodata & -21.03  & -24.19 \nl
 11119$+$3257    &\nodata & 37      &     2.9 &    0.6 &    5.5 &    1.1 &   E  &  2.6   &1.7     & 19.2    &14.0    & -22.71  & -27.14 \nl
 11130$-$2659    &\nodata & \nodata &    14.5 &    6.7 &    1.9 &    0.8 &   A  &\nodata &\nodata & \nodata &\nodata & -21.74  & -24.41 \nl
 11180$+$1623 E  &\nodata & \nodata &     5.9 &    0.6 &    3.5 &    4.4 &   E  &  4.6   &\nodata & 21.4    &\nodata & -21.45  & -24.65 \nl
 11223$-$1244    &\nodata & \nodata &    29.1 &   14.3 &   13.9 &   16.1 &   A  &\nodata &\nodata & \nodata &\nodata & -22.93  & -26.01 \nl
 11387$+$4116    &\nodata & 68      &     4.6 &   18.8 &    1.9 &    1.3 &  E/D &  3.5   &3.0     & 20.5    &17.4    & -21.52  & -24.66 \nl
 11506$+$1331    &\nodata & 59      &    11.3 &   16.4 &   11.4 &   17.4 &   A  &\nodata &\nodata & \nodata &\nodata & -21.41  & -24.37 \nl
 11582$+$3020    &\nodata & \nodata &     0.9 &    0.8 &    0.3 &    0.7 &  E/D &  7.1   &\nodata & 22.5    &\nodata & -21.53  & -24.88 \nl
 11598$-$0112    &\nodata & 69      &     2.8 &    8.2 &   15.6 &    8.4 &  E/D &  4.2   &3.0     & 20.3    &16.8    & -22.69  & -25.41 \nl
 12018$+$1941    &\nodata & \nodata &     3.1 &    0.9 &    1.0 &    0.7 &  E/D &  4.6   &\nodata & 21.1    &\nodata & -21.80  & -24.86 \nl
 12032$+$1707    &\nodata & \nodata &    11.2 &   24.9 &   15.0 &   13.1 &   A  &\nodata &\nodata & \nodata &\nodata & -22.22  & -25.79 \nl
 12072$-$0444    &\nodata & 55      &    23.7 &    7.2 &   19.8 &    3.4 &   E  &  6.3   &\nodata & 21.4    &\nodata & -22.30  & -24.92 \nl
 12112$+$0305    &\nodata & \nodata &     6.2 &    3.8 &   17.7 &   11.5 &   A  &\nodata &\nodata & \nodata &\nodata & -21.51  & -24.71 \nl
 12127$-$1412    &\nodata & \nodata &     4.1 &    2.2 &    3.0 &    0.5 &  E/D &  4.9   &\nodata & 21.5    &\nodata & -21.08  & -25.63 \nl
 12265$+$0219    &     68 &      53 &     7.9 &    3.2 &   28.8 &    2.6 &   E  &  2.5   &2.0     & 16.1    &12.9    & -25.22  & -28.41 \nl
 12359$-$0725 N  &\nodata & \nodata &     2.2 &    0.5 &    3.7 &    2.8 &  E/D &  3.4   &\nodata & 21.5    &\nodata & -20.56  & -23.87 \nl
 12359$-$0725 S  &\nodata & \nodata &    10.9 &    3.2 &    0.0 &    0.0 &   E  &  4.6   &\nodata & 22.3    &\nodata & -20.54  & -22.86 \nl
 12447$+$3721    &\nodata & \nodata &     5.2 &   12.5 &    0.7 &    0.5 &   A  &\nodata &\nodata & \nodata &\nodata & -21.00  & -23.90 \nl
 12540$+$5708    &     61 &      58 &    55.6 &    7.9 &   29.9 &    3.5 &   E  &  3.3   &\nodata & 19.8    &\nodata & -22.52  & -25.85 \nl
 13106$-$0922    &\nodata & \nodata &     0.8 &    5.3 &    2.0 &    2.5 &  E/D &  6.4   &\nodata & 21.4    &\nodata & -21.61  & -24.39 \nl
 13218$+$0552    &\nodata & \nodata &     5.3 &    0.7 &   12.6 &    0.7 &   E  &  5.4   &4.0     & 21.3    &15.5    & -22.07  & -27.35 \nl
 13305$-$1739    &\nodata & \nodata &    44.3 &    8.7 &    0.8 &    1.8 &   E  &  3.9   &\nodata & 19.4    &\nodata & -22.76  & -25.82 \nl
 13335$-$2612    &\nodata & \nodata &     0.5 &    0.1 &    5.9 &    7.8 &   A  &\nodata &\nodata & \nodata &\nodata & -21.74  & -24.80 \nl
 13342$+$3932    &\nodata & 44      &     7.8 &    1.5 &    4.3 &    3.6 &   E  &  6.0   &5.2     & 21.1    &17.9    & -22.96  & -26.43 \nl
 13428$+$5608    &     28 &      38 &    76.7 &   14.8 &    4.8 &    0.6 &   E  &  4.3   &\nodata & 20.4    &\nodata & -22.33  & -24.67 \nl
 13443$+$0802 NE &\nodata & \nodata &    14.7 &   36.1 &   16.8 &   27.5 &   A  &\nodata &\nodata & \nodata &\nodata & -22.51  & -25.93 \nl
 13443$+$0802 SW &\nodata & \nodata &     0.9 &    0.1 &    1.1 &    2.6 &  E/D &  4.3   &\nodata & 21.7    &\nodata & -21.16  & -23.64 \nl
 13451$+$1232    &\nodata & \nodata &    36.3 &   15.7 &    6.0 &   15.0 &   A  &\nodata &\nodata & \nodata &\nodata & -23.32  & -25.87 \nl
 13454$-$2956 N  &\nodata & \nodata &     1.2 &    0.6 &    1.0 &    0.1 &   E  &  6.9   &\nodata & 22.2    &\nodata & -21.26  & -25.18 \nl
 13454$-$2956 S  &\nodata & \nodata &     2.2 &    1.0 &    0.4 &    0.4 &  E/D &  6.8   &\nodata & 22.2    &\nodata & -21.23  & -25.26 \nl
 13469$+$5833    &\nodata & \nodata &    65.0 &   14.5 &    7.5 &   14.3 &   A  &\nodata &\nodata & \nodata &\nodata & -21.95  & -25.14 \nl
 13509$+$0442    &\nodata & 57      &    16.8 &   17.6 &    2.0 &    1.0 &   A  &\nodata &\nodata & \nodata &\nodata & -21.02  & -24.70 \nl
 13539$+$2920    &\nodata & \nodata &     7.4 &   18.8 &   11.1 &    8.0 &   A  &\nodata &\nodata & \nodata &\nodata & -21.47  & -25.02 \nl
 14053$-$1958    &\nodata & \nodata &    15.1 &    7.1 &    1.3 &    1.7 &  E/D &  6.3   &\nodata & 22.0    &\nodata & -21.39  & -24.64 \nl
 14060$+$2919    &\nodata & \nodata &     9.8 &   16.5 &    7.2 &    1.5 &  E/D &  5.7   &\nodata & 21.0    &\nodata & -21.98  & -25.21 \nl
 14070$+$0525    &\nodata & \nodata &     0.8 &    2.3 &    0.7 &    1.3 &  E/D &  6.1   &\nodata & 21.5    &\nodata & -22.25  & -25.78 \nl
 14121$-$0126    &\nodata & \nodata &    31.8 &   47.8 &   24.2 &   24.8 &   A  &\nodata &\nodata & \nodata &\nodata & -21.79  & -25.51 \nl
 14197$+$0813    &\nodata & 71      &     9.2 &   25.7 &    0.2 &    0.2 &   A  &\nodata &\nodata & \nodata &\nodata & -22.21  & -24.60 \nl
 14202$+$2615    &\nodata & \nodata &     9.7 &    5.1 &    5.2 &    2.4 &   A  &\nodata &\nodata & \nodata &\nodata & -22.31  & -25.80 \nl
 14252$-$1550    &\nodata & \nodata &     9.6 &   19.9 &   15.3 &   22.3 &   A  &\nodata &\nodata & \nodata &\nodata & -21.33  & -24.98 \nl
 14348$-$1447    &\nodata & \nodata &     5.7 &   14.5 &   10.7 &   12.9 &   A  &\nodata &\nodata & \nodata &\nodata & -22.36  & -25.34 \nl
 14394$+$5332 E  &     23 &      23 &    23.4 &    6.5 &    8.3 &    2.8 &   E  &  3.2   &\nodata & 20.1    &\nodata & -21.77  & -25.23 \nl
 14394$+$5332 W  &\nodata & \nodata &     3.4 &    6.4 &    1.8 &    3.1 &   D  &\nodata &\nodata & \nodata &\nodata & -21.25  & -23.91 \nl
 14485$-$2434    &\nodata & 73      &     9.7 &    2.1 &    2.9 &    1.1 &   E  &  5.0   &\nodata & 21.1    &\nodata & -21.46  & -24.45 \nl
 15001$+$1433    &\nodata & \nodata &    14.1 &   26.4 &   23.1 &    7.2 &   A  &\nodata &\nodata & \nodata &\nodata & -22.64  & -25.68 \nl
 15043$+$5754    &\nodata & \nodata &    17.1 &   60.0 &   10.5 &   23.4 &   A  &\nodata &\nodata & \nodata &\nodata & -21.21  & -25.67 \nl
 15130$-$1958    &\nodata & \nodata &    18.1 &    3.8 &    5.2 &    2.1 &   E  &  2.9   &\nodata & 20.6    &\nodata & -20.94  & -25.23 \nl
 15206$+$3342    &     48 &      53 &     4.6 &    1.1 &    6.5 &   14.9 &  E/D &  2.9   &\nodata & 19.9    &\nodata & -21.99  & -24.82 \nl
 15225$+$2350    &\nodata & 62      &    25.3 &    4.0 &    1.3 &    0.7 &   A  &\nodata &\nodata & \nodata &\nodata & -21.47  & -24.35 \nl
 15327$+$2340    &\nodata & \nodata &    10.8 &   23.6 &   11.3 &    3.1 &   E  &  4.9   &2.1     & 20.8    &17.1    & -21.39  & -24.41 \nl
 15462$-$0450    &     74 &      43 &     2.4 &    2.9 &    2.5 &    4.7 &  E/D &  4.7   &1.5     & 21.6    &15.8    & -20.96  & -25.04 \nl
 16090$-$0139    &\nodata & \nodata &     2.4 &    4.6 &    6.4 &    1.1 &  E/D &  5.2   &\nodata & 21.3    &\nodata & -21.95  & -25.25 \nl
 16156$+$0146    &\nodata & \nodata &    23.3 &   51.8 &   21.0 &   21.8 &   A  &\nodata &\nodata & \nodata &\nodata & -20.87  & -23.91 \nl
 16300$+$1558    &\nodata & \nodata &     5.2 &    1.0 &    1.9 &    8.5 &  E/D &  7.2   &5.8     & 21.5    &18.0    & -22.63  & -25.82 \nl
 16333$+$4630    &\nodata & \nodata &    16.2 &   22.1 &   40.2 &   38.9 &   A  &\nodata &\nodata & \nodata &\nodata & -22.11  & -25.42 \nl
 16468$+$5200    &\nodata & \nodata &    37.5 &   90.6 &    4.2 &    7.2 &   A  &\nodata &\nodata & \nodata &\nodata & -21.51  & -24.32 \nl
 16474$+$3430    &\nodata & \nodata &    68.8 &  114.6 &   10.2 &   19.2 &   A  &\nodata &\nodata & \nodata &\nodata & -22.07  & -25.34 \nl
 16487$+$5447    &\nodata & \nodata &    13.8 &   27.6 &   20.6 &   45.0 &   A  &\nodata &\nodata & \nodata &\nodata & -21.87  & -24.69 \nl
 17028$+$5817 W  &\nodata & \nodata &     3.2 &    2.1 &    7.9 &    1.5 &  E/D &  5.3   &\nodata & 22.0    &\nodata & -21.07  & -24.63 \nl
 17028$+$5817 E  &     31 &      31 &     4.3 &    1.5 &    2.3 &    8.2 &  E/D &  3.1   &\nodata & 21.1    &\nodata & -20.58  & -23.82 \nl
 17044$+$6720    &\nodata & 70      &     0.8 &    4.2 &    5.5 &    2.1 &  E/D &  5.2   &\nodata & 21.8    &\nodata & -21.81  & -24.23 \nl
 17068$+$4027 E  &\nodata & \nodata &    15.7 &    9.5 &    2.9 &    7.7 &   A  &\nodata &\nodata & \nodata &\nodata & -21.47  & -24.80 \nl
 17179$+$5444    &\nodata & 65      &    12.4 &    5.4 &    4.0 &    0.7 &  E/D &  5.4   &\nodata & 21.2    &\nodata & -22.35  & -25.05 \nl
 20414$-$1651    &     13 &      64 &    11.4 &    3.7 &    0.1 &    1.8 &  E/D &  3.6   &\nodata & 21.0    &\nodata & -21.01  & -23.91 \nl
 21208$-$0519    &\nodata & \nodata &    34.9 &   78.7 &   79.2 &  130.0 &   A  &\nodata &\nodata & \nodata &\nodata & -22.55  & -25.65 \nl
 21219$-$1757    &     91 &      63 &    18.0 &    2.8 &   14.9 &    0.5 &   E  &  4.1   &\nodata &20.4     &\nodata & -22.17  & -26.06 \nl
 21329$-$2346    &\nodata & \nodata &     4.6 &    1.8 &    3.6 &    1.6 &   E  &  6.7   &\nodata &22.5     &\nodata & -21.66  & -24.52 \nl
 21477$+$0502 E  &\nodata & \nodata &    15.7 &   15.6 &    5.3 &   14.6 &   A  &\nodata &\nodata & \nodata &\nodata & -21.86  & -24.94 \nl
 21477$+$0502 W  &\nodata & \nodata &     2.6 &    1.2 &    3.6 &    5.3 &  E/D &  4.6   &\nodata &23.1     &\nodata & -20.37  & -23.42 \nl
 22088$-$1831    &\nodata & \nodata &    16.1 &    1.9 &    2.2 &    3.8 &   A  &\nodata &\nodata & \nodata &\nodata & -22.18  & -25.21 \nl
 22206$-$2715    &\nodata & \nodata &    32.9 &   77.9 &   25.7 &   33.1 &   A  &\nodata &\nodata & \nodata &\nodata & -22.18  & -25.21 \nl
 22491$-$1808    &\nodata & \nodata &    21.6 &   26.4 &    5.3 &    1.1 &   A  &\nodata &\nodata & \nodata &\nodata & -21.90  & -24.56 \nl
 22541$+$0833 NW &\nodata & \nodata &     9.9 &    2.5 &    4.9 &    1.2 &   E  &  3.3   &2.4     &21.0     &16.7    & -21.44  & -24.96 \nl
 22541$+$0833 SE &     60 &      60 &     6.1 &    2.3 &    2.2 &    3.2 &   E  &  3.3   &\nodata &21.5     &\nodata & -20.34  & -23.56 \nl
 23060$+$0505    &\nodata & 81      &     1.8 &    2.1 &    9.8 &    0.1 &   E  &  6.2   &4.6     &21.0     &16.8    & -22.92  & -25.29 \nl
 23129$+$2548    &\nodata & \nodata &     2.8 &    9.6 &    0.9 &    1.0 &  E/D &  4.3   &\nodata &21.5     &\nodata & -21.63  & -24.68 \nl
 23233$+$2817    &\nodata & 76      &     4.1 &   12.1 &    2.7 &    8.1 &   D  &\nodata &\nodata & \nodata &\nodata & -22.02  & -24.89 \nl
 23234$+$0946    &\nodata & \nodata &    31.4 &   96.0 &   16.9 &   14.2 &   A  &\nodata &\nodata & \nodata &\nodata & -21.84  & -25.25 \nl
 23327$+$2913 S  &\nodata & \nodata &     4.4 &    3.4 &    2.5 &    3.4 &  E/D &  4.9   &\nodata &21.5     &\nodata & -21.71  & -24.60 \nl
 23327$+$2913 N  &\nodata & \nodata &    12.5 &    3.7 &    3.9 &    8.3 &   E  &  5.9   &\nodata &22.5     &\nodata & -21.35  & -24.08 \nl
 23389$+$0300    &\nodata & \nodata &    15.7 &   37.9 &    8.7 &   21.4 &   A  &\nodata &\nodata & \nodata &\nodata & -20.99  & -24.50 \nl
 23498$+$2423    &\nodata & \nodata &    13.7 &    9.2 &   36.4 &   21.2 &   A  &\nodata &\nodata & \nodata &\nodata & -22.52  & -26.65 \nl
\tablenotetext{} {{\it Col 1:}\ Object name.\ See text for details.}
\tablenotetext{} {{\it Col 2:}\ R-band \% of PSF relative to the galaxy.}
\tablenotetext{} {{\it Col 3:}\ $K^\prime$-band \% of PSF relative to the galaxy.}
\tablenotetext{} {{\it Col 4:}\ R-band reduced $\chi^2$ value fitted to the exponential profile.}
\tablenotetext{} {{\it Col 5:}\ R-band reduced $\chi^2$ value fitted to the de Vaucouleurs's $r^{1/4}$ law profile.}
\tablenotetext{} {{\it Col 6:}\ $K^\prime$-band reduced $\chi^2$ value fitted to the exponential profile.}
\tablenotetext{} {{\it Col 7:}\ $K^\prime$-band reduced $\chi^2$ value fitted to the de Vaucouleurs's $r^{1/4}$ law profile.}
\tablenotetext{} {{\it Col 8:}\ Best fitted profile (A: ambiguous, E: elliptical-like, D:disk-like, E/D: border-line object between E and D.)}
\tablenotetext{} {{\it Col 9:}\ R-band seeing-deconvolved half-light (effective) radius $r_e$ for E and E/D objects.}
\tablenotetext{} {{\it Col 10:}\ $K^\prime$-band seeing-deconvolved half-light (effective) radius $r_e$ for E and E/D objects.}
\tablenotetext{} {{\it Col 11:}\ R-band effective surface brightness $\mu_e$,
not corrected for cosmological dimming. }
\tablenotetext{} {{\it Col 12:}\ $K^\prime$-band effective surface brightness $\mu_e$, not corrected for cosmological dimming.}
\tablenotetext{} {{\it Col 13:}\ Absolute R magnitude for the host galaxy.}
\tablenotetext{} {{\it Col 14:}\ Absolute $K^\prime$ magnitude for the host galaxy.}
\enddata
\end{deluxetable}